\documentclass[iicol,pdflatex,sn-mathphys-num]{sn-jnl}% Math and Physical Sciences Numbered Reference Style 
\usepackage{braket}
\usepackage{units}

\usepackage{bm}
\usepackage{color}
\usepackage{simplewick}
\usepackage[version=3]{mhchem}
\usepackage{wasysym}
\usepackage{physics}
\usepackage{cleveref}
\usepackage{tikz}
\usetikzlibrary{positioning}

\usepackage{minitoc}

\usepackage{graphicx}%
\usepackage{multirow}%
\usepackage{amsmath,amssymb,amsfonts}%
\usepackage{amsthm}%
\usepackage{mathrsfs}%
\usepackage[title]{appendix}%
\usepackage{xcolor}%
\usepackage{textcomp}%
\usepackage{manyfoot}%
\usepackage{booktabs}%
\usepackage[normalem]{ulem}
\newcommand{\stkout}[1]{\ifmmode\text{\sout{\ensuremath{#1}}}\else\sout{#1}\fi}

\DeclareMathOperator{\sign}{sign}

\newcommand{\wrot}{\omega_{\text{rot}}}

\newcommand{\wtilde}{\tilde{\omega}}

\renewcommand{\vec}[1]{{\boldsymbol #1}}

\newcommand{\NEW}[1]{{#1}}
\newcommand{\NEWW}[1]{{#1}}%{{\color{blue}#1}}
\newcommand{\NEWWW}[1]{{#1}}%{{\color{blue}#1}}

\newcommand{\SignM}[1]{{\color{red}#1}}

\crefname{figure}{Fig.}{Fig.}
\Crefname{figure}{Fig.}{Fig.}
\crefname{equation}{Eq.}{Eq.}
\Crefname{equation}{Eq.}{Eq.}
\crefname{section}{Sec.}{Sec.}
\Crefname{section}{Sec.}{Sec.}
\crefname{appendix}{App.}{App.}
\Crefname{appendix}{App.}{App.}

\begin{document}

\title{\NEWW{Propelling Ferrimagnetic Domain Walls by Dynamical Frustration}}
%\title{Active Magnetic Matter:  Propelling Ferrimagnetic Domain Walls by Dynamical Frustration}

%\title{Propelling Ferrimagnetic Domain Walls by Dynamical Frustration}
%\title{A Route to Active Magnetic Matter via Self-Propelled Domain Walls in a Driven Ferrimagnet}

%\title{Creating Active Magnetic Matter in a Ferrimagnet by Dynamical Frustration}

%\title{Creating Active Magnetic Matter using Self-Propelled Domain Walls in a Driven Ferrimagnet}


\author*[1]{\fnm{Dennis} \sur{Hardt}}\email{hardt@thp.uni-koeln.de}

\author[1]{\fnm{Reza} \sur{Doostani}}%\email{iiauthor@gmail.com}

\author[1]{\fnm{Sebastian} \sur{Diehl}}%\email{iiiauthor@gmail.com}

\author[1,2]{\fnm{Nina} \spfx{del} \sur{Ser}}%\email{iiiauthor@gmail.com}

\author[1]{\fnm{Achim} \sur{Rosch}}
%\email{iiiauthor@gmail.com}

\affil[1]{\orgdiv{Institute for Theoretical Physics}, \orgname{University of Cologne}, \orgaddress{\city{Cologne}, \postcode{50937}, \country{Germany}}}

\affil[2]{\orgdiv{Division of Physics, Mathematics and Astronomy}, \orgname{California Institute of Technology}, \orgaddress{\street{1200 E California Blvd}, \city{Pasadena}, \postcode{91125}, \state{CA}, \country{USA}}}

% PACS numbers have now been replaced by PhySH
%..///

% ---------------------------------------------------------------------------------

\abstract{\NEWW{Many-particle systems driven out of thermal equilibrium  can show properties qualitatively different from any thermal state.
Here, we study a ferrimagnet in  a weak oscillating magnetic field. In this model, domain walls are not static, but are shown to move actively in a direction chosen by spontaneous symmetry breaking. Thus they act like self-propelling units. Their collective behaviour is reminiscent of other systems with actively moving units studied in the field of `active matter', where, e.g., flocks of birds are investigated.
The active motion of the domain walls emerges from `dynamical frustration'. The antiferromagnetic $xy$-order rotates clockwise or anticlockwise, determined by the sign of the ferromagnetic component. This necessarily leads to frustration at a domain wall, which gets resolved by propelling the domain wall with a velocity proportional to the square root of the driving power across large parameter regimes. This motion and strong hydrodynamic interactions lead to a linear growth of the magnetic correlation length over time, much faster than in equilibrium.  The dynamical frustration furthermore makes the system highly resilient to noise. The correlation length of the weakly driven one-dimensional system can be orders of magnitude larger than in the corresponding equilibrium system with the same noise level. 
}
}

%possible keywords for nature
%\keywords{keyword1, Keyword2, Keyword3, Keyword4}

\maketitle
\NEWW{Realising new states of matter out of thermal equilibrium in a solid is a challenging task, as solids tend to relax rapidly towards equilibrium. However, by laser or microwave irradiation it has been possible to realise, e.g., Bose-Einstein condensates (BEC) of excitons, polaritons, or magnons \cite{excitonPolariton2002,BoseEinstein_pumping,BECexciton_review2022}. These condensates spontaneously break time-translation invariance and realise novel non-equilibrium phase transitions \cite{BECexciton_review2022,BECKPZDiehl}. Nevertheless, the differences to equilibrium systems remain relatively subtle in such systems.

The goal of this paper is to show that a simple model of a driven magnet shows a qualitatively different type of non-equilibrium dynamics. 
We consider a ferrimagnet that hosts both antiferromagnetic (AFM) order in the $xy$-plane and ferromagnetic (FM) order in the $z$-direction \cite{Kim2022,ZelleUniversalPhaseTransition}.  This magnet has a continuous spin-rotation symmetry. We drive it out of thermal equilibrium by, e.g., external radiation.
Even for weak perturbations, the domain walls in this system obtain a finite velocity and actively move through the system. Importantly, the direction of motion is not imprinted by the setup, but arises from spontaneous symmetry breaking. These moving domain walls strongly interact with each other by hydrodynamics interactions.

The dynamics of actively moving, self propelled units has been intensively investigated in the field of 
active matter } \cite{reviewActive,reviewDryActive,reviewOfreviews,trafficactive,BacteriaActiveMatter,FlocksTonerTuRamaswamy,VicsekModel,%Weber_2019,CytoskeletonActiveMatter,
Elgeti_2015,ArtificialActiveMatter,lightDrivenReview,Wysocki_2016,spinners1,spinners2,Tonerlongrangeorder2D,Fruchart2021}, which studies systems very different from solids,  such as 
cars on a busy street~\cite{trafficactive}, self-propelling bacteria \cite{BacteriaActiveMatter}, flocks or herds of animals~\cite{FlocksTonerTuRamaswamy,VicsekModel}, %systems with active chemical reactions \cite{Weber_2019}, moving components of the cytoskeleton of living cells~\cite{CytoskeletonActiveMatter} 
or artificial self-propelled particles (synthetic systems) \cite{ArtificialActiveMatter,lightDrivenReview,Elgeti_2015,Wysocki_2016}. They have in common that the constituents locally use energy to induce, e.g., motion. Thus, detailed balance is broken on a local level, and systems of this kind are intrinsically far from equilibrium.
Such active matter systems show collective behaviours with no equilibrium counterparts.
One familiar example is a traffic jam. Another concerns the synchronisation of the orientation of birds flying in large flocks. 
A model introduced by Vicsek et al. in 1995 \cite{VicsekModel} to describe flocking birds has been argued to exhibit spontaneous breaking of rotational symmetry in a two-dimensional system with noise, which is ruled out by the Mermin-Wagner theorem in equilibrium systems \cite{Tonerlongrangeorder2D}. Furthermore, new types of phase transitions emerge in such systems \cite{Fruchart2021,ZelleUniversalPhaseTransition,daviet2023nonequilibrium}.

\begin{figure}
    \centering
           \begin{tikzpicture}
    \draw (0, 0) node[inner sep=0] (a) {  \includegraphics[width=0.9\linewidth]{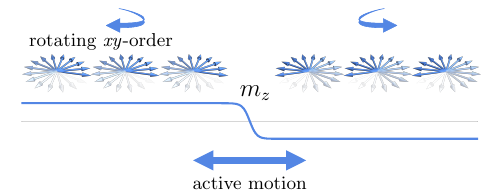}};
    \node[
      above left=-0.5cm and -0.35cm of a] {a)};
    \end{tikzpicture}
    \begin{tikzpicture}
    \draw (0, 0) node[inner sep=0] (a) {\includegraphics[height=0.405\linewidth]
    {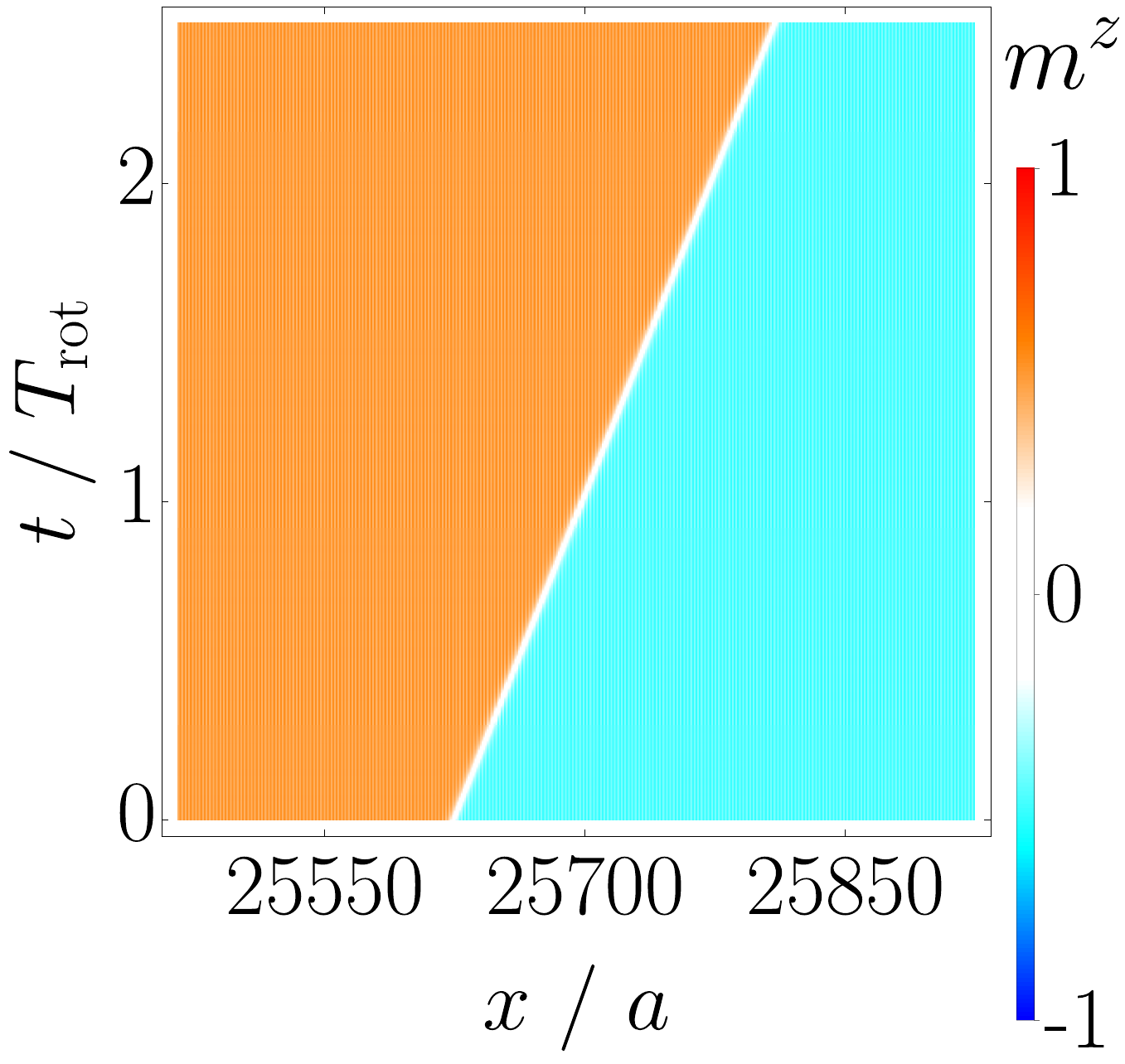}};
    \node[
      above left=-0.5cm and -0.35cm of a] {b)};
    \end{tikzpicture}
    \begin{tikzpicture}
    \draw (0, 0) node[inner sep=0] (b) {\includegraphics[height=0.405\linewidth]{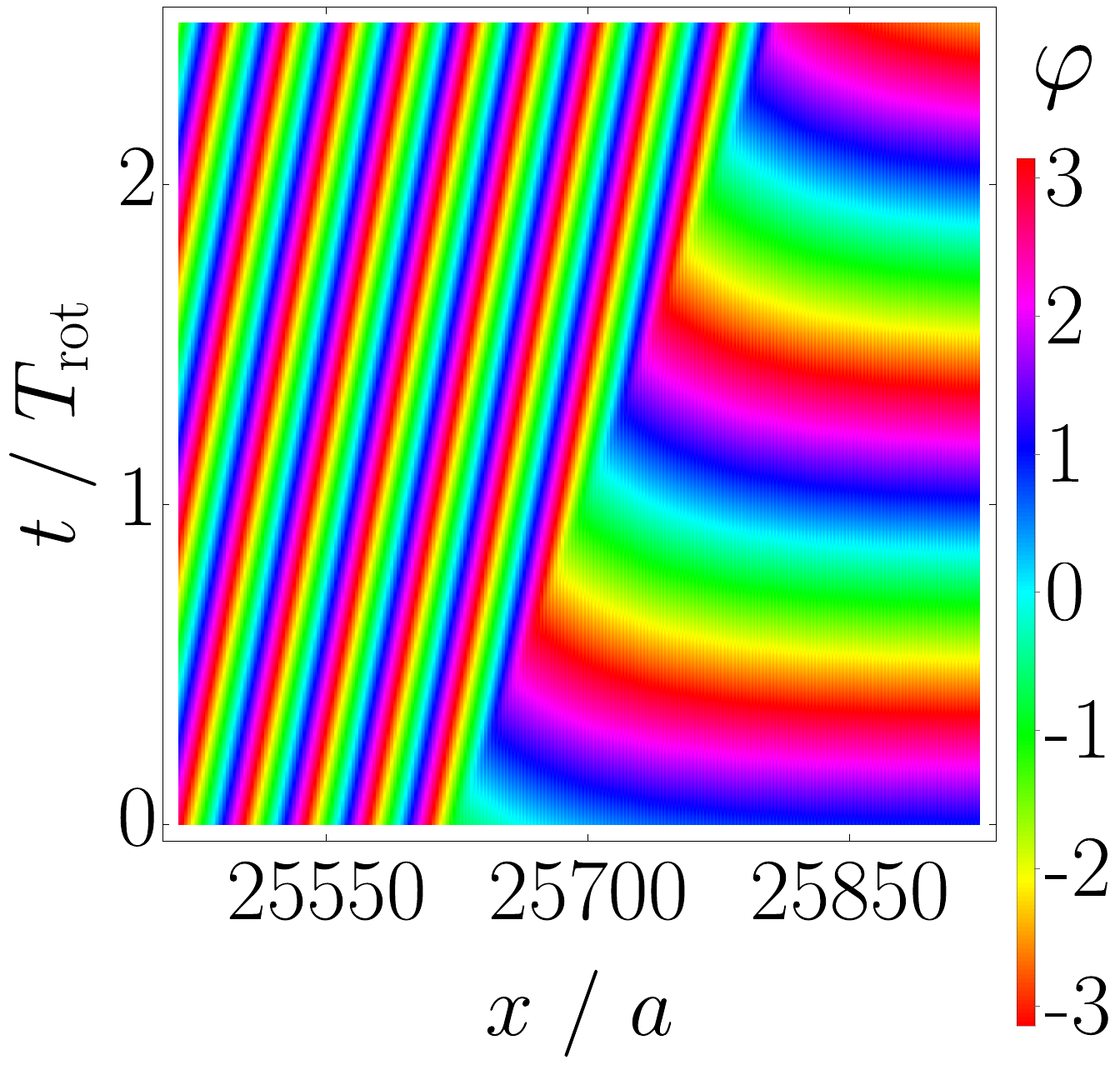}};
    \node[
      above left=-0.5cm and -0.35cm of b] {c)};
    \end{tikzpicture}
    \caption{Moving domain wall. a) Schematic picture of a domain wall in a driven ferrimagnet. The (staggered) in-plane magnetisation rotates in opposite directions on the left and right side, leading to dynamical frustration. This can lead to  an active motion of the domain wall either to the left or right. Data from a numerical simulation of a right-moving domain wall are displayed in panels b) and c). b) shows the magnetisation $m^z(x,t)$, while c) shows the phase $\varphi(x,t)$, describing the orientation of the staggered in-plane magnetisation. 
    A substantial phase gradient  $\partial_x \varphi$ is built up behind the moving domain. 
    Parameters: $J=1, \Delta=0.8,\delta_2=-0.6,\delta_4=1,g_1=1,g_2=0.1,\alpha=0.1,B_0=0.15,\omega=3.6$ resulting in a rotation period $T_\text{rot}\approx 612$, used to scale the vertical axis for a system of 40,000 spins.
     }
    \label{fig:movingDW}
\end{figure}
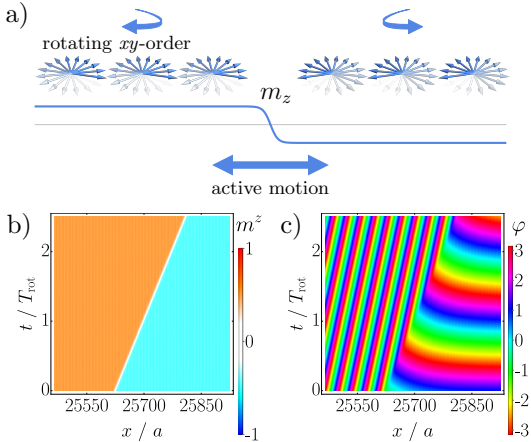
\NEWW{Several recent theoretical studies have investigated the link between solid-state physics and active matter}. Lindblad operators \cite{khasseh2023active,zheng2024mimicking}, non-Hermitian Hamilton operators \cite{nonHerm22,nonUnitaryQW23}  or `active spins' with unusual dynamics \cite{activeSpin,activeSpin2} have been used to obtain quantum analogs of active matter and active matter coupled to magnetic order. 

Here we use a different route towards realising active magnetic matter. Our model starts from a realistic description of a driven ferrimagnet, which \NEWW{simultaneously may show properties reminiscent of active matter. Thereby, we make use of two general principles: `activation of Goldstone modes' \cite{Nina_screw,Nina_jellyfish} and a mechanism which we call `dynamical frustration'. }
Goldstone modes have the advantage that they can be `activated' by arbitrarily small perturbations.
%As a concrete example, we consider  
% a ferrimagnet, hosting both antiferromagnetic (AFM) order in the $xy$-plane and ferromagnetic (FM) order in the $z$-direction \cite{Kim2022,ZelleUniversalPhaseTransition}.  This magnet with  a continuous spin-rotation symmetry is weakly driven  out of thermal equilibrium by, e.g., external radiation. 
This can be modelled by an equation for the angle $\varphi$, which describes the orientation of the staggered in-plane magnetisation of our ferrimagnet. 
Symmetries allow us to consider the following term in the equation of motion of $\varphi$,
\begin{align}
\partial_t \varphi = - \gamma \,m^z \label{eq:activateG},
\end{align}
where $m^z$ is the FM magnetisation in the $z$-direction and $\gamma$ is a prefactor. In thermal equilibrium, $\gamma=0$ and $\varphi=\text{const.}$, which follows from the fact that a perpetuum mobile is forbidden. This can also be formally derived from symmetries of the Keldysh action encoding thermal equilibrium \cite{Sieberer15equil}. Out of equilibrium, however, there is no reason why $\gamma$ should vanish, and indeed one finds that $\gamma$ grows linearly in the power used to drive the system out of equilibrium \cite{Nina_screw,ZelleUniversalPhaseTransition}. Thus, the Goldstone mode of the ferrimagnet is activated, $\varphi=\gamma m^z t+\varphi_0$. A detailed theory of the activation of Goldstone modes in helimagnets and single skyrmions is presented in Refs.~\cite{Nina_screw,Nina_jellyfish}. In Ref.~\cite{ZelleUniversalPhaseTransition}, it was shown that the coupling $\gamma$ in Eq.~\eqref{eq:activateG} drastically changes the phase transition from an $xy$-AFM phase into the ferrimagnetic phase by rendering it first-order. Experimentally, rotations of out-of-equilibrium magnets are, for example, routinely observed in so-called spin-torque oscillators~\cite{spintorque_chen,spintorque_Manchon}. Ref. \cite{SkyrmionGoldstoneExp} discusses how a rotation of a skyrmion lattice can be induced by laser pulses. Furthermore, Bose-Einstein condensates of magnons \cite{BoseEinstein_pumping} also lead to oscillating and rotating magnetisation patterns. \NEW{Models of self-rotating particles with hydrodynamic interactions have also been intensively studied \cite{spinners1,spinners2} in the context of biological systems and models for synchronisation. An important difference here is that in our systems the spins are localised. \NEWW{The rotation in time of the local magnetisation shows that detailed balance is broken on a local level, a necessary (but not sufficient) condition to obtain `active matter'.}  }

\NEWW{To obtain self-propelled domain walls}, we need a second concept: `dynamical frustration', which occurs in our system at domain walls in magnetisation $m_z$, see Fig.~\ref{fig:movingDW}a).
On the left and right sides of the domain wall, the (staggered) $xy$-order rotates in opposite directions, according to Eq.~\eqref{eq:activateG}. `Dynamical frustration' describes that the in-plane magnetisation directions periodically point in opposite 
directions, costing a large amount of energy. We will show that for a {\em static} domain wall, due to this frustration no smooth, singularity-free interpolation of the angle $\varphi(x,t)$ as a function of space and time
 exists across the domain wall. However, \NEW{ this frustration can be lifted if the domain wall is \emph{moving}}. We will indeed show that, across wide parameter regimes, we obtain an active motion of the domain wall. \NEW{Importantly,} the direction of motion, left or right, arises from spontaneous symmetry breaking, making it similar to self-driven particles. 
 \NEW{The domain wall thus behaves as if having a `motor' attached to it which can point in two different directions, reminiscent of actively moving particles. This makes our system very different from a significant number of previously studied systems where a {\em directed} motion of domain walls is induced by (spin-) currents, laser pulses or static or oscillating fields, see, e.g., Refs.~\cite{spintorque_Manchon,Tserkovnyak1,Tserkovnyak2,Kwon21RotDW,domainWallKyungJin}.}

 \NEWW{Another example of dynamical frustration   arises when lattices of magnetic skyrmions \cite{skyrmion0,rotatingSkyrmion0} are imaged by an electron microscope \cite{seki}. Here, the skyrmion lattice starts to rotate \cite{rotatingSkyrmion1} due to torques induced by the imaging process and breaks into domains \cite{rotatingSkyrmion2}. In this case, the relative rotation of the domains is a source of dynamical frustration that induces long-range forces and a collective motion of dislocations, which are studied both experimentally and theoretically in Ref.~\cite{rotatingSkyrmion2}.  }

\bmhead{Results}\ \\
{\em Model -- }
We consider a 1D classical spin model, $|\bm{S}|=1$, with the following Hamiltonian
\begin{align}        H=&J\sum_{i} \left(S^x_iS^x_{i+1}+S^y_iS^y_{i+1}-\Delta S^z_i S^z_{i+1}  \right)\nonumber \\
        &+\sum_i \left(\frac{\delta_2}{2} \left. S^z_i \right.^2 +\frac{\delta_4}{4} \left. S^z_i \right.^4 -g_i B_z(t) S^z_i\right),\label{eq:discreteFerrimagnetHamiltonian}
\end{align}
where $J,\Delta>0$ impose AFM couplings in the $x$- and $y$-directions, but FM interactions for the $z$-component. 
\NEW{Our main results apply also to easy-cone ferromagnets obtained for $J<0, \Delta >0$ as long as long-ranged dipolar interaction can be neglected.}
$\delta_2$ and $\delta_4>0$ introduce a uniaxial anisotropy. For $\delta_2<2J(\Delta-1)$ and $B_z=0$, one obtains a ferrimagnetic ground state with a constant $S_i^z=\pm m^z_0$, where $m^z_0=\sqrt{\frac{-\delta_2-2J(1-\Delta)}{\delta_4}}\geq 0$. The $g$-factors $g_i$ are chosen differently for even and odd sites. Such staggered $g$ factors naturally occur in magnets with, e.g., a screw axis \cite{Affleck99} and are used here to avoid that $B_z(t)$ couples to the conserved magnetisation only.
The dynamics of the system is described by the stochastic Landau-Lifshitz-Gilbert equation, which includes a phenomenological damping term $\alpha$ and a corresponding noise term proportional to the temperature $T$, see methods. We first  study the noiseless case, $T=0$.

To drive the system out of thermal equilibrium, we consider the effect of a rapidly oscillating magnetic field, $B_z(t)=B_0 \cos(\omega t)$, $\omega \sim J$. In the ferrimagnetic phase, $0<|m_0^z|<1$, this oscillation `activates' the Goldstone mode. The staggered $xy$-order, described by the angle $\varphi$, begins to precess collectively, 
\begin{equation}\label{eq:phiLin}
    \varphi(t)= - \wrot  t+\varphi_0, 
    \quad \wrot \approx \gamma m^z, \quad \gamma\propto B_0^2,
\end{equation} 
where the formulas are valid for small $B_0$ and a small FM magnetisation. The fields $m^z$ and $\varphi$ are defined by  $\langle S_i^z\rangle=m^z(x)$, $\langle (S_i^x,S_i^y)\rangle = (-1)^i \sqrt{1-(m^z)^2} (\cos(\varphi(x)),\sin(\varphi(x))$. Here, $x$ is the coarse-grained position variable, with $x=i a$ at the position of spin ${\bf S}_i$. $\langle \dots \rangle$ denotes an average over the oscillation period of the external field (in plots, however, for simplicity's sake we show stroboscopic images instead, see methods).  
An analytical calculation of  $\wrot$ using second order perturbation theory in $B_0$ for arbitrary $m^z$ is given in the supplementary material, App. \ref{app:GostoneModeActivation}. The precise way in which non-equilibrium is implemented is expected to be irrelevant as long as the $xy$-symmetry remains intact. In an experimental system one could, e.g., also use a laser to create electronic excitations. In this case, no staggered $g$-factors are needed and one expects that $\gamma$ is proportional to the power of the laser.

{\em Moving domain walls -- } \label{sec:antiferrimagnet}
In our system, the $xy$-magnetisation rotates in the clockwise or anticlockwise direction when the FM magnetisation points up or down, respectively. As argued in the introduction, this leads to dynamical frustration at the centre of the domain wall where $m^z$ changes sign. 
Fig. \ref{fig:movingDW}(b,c) shows the result of a numerical simulation of such a domain wall. Depending on the initial conditions, the domain wall moves either to the right or to the left with a constant velocity. Far from the domain wall boundary, the angle $\varphi$ develops a finite slope which builds up behind the moving domain wall.

\begin{figure}
    \centering
    \includegraphics[width=0.95 \linewidth]{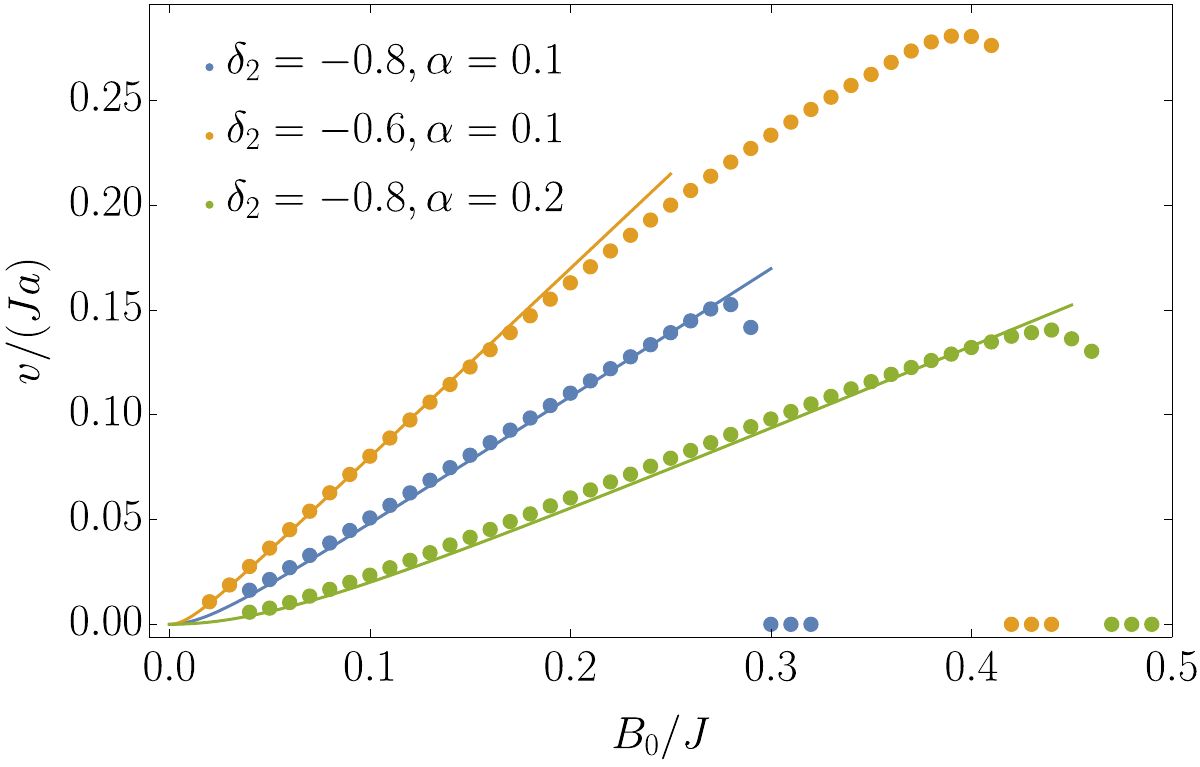}
    \caption{Velocity $v$ of a domain wall as function of the amplitude of the oscillating field $B_0$, for different values of anisotropy $\delta_2$ and damping $\alpha$. Across a wide parameter range, $v$ is linear in $B_0$ and thus proportional to $\sqrt{|\wrot|}$, see Eq.~\eqref{eq:vFast}. Lines: analytical calculation of the domain wall velocity, Eq.~\eqref{eq:velocityFull}, without fitting parameters. Parameters: $J=1, \Delta=0.8,\delta_2=-0.6$ and $-0.8$, $\delta_4=1,g_1=1,g_2=0.1,\alpha=0.1 $ and $0.2,\omega=3.6$ for a system of 40,000 spins. Numerical errors are smaller than the size of the symbols.}
    \label{fig:velicityDW}
\end{figure}

To develop an analytical theory for the velocity of the moving domain wall, we first consider the limit of vanishing damping, $\alpha \to 0$. \NEW{The following calculation uses the interplay of spin (super) currents and domain wall motion as studied previously by Kim and Tserkovnyak~\cite{Tserkovnyak1}, see also Refs.~\cite{Tserkovnyak2,domainWallKyungJin}.}
For a  domain wall moving with the velocity $v$, we make the ansatz
\begin{equation}\label{eq:ansatz1}
    \begin{aligned}
    m^z(x,t)&=m^z(x-v t),\\
    \varphi(x,t)&=\Phi(x - v t) + \tilde \omega t,        
    \end{aligned}
\end{equation}
see App.~\ref{appD:subsecMovingDomainWallAnsatz}  for details. \NEW{This ansatz captures the magnetisation profile shown in Figs.~\ref{fig:movingDW}b) and c), obtained in the long-time limit. The term $\tilde \omega t$ is needed to capture the asymmetry between the left and the right side. Note that both the direction of $v$ and the sign of the phase gradients arise from spontaneous symmetry breaking by tiny asymmetries in the initial condition.} We need only one extra input to solve the steady-state problem: 
for vanishing damping, the magnetisation $m^z$ is conserved,
$\partial_t m^z +\partial_x j^z =0$ and the spin (super) current is proportional to the gradient of the phase \cite{spincurrent_Halperlin_Hohenberg,superfluid_sonin,Tserkovnyak1,Tserkovnyak2}, \begin{align}    j^z= \rho_s \partial_x \varphi\label{eq:spincurrent},
\end{align}
where $\rho_s \approx J a^2 (1-(m_0^z)^2)$ is the spin stiffness and $a$ the lattice constant.
Substituting in Eq.~\eqref{eq:ansatz1}, and integrating the continuity equation over space from $-x_0$ to $x_0$, one obtains
\begin{align}\label{eq:boundary}
    -v \, m^z \Bigr|_{-x_0}^{x_0}+\rho_s \partial_x \Phi\Bigr|_{-x_0}^{x_0}=0.
\end{align}
\begin{figure*}[!htb]
    \centering    \includegraphics[width=0.95\linewidth]{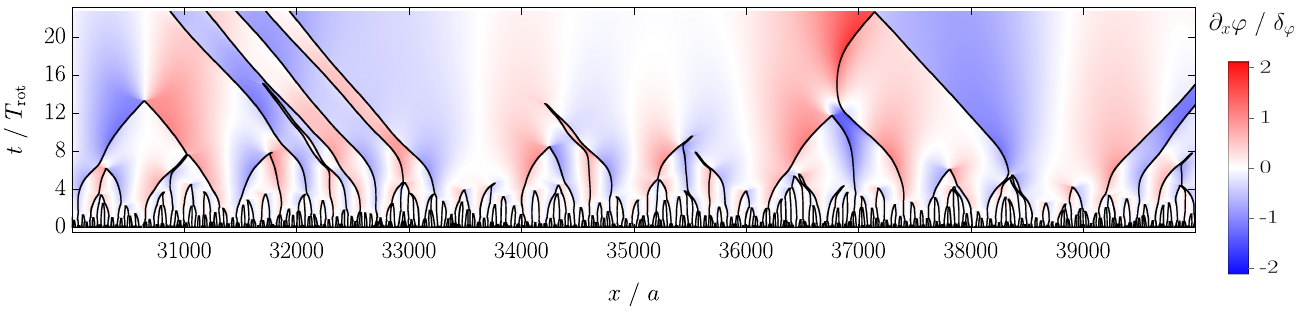}
    \caption{Worldlines of domain walls after a quench from the AFM into the ferrimagnetic phase driven by a small oscillating field $B_0=0.15\, J$. Worldlines end either when a left-moving and a right-moving domain collide or when the distance between two domain walls  shrinks to zero.
    Colour: amplitude of $\partial_x \varphi$ in units of $\delta_\varphi$, given by the difference of phase gradients  $\delta_\varphi=\partial_x \varphi^\text{left}-\partial_x \varphi^\text{right}\approx0.176$ across  a single domain wall, see Fig.~\ref{fig:movingDW}(b,c). The motion of domain walls is subject to a long-range hydrodynamic interaction mediated by $\partial_x \varphi$. Parameters: same as in Fig.~\ref{fig:movingDW}, initial state: ordered $xy$-AFM with a small random $S^z_i=\pm 0.1$, see methods.
     }
    \label{fig:worldlines}
\end{figure*}
Far from the domain wall, the magnetisation takes its bulk value $m^z=\pm m^z_0$  and the spins rotate with 
$
\partial_t \varphi = -v \partial_x \Phi+\tilde \omega = \mp |\wrot|$, Eq.~\eqref{eq:phiLin}.
Therefore, we obtain  from Eq. \eqref{eq:boundary} directly $-2 v m^z_0+2 \rho_s \frac{|\wrot|}{v}=0$ or, equivalently,
\begin{align}
    v=\pm \sqrt{ \frac{\rho_s|\wrot|}{m^z_0}}\approx \pm \sqrt{\rho_s \gamma} \quad \text{for } \alpha \to 0, \label{eq:vFast}
\end{align}
\NEW{where the sign of the velocity is determined by the sign of the jump in the gradients, $\partial_x \Phi(x_0)-\partial_x \Phi(-x_0)$.
Eq.~\eqref{eq:vFast}} is a noteworthy result for four reasons: (i) It has been derived  without any knowledge of the shape of the domain wall. All microscopic parameters entering the equation are known analytically for small $B_0$. (ii) The direction of motion has to be chosen by spontaneous symmetry breaking and is, for example, determined by small fluctuations of $\varphi$ in the initial state. 
(iii) As $\wrot$ is linear in $m^z_0$, the velocity of domain walls remains finite even for $m^z_0 \to 0$, i.e., when the phase transition from an $xy$-ordered phase to the ferrimagnetic phase is approached. Indeed, in a previous analysis \cite{ZelleUniversalPhaseTransition}, we found that this velocity characterises the field theory of the critical point in higher dimensions. 
Finally, (iv), the velocity is \emph{very} fast. $v$ is proportional to the square root of $\wrot$, making it linear in the amplitude $B_0$ of the oscillating field or, equivalently, proportional to the square root of the power used to drive our system out of thermal equilibrium. For weak $B_0$, this makes $v$ much larger than, e.g., $\wrot \xi_0$, where $\xi_0$ is the width of the domain wall. 
Actively moving phase boundaries have also been shown to exist in models containing mixtures of active  and passive Brownian particles~\cite{Wysocki_2016}.

Eq.~\eqref{eq:vFast} is valid for $\alpha \to 0$ for finite $\wrot$. The calculation for finite
$\alpha$ is more challenging. In the supplementary material, App.~\ref{app:DomainWallVelocity}, we derive an approximate formula for the velocity, 
\begin{align}
v\approx& \pm \left(\sqrt{\left( \frac{\alpha \rho_s }{\xi_0 \eta}\right)^2+\frac{\rho_s |\wrot|}{m_0^z} }- \frac{\alpha \rho_s}{\xi_0 \eta} \right) \label{eq:velocityFull}
\\  \approx& \pm \left\{ 
\begin{array}{cc}
     \sqrt{\displaystyle  \frac{\rho_s|\wrot|}{m^z_0}} &  \text{for }\  \displaystyle \frac{|\wrot|}{m_0^z} \gg \alpha^2 \frac{ \rho_s}{\xi_0^2} \\[3mm]
\displaystyle \frac{ \xi_0 \eta |\wrot|}{ 2 \alpha m_0^z}     & \text{for }\  \displaystyle \frac{|\wrot|}{m_0^z} \ll \alpha^2 \frac{ \rho_s}{\xi_0^2},
\end{array}\right. \nonumber
\end{align}
where  $\xi_0$ is the width of the domain wall in equilibrium, $\xi_0=\frac{a \sqrt{2 J \Delta/\delta_4}}{m^z_0} $, and $\eta$ is a dimensionless numerical factor, with $\eta=3$ close to the ferrimagnetic phase transition.
For $\wrot \to 0$, the velocity of the domain wall is linear in $\wrot$. In this limit, the domain wall moves by a large distance $\frac{\pi \eta \xi_0}{\alpha m_0^z}\gg\xi_0$ during each rotation, inversely proportional to the (often very small) Gilbert damping $\alpha$ and the average magnetisation $m_0^z<1$.  \NEW{In this regime, a similar formula has previously been obtained for a domain wall in a ferrimagnet driven by an external rotating field \cite{Kwon21RotDW}, but in this case the direction of motion is not chosen spontaneously but imprinted externally.}

In Fig.~\ref{fig:velicityDW}, we show the velocity of the domain wall obtained from numerical simulations, together with Eq.~\eqref{eq:velocityFull}. 
Taking into account that there is no fitting parameter and the approximate nature of the derivation of Eq.~\eqref{eq:velocityFull}, the agreement for $B_0 \apprle 0.2$ is very good.
The analytical formula Eq.~\eqref{eq:velocityFull} does not include the renormalisation of $m_0^z$ and $\rho_s$ for large $B_0$, which may explain small deviations for larger $B_0$.

Above a critical strength of the driving field, $B_0>B_c$, the moving domain wall solution ceases to exist. Instead, one obtains localised domain walls. In the supplementary material, App.~\ref{App:localisedDWs_topo}, we show that topology enforces in this case the presence of singular configurations of $\varphi(x,t)$ (phase slips or, equivalently, space-time vortices \cite{He2017}), which manifest as sudden bursts of spin excitations occurring periodically once per rotation period, $T_\text{rot}=2 \pi/\wrot$.

\begin{figure*}[!ht]
    \centering
    \begin{tikzpicture}
    \draw (0, 0) node[inner sep=0] (a) {
   \includegraphics[height=0.3\linewidth]{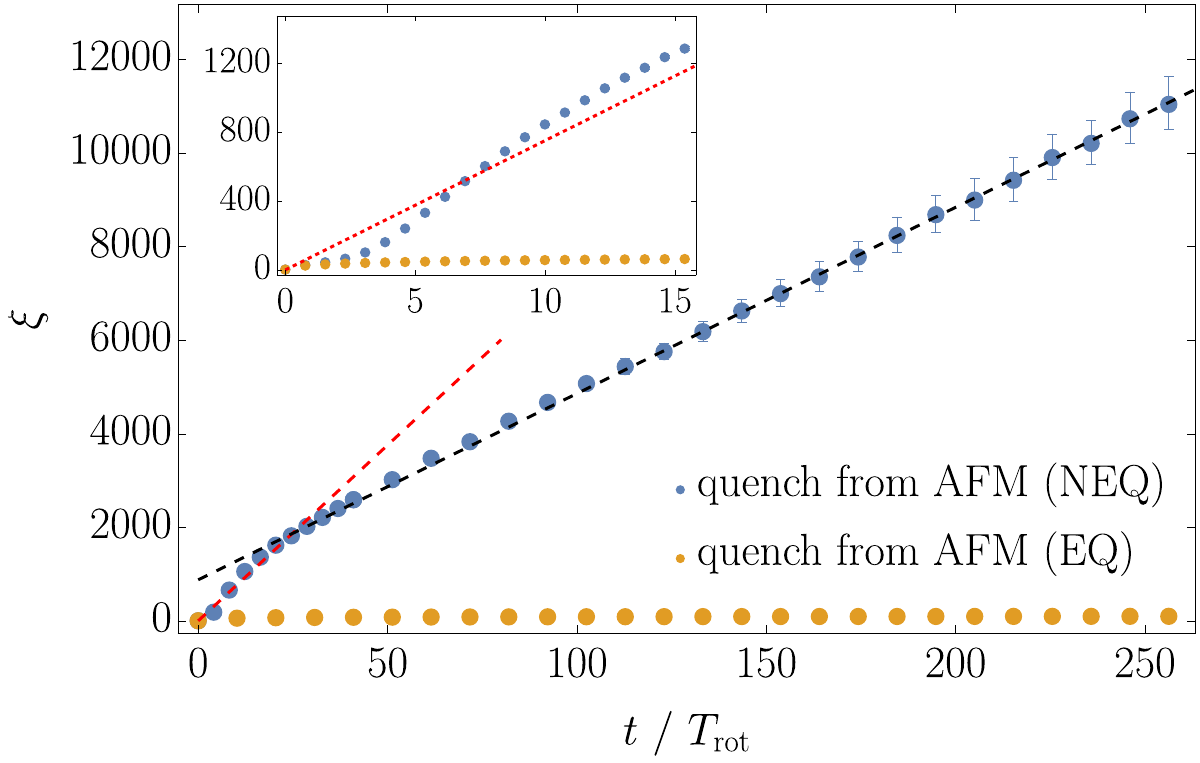}};
    \node[
      above left=-0.5cm and -0.35cm of a] {a)};
    \end{tikzpicture}
   \hspace{0.5cm}
    \begin{tikzpicture}
    \draw (0, 0) node[inner sep=0] (b) {
        \includegraphics[height=0.3\linewidth]{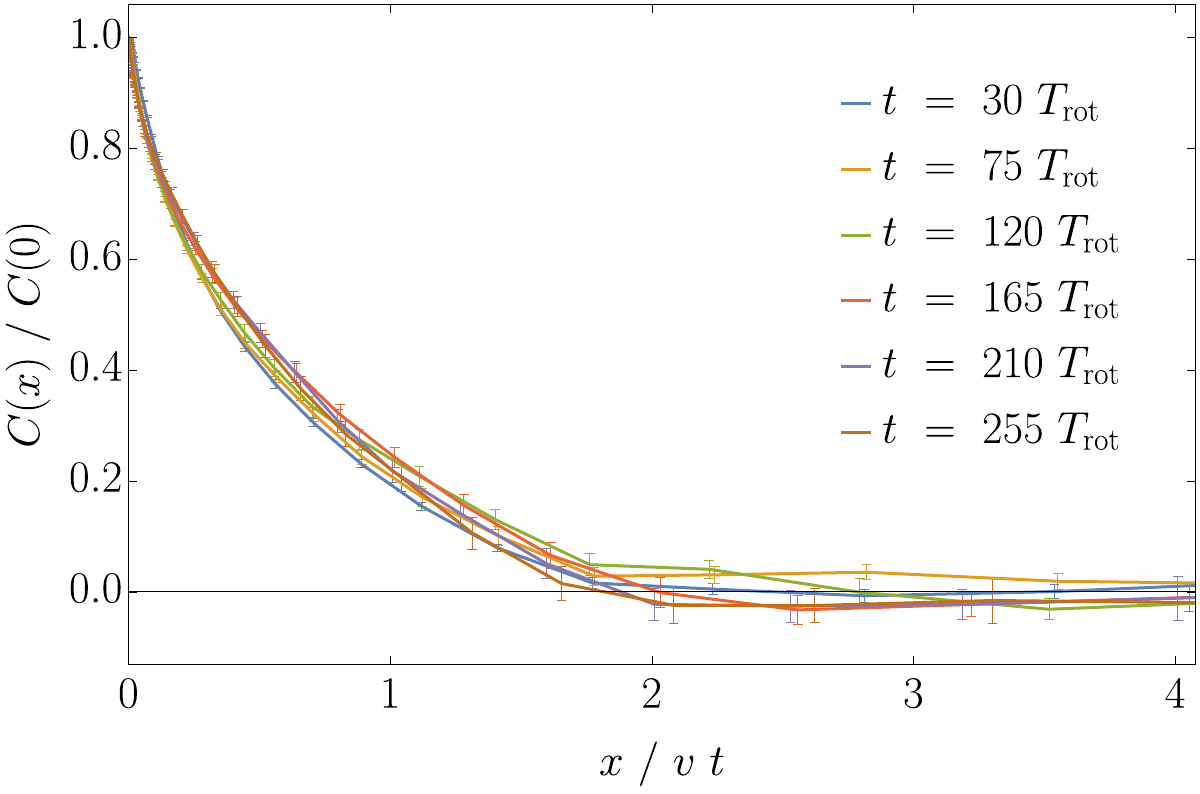}};
    \node[
      above left=-0.5cm and -0.35cm of b] {b)};
    \end{tikzpicture}
    \caption{ Buildup of FM correlations after a quench from an AFM-ordered phase into the ferrimagnetic phase in the absence of noise, $T=0$, see also Fig.~\ref{fig:worldlines}. Panel a): Correlation length $\xi=1/n_\text{DW}$ as function of time after a quench from an AFM-ordered phase into the ferrimagnetic phase both for a driven and a non-driven system, marked by NEQ (non-equilibrium) and EQ (equilibrium) in the legend, respectively.  The inset shows that the short-time dynamics of the driven and non-driven system is identical, but after a few rotation periods $T_\text{rot}$, the driven system shows a very fast increase of the correlation length. In the long-time limit, $\xi$ grows linearly in time ($ \sim 0.065 \ t$, black dashed line). For comparison, $v t$, where $v$  is the velocity of a single domain wall, is also shown as a dashed red line. The linear growth of correlations with time can also be seen directly from a scaling plot of the equal-time correlation function $C(x)=\langle S^z_j S^z_{j+x/a}\rangle$ (averaged over $j$), which is shown (for even $x/a$) in panel b) as function of $x/(v t)$. Here $v$ is the velocity of a single domain wall. The plot shows that the maximal speed is $2 v$, arising from two domain walls moving in opposite directions. Parameters: as in Fig.~\ref{fig:worldlines}, average over 20 initial states in simulations with 500,000 spins each (15 initial states for the equilibrium states). Error bars denote the corresponding standard deviation of the mean. 
    }
\label{fig:velocityDist_and_correlationlength}
\end{figure*}

\begin{figure*}[ht!]
    \begin{tikzpicture}
    \draw (0, 0) node[inner sep=0] (a) {
    \includegraphics[width=\linewidth]{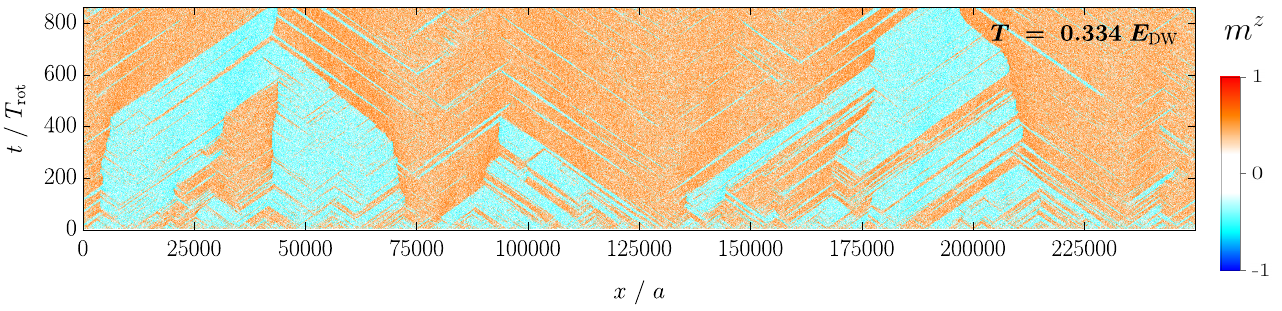}};
    \node[
      above left=-0.5cm and -0.35cm of a] {a)};
    \end{tikzpicture}\\
    \begin{tikzpicture}
    \draw (0, 0) node[inner sep=0] (b) {
    \includegraphics[width=\linewidth]{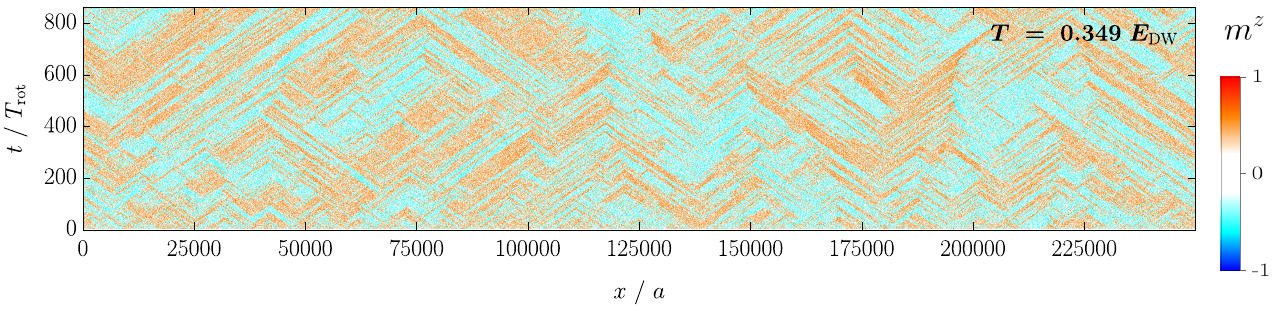}};
    \node[
      above left=-0.5cm and -0.35cm of b] {b)};
    \end{tikzpicture}\\
    \begin{tikzpicture}
    \draw (0, 0) node[inner sep=0] (c2) {
    \includegraphics[width=0.95\linewidth]{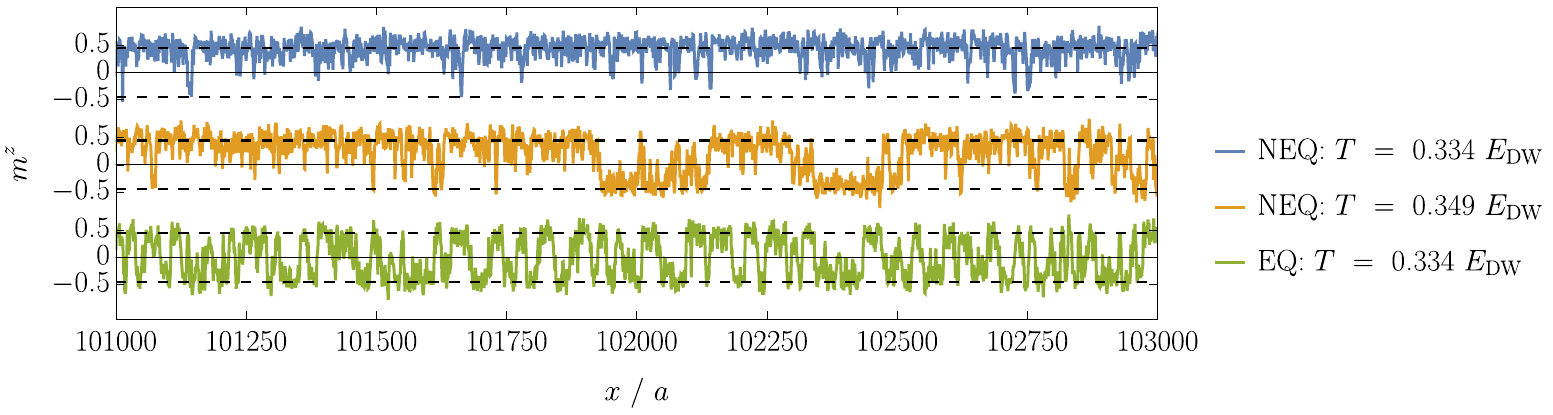}};
    \node[
      above left=-0.5cm and -0.35cm of c2] {c)};
    \end{tikzpicture}
    
    \caption{Panels (a,b): Magnetisation as function of $x$ and $t$ after a quench into the ferrimagnetic phase for two slightly different temperatures, and thus different noise levels, for systems of 250,000 spins (other parameters and colour scale as in Fig.~\ref{fig:movingDW}).
Panel c): $m^z(x)=S^z_i$ at the final timestep of the simulations shown in panels a) and b), for a small region with 2,000 spins. The lower green curve shows that an equilibrium system has a much shorter correlation length at similar noise levels.} 
    \label{fig:magNoise}
\end{figure*}
{\em Dynamics of ordering -- } 
To investigate the dynamics of ordering, we consider a quench starting from an initial state with long-ranged AFM order perturbed by a small random component of the magnetisation in the $z$-direction. An example for the dynamics after such a quench is shown in Fig.~\ref{fig:worldlines}.
The domain walls move and when two of them meet, they get annihilated.
Thus, the correlation length, defined as the inverse of the domain wall density, $\xi=1/n_\text{DW}$, grows rapidly, see Fig.~\ref{fig:velocityDist_and_correlationlength}.
At intermediate times, the correlation length grows approximately with the speed of a single domain wall, $\xi\approx v t$. In the long time limit, $\xi$ also grows linearly in time,
\begin{align}
    \xi(t) \approx \mu\, v t \quad \text{for } t \to \infty, \label{eq:xiLinear}
\end{align}
where $\mu \lesssim 1$ is a numerical constant. $\mu\, vt$,  with $\mu\approx 0.53$ for the chosen parameters, is shown as a dashed black line in Fig.~\ref{fig:velocityDist_and_correlationlength}a). \NEW{A linear growth of correlations has been reported before for active-matter models with self-propelled particles \cite{selfP1,selfP2}.}

This very efficient and fast growth of the correlation length is {\em not} what is expected from a theory of independently moving and pairwise annihilating domain walls. Consider a 1D model, where particles with velocity $\pm v$ annihilate each other whenever two of them collide \cite{2velocityModel}. In such a model,  within time $t$ roughly $N_L=N \pm \sqrt{N}$ left-moving and $N_R=N\pm \sqrt{N}$ right-moving particles interact with each other, where 
$N\sim v t/d_0$ if $d_0$ is the initial distance of the particles. After a time $t$, most of them will have annihilated, but this process leaves approximately $|N_R-N_L|\sim \sqrt{N}$ particles in an area of size $v t$. Thus, in this model the correlation length grows relatively slowly with
\begin{align}
    \xi \sim \sqrt{d_0 v t }  \quad \text{for independently moving domains}
\end{align}
for $v t \gg d_0$~\cite{2velocityModel,2velocityModel_Krug}, in contrast to our finding.

Therefore, we conclude that the rapid growth of $\xi$ can only arise from a strong interaction of the domain walls, which is mediated by the gradients of the $xy$-order imprinted by dynamical frustration. This is shown by the following qualitative argument: within a FM domain, the angle $\varphi$ grows linearly in time, $\varphi \approx \pm \wrot t$, with opposite signs for up and down domains. If domains  have the typical size $\xi(t)$, this implies that the typical gradient of $\varphi$ is of the order of 
\begin{align}
|\partial_x \varphi| \sim \frac{2 |\wrot| t}{\xi(t)}.
\end{align}
If we now demand that gradients remain bounded for $t\to \infty$, then it is plausible that $\xi(t)$ has to grow linearly in $t$, which explains our numerical observation, Eq.~\eqref{eq:xiLinear}. This derivation is not rigorous, as it is possible to construct solutions of, e.g., equally spaced domain walls with a time-independent distance. However, in our simulations such configurations are not stable due to their interactions with an inhomogeneous $\partial_x \varphi$-background.

\begin{figure*}[!ht]
    \centering
    \includegraphics[width= 0.64 \linewidth]{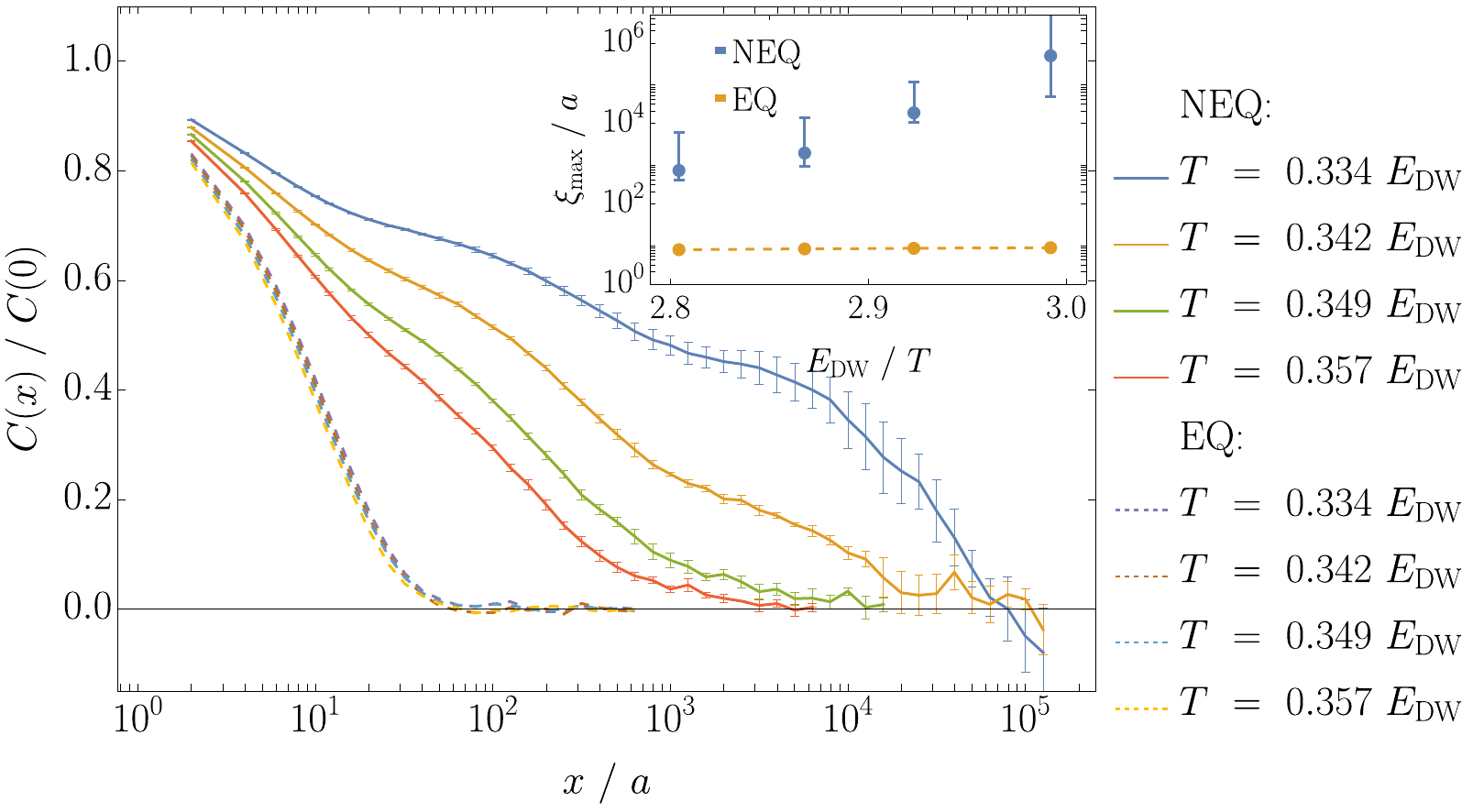}
     \caption{Equal-time correlation function $C(x)=\langle m^z(x_0+x) m^z(x_0)\rangle$  in the presence of thermal noise for systems of the size of 250,000 spins (parameters as in Fig.~\ref{fig:movingDW}, only even $x/a$ are shown, error bars denote the standard deviation of the mean). The data is measured at $t\approx 520{,}000 / J \approx 850 \,T_\text{rot}$ after a quench from an AFM state. The error bars are standard deviation of the mean obtained by averaging over four runs. With the exception of the curve for the lowest $T$ (blue), the system has obtained a stationary state at that time, see supplementary material, App.~\ref{app:ferriquench}.
    Note the logarithmic scale on the $x$-axis, needed due to a broad distribution of domain sizes, see Fig.~\ref{fig:magNoise}.
    Compared to the non-driven system in thermal equilibrium (dashed lines), the correlation length becomes many orders of magnitude larger for $T\lesssim 0.025$. Inset: The correlation length $\xi_\text{max}$ grows faster than exponential with $1/T$. For a discussion of error bars in the inset, see supplementary material, App.~\ref{app:ferriquench}. }
    \label{fig:finiteTempCorel}
\end{figure*}

Microscopically, the necessary interactions
are mediated by $\varphi$. More precisely, the domain walls interact with $\partial_x^2 \varphi$, as can be seen directly from the relevant Ginzburg-Landau theory, see methods. A negative $\partial_x^2 \varphi$ leads to  $\partial_x j^z<0$, see Eq.~\eqref{eq:spincurrent}, and thus to an increase of $m^z$, which acts as a repulsive force for an approaching domain with $m^z<0$.
Here, the main source of large $\partial_x^2 \varphi$ are cases where two domain walls meet and annihilate. As can be seen in Fig.~\ref{fig:worldlines}, each annihiliation event leaves behind a characteristic trace in $\partial_x^2 \varphi$ where $\partial_x \varphi$ changes sign. Close to $x=37{,}000 \, a$ in Fig.~\ref{fig:worldlines}, a left-moving down domain wall changes its direction of motion when approaching a region with $\partial_x^2 \varphi <0$. After transforming into a right-moving domain wall, it finally annihilates with another left-moving domain wall. Importantly, the interactions mediated by $\partial_x^2 \varphi$ are strongly retarded, as it takes a very long time proportional to $\xi^2$ for $\partial_x^2 \varphi$ to decay by diffusion. This is much shorter than the time of order $\xi/v$ needed for the approach of other domain walls, see supplementary material App.~\ref{app:hydro}
for a hydrodynamic view of these effects.

{\em Resilience to noise --} Finally, we investigate the influence of thermal noise (described by a random magnetic field $\vec \xi_i(t)$, see methods) on our findings.  In a 1D system in thermal equilibrium, such a noise term destroys long-ranged order and induces domain walls with a typical distance $\xi_\text{th} \sim e^{E_{\text{DW}}/{k_B T}}$, where $E_{\text{DW}}$ is the domain wall energy at $T=0$, and the temperature $T$ controls the noise strength. For the parameters chosen in our figures, $E_\text{DW}=0.076 \, J$. In the following, we will therefore always use $T/E_\text{DW}$ to specify temperatures.
Figs.~\ref{fig:magNoise}a) and b) show the magnetisation of a driven ferrimagnet as function of space and time after a quench from the AFM phase for two different temperatures. As in the noiseless case studied above,  domain walls actively move with a velocity approximately $10\,\%$ larger than in the noiseless case, see supplementary material, App.~\ref{app:spacetime}. It is surprising that the velocity is affected so little, taking into account that the  $xy$-correlation length $\xi_\text{AFM}$ in the noisy and driven system becomes tiny, $\xi_\text{AFM} \lesssim 30$, see supplementary material, App.~\ref{app:extraResults_shortxyorder}.

Fig.~\ref{fig:magNoise}a) shows the formation of giant domains  with a size exceeding $10^5$ spins. This has to be compared to an equilibrium system, where for the same noise level the correlation length is less than $20$ sites.

In Fig.~\ref{fig:finiteTempCorel}  we show the steady-state correlation function for both the equilibrium and non-equilibrium systems. In the driven system, the correlation function decays very slow in space, roughly with a linear slope on a logarithmic scale, $C(x)=C(0) (1-\ln[x/a]/\ln[\xi_\text{max}(T)/a]$, showing that the system hosts domains with very different sizes.  $\xi_\text{max}(T)$ can be interpreted as the largest correlation length in the system, see supplementary material, App.~\ref{app:ferriquench}, for details. $\xi_\text{max}(T)$ grows much faster than exponential with $1/T$, see inset of Fig.~\ref{fig:finiteTempCorel}, showing that the rapid growth does not arise from a thermal process. \NEWWW{A direct comparison of the space-time correlations of the driven and non-driven systems, see supplementary material, App.~\ref{app:spacetime}, shows the qualitative change of dynamical properties of the driven system.}

How can even a relatively weak oscillating field increase the correlation length by many orders of magnitude and why is the driven system so resilient to noise?  In the regime investigated here, fluctuations are not small and even within an ordered domain, they often induce a sign change, see the upper curve in  Fig.~\ref{fig:magNoise}c). Locally, fluctuations are very similar to the thermal system. The difference is, however, that the driven system is very efficient in  suppressing the generation of defects and also in healing some of the most frequently occurring defects.
The main origin of this healing effect is the dynamical frustration of domains rotating in opposite direction. In the supplementary material, App.~\ref{app:resilience}, we discuss and investigate the detailed microscopic mechanisms behind this observation.

\NEWW{One may ask whether one can (or indeed should) use the term `active magnetic matter' to describe the system studied  here. As discussed in the introduction, active matter is usually defined as being composed of large numbers of `active agents', each breaking detailed balance on a local level and giving rise to collective non-equilibrium phenomena. In our case, one can either view each spin or each domain wall as one of the active agents. Each spin in our system is rotating either clockwise or counter-clockwise, thus breaking detailed balance on  a local level.
Similarly, a domain wall moves actively in space, either to the right or to the left side. Furthermore, we have shown that the collective dynamics of domain walls is dominated by long-range interactions and cannot be understood from their individual motion. 
 In our case of an active magnetic system, the external drive is implemented in a continuous way by an oscillating external field. This is also the case in more traditional active systems, for example of biological origin. There, some type of feeding is required, but continuous drive is not needed in most biological systems, where energy can be stored. Other examples of continuously driven active matter systems
use light or ultrasound as a stimulus 
\cite{reviewOfreviews,lightDrivenReview,BiVO4activeMatter,soundspecificreview}.
%The overarching characteristic shared by our active magnetic system and more traditional active matter systems is the external driving. In the magnetic case, this is implemented in a continuous way by an oscillating external field; while, for example, in biological active systems, the feed of the self-propelled agents typically has a more complex temporal profile.  
% In our case of active magnetic systems, the external drive implemented in a continuous way by an oscillating external field, but also in more traditional active systems, for example of biological origin, some type of feeding is required. Yet, in both cases, the common characteristic is an external driving.
}

In conclusion, we show that a simple oscillating field transforms a ferrimagnet into an `active magnet'. This is a state of matter with properties very different to those of equilibrium systems. Ferrimagnets recently attracted considerable interest in the field of spintronics \cite{Kim2022}, as they combine the advantages of FM and AFM systems. Here, we mainly need the rotation of an $xy$-order parameter in a driven system to be controlled by the orientation of a FM component perpendicular to it. The rotation of the $xy$-order parameter leads to an effect which we call `dynamical frustration', resulting in an efficient propulsion of the domain wall, which now moves actively either to the left or to the right by spontaneous symmetry breaking. The net effect is a very fast buildup of FM correlations on huge length scales indicating a  high degree of robustness of the system to thermal noise due to an efficient non-equilibrium self-healing mechanism.  \NEW{ While our study has focused on one-dimensional systems, we show in App.~\ref{app:2D_results} that key results such as the active motion of domain walls and the rapid buildup of correlations also apply in two dimensions.} \NEW{The most promising route for an experimental realisation of our model is to use widely studied magnetic multilayer systems, whose properties (including magnetic anisotropies) can largely be tuned by layer thickness and the combination of different materials \cite{multilayer1,multilayer2}. For example, \'Sl\c{e}zak {\it et al.} ~\cite{artifialFerrimagnet2021} recently demonstrated direct coupling of an easy-plane antiferromagnetic NiO layer to an adjacent ferromagnet Fe layer, providing a direct route towards a room-temperature realisation of our model.} \NEWW{Experimentally, it should be rather straightforward to observe the motion of domain walls using standard imaging methods based on the Faraday or Kerr effect, which are routinely used to observe the motion of domain walls \cite{MOKE}.}

A surprising aspect of our results is that very weak perturbations have a profound effect on the qualitative and quantitative behaviour of the system. For the chosen parameters, the speed of rotation of the magnetisation was tiny, $\wrot J \approx \pm 0.01$. This tiny perturbation changes the dynamics of the system, affects the correlations both on small and large length scales, see Fig.~\ref{fig:finiteTempCorel}, and can increase the correlation length by orders of magnitude. In equilibrium systems, tiny perturbations (if they are `relevant' in the renormalisation-group sense) can also sometimes strongly affect properties of correlation functions, but only at very long distances. We are not aware of any equilibrium systems in which small perturbations cause such a profound change of magnetic correlations on all length scales. Our driven system is different, as it induces active motion of domain walls.

In our study, we have investigated a minimal model of a driven magnet. 
We anticipate that it is possible to activate (approximate) Goldstone modes in a large class of systems and to use the principle of dynamical frustration to realise actively moving defects. \NEW{For example, basic features of our models can be realised directly by coupling a superconducting layer to a ferroelectric or a polar metal in the presence of external radiation using, e.g., oxide heterostructures \cite{mannhardt07}.} Given the enormous variety of complex symmetries and associated topological defects available in solid-state systems, we anticipate numerous opportunities to realise systems which may be identified as active magnetic matter or, more broadly, active quantum matter. An important aspect of this challenge will be finding experimental realisations of such systems. 

\section*{Methods}
The dynamics of the ferrimagnetic system, Eq.~\eqref{eq:discreteFerrimagnetHamiltonian},  is obtained from the (stochastic) Landau-Lifshitz-Gilbert (LLG) equation
\begin{align}\label{eq:LLG}
    \frac{d \vec S_i}{d t}= \vec S_i \times \left(-\frac{d H}{d \vec S_i}+\vec \xi_i(t)\right)+\alpha \vec S_i \times \frac{d \vec S_i}{d t},
\end{align}
with  the  dimensionless damping constant $\alpha$. $\vec \xi_i(t)$ is a Gaussian random noise term  with $\left\langle \xi_i(t)\xi_j(t')\right\rangle=2 \alpha k_B T \delta_{i,j} \delta(t-t')$ which arises from thermal fluctuations. $H$ is given in Eq.~\eqref{eq:discreteFerrimagnetHamiltonian}.

In order to simulate the micromagnetic system, we use the GPU-accelerated open-source software Mumax3 \cite{MumaxRef,MumaxRefThermal}. We simulate one-dimensional systems with sizes ranging from 20,000 to 500,000 spins with open boundary conditions for simulations of a single domain wall and periodic boundary conditions for all other simulations. \NEW{In the supplementary material we also show simulations for two-dimensional systems with up to $10^8$ spins but only for shorter runtimes.} For quenches from the AFM phase, we use initial states with $S_i^z$ drawn from a uniform distribution in the interval of $(-0.1,0.1)$, $S_i^x=(-1)^i \sqrt{1-(S_i^z)^2}$ and $S_i^y=0$, i.e., $\varphi=0$.

In all figures displaying the time evolution, we show stroboscopic images, obtained at integer multiples of the driving period $2 \pi/\omega$. 

For our analytical arguments, we mainly  use the LLG Eq.~\eqref{eq:LLG} either directly or evaluated in a continuum limit, see supplementary material, App.~\ref{app:GostoneModeActivation} and \ref{app:DomainWallVelocity}, for details. Alternatively,  one can analyse the appropriate (noisy) Ginzburg-Landau theory
describing the ferrimagnet close to the transition to the AFM \cite{ZelleUniversalPhaseTransition}. This Ginzburg Landau theory is formulated in the continuum limit for the FM magnetisation $m^z(x,t)$
and the angle $\varphi(x,t)$ parametrising the orientation of the AFM $xy$-order. A  Taylor expansion in $m^z$, derivatives of $\varphi$ and the number of gradients  gives to leading order 
\begin{align} \label{eq:GL}
        \alpha \dot{\varphi} = - \alpha \gamma m^z &+ \rho_s \partial_x^2 \varphi + \dot{m}^z + \xi_\varphi, \nonumber\\
        \alpha \dot{m}^z = - r m^z -&u (m^z)^3 +D \partial_x^2 m^z{} \\
        - &K \partial_x^2 \varphi-\dot{\varphi}+\xi_m,  \nonumber
\end{align}
where $\xi_{\varphi/m}$ describe Gaussian white noise with   $\langle \xi_{\varphi/m} \rangle = 0$ and  $\langle \xi_{\varphi/m} \xi_{\varphi/m} \rangle =2 \alpha T \delta(t-t') \delta(x-x')$. The noise strength is fixed by the damping terms ($\sim \alpha$) and temperature. The equilibrium system is obtained by setting $\gamma=K=0$. All parameters of Eq.~\eqref{eq:GL} are known analytically with the exception of $K$, with $r=\delta_2 + 2J (1-\Delta)$, $u=\delta_4$, $\rho_s=J a^2$, $D=J \Delta \, a^2$. $\gamma$, responsible for the rotation of the $xy$-order, is proportional to $B_0^2$ for small $B_0$, and is computed in the supplementary material, App.~\ref{app:GostoneModeActivation}.  Note that the effective field theory cannot be used to quantitatively describe the space-time vortices discussed in the supplementary material,  App.~\ref{App:localisedDWs_topo}, as the magnetisation reaches $\pm 1$ at the vortex core, which is outside of the range of validity of Eq.~\eqref{eq:GL}. 

From the first line in Eq.~\eqref{eq:GL}, one recovers Eq.~\eqref{eq:activateG} in the ferrimagnetic phase for $\partial_x^2 \varphi=0$ and stationary $m^z$. Similarly, for $\alpha \to 0$ one obtains a continuity equation for $m_z$, with the current given by Eq.~\eqref{eq:spincurrent}.

%\section*{Discussion}\label{sec12}

%Discussions should be brief and focused. In some disciplines use of Discussion or `Conclusion' is interchangeable. It is not mandatory to use both. Some journals prefer a section `Results and Discussion' followed by a section `Conclusion'. Please refer to Journal-level guidance for any specific requirements. 

%\section*{Conclusion}\label{sec13}

%Conclusions may be used to restate your hypothesis or research question, restate your major findings, explain the relevance and the added value of your work, highlight any limitations of your study, describe future directions for research and recommendations. 

%In some disciplines use of Discussion or 'Conclusion' is interchangeable. It is not mandatory to use both. Please refer to Journal-level guidance for any specific requirements. 

\backmatter

%\bmhead{Supplementary information}

%If your article has accompanying supplementary file/s please state so here. 

%\section*{Declarations}

\bmhead{Data availability} 

The supporting data generated in this study have been deposited in a Zenodo database under accession code \cite{data_zenodo}.

%The supporting data and codes for this article are available from Zenodo \cite{data_zenodo}.

\bmhead{Code availability} All data shown was generated with Mumax3 as described in the methods section. We provide one file to run such simulations as an example on Zenodo \cite{data_zenodo}.

\bmhead{Acknowledgements}

We acknowledge useful discussions with Romain Daviet, Gerhard Gompper, Joachim Krug, and Carl Zelle. We thank the  German Research Foundation for financial support via CRC 1238 (Project-ID 277146847, project C04) and CRC/TRR 183 (Project-ID 277101999, project A01). Furthermore, we thank the Regional Computing Center of the University of Cologne (RRZK) for providing computing time on the DFG-funded (Funding number: INST 216/512/1FUGG) High Performance Computing (HPC) system CHEOPS as well as technical support.

\bmhead{Author contributions}
D.H., N.d.S and A.R. developed the analytical theory with N.d.S deriving the theory of domain wall velocities at finite $\alpha$. D.H. and R.D. performed simulations using the effective Ginzburg Landau theory (D.H.) and Mumax (R.D.). S.D. and A.R. designed the study. All authors contributed to the manuscript.

\bmhead{Competing interests}
The authors declare no competing interests.

\bibliography{Library}

\appendix

\onecolumn

\section*{\Large Supplementary Material to `\NEWW{Propelling Ferrimagnetic Domain Walls by Dynamical Frustration}'} 

\doparttoc
\faketableofcontents
\renewcommand \thepart{}
\renewcommand \partname{}
\part{}
\parttoc

\renewcommand{\theequation}{S\arabic{equation}}
\setcounter{equation}{0}
\renewcommand{\thefigure}{S\arabic{figure}}
\renewcommand{\theHfigure}{S\arabic{figure}}
\setcounter{figure}{0}
\renewcommand{\thetable}{S\arabic{table}}
\setcounter{table}{0}

\section{Topology of Dynamically Frustrated Domain Walls}
\subsection{Localised Domain Walls and Their Topology} \label{App:localisedDWs_topo}

In this section, we discuss the properties of non-moving domain walls. We observe those primarily when the strength of the driving oscillating field is large.
Moving domain walls only exist below a critical  field $B_c$, see Fig.~\ref{fig:velicityDW} of the main text. For $B<B_c$, we obtain a coexistence of moving and localised domain walls, while for $B_0>B_c$ only localised single domain walls exist. For smaller $B_0$, however, more and more fine-tuning of initial  conditions is required to obtain localised solutions.

\begin{figure}[h]
    \begin{minipage}{0.95\linewidth} 
    \centering 
    \begin{tikzpicture}
    \draw (0, 0) node[inner sep=0] (c) {\includegraphics[height=0.27\linewidth]
    {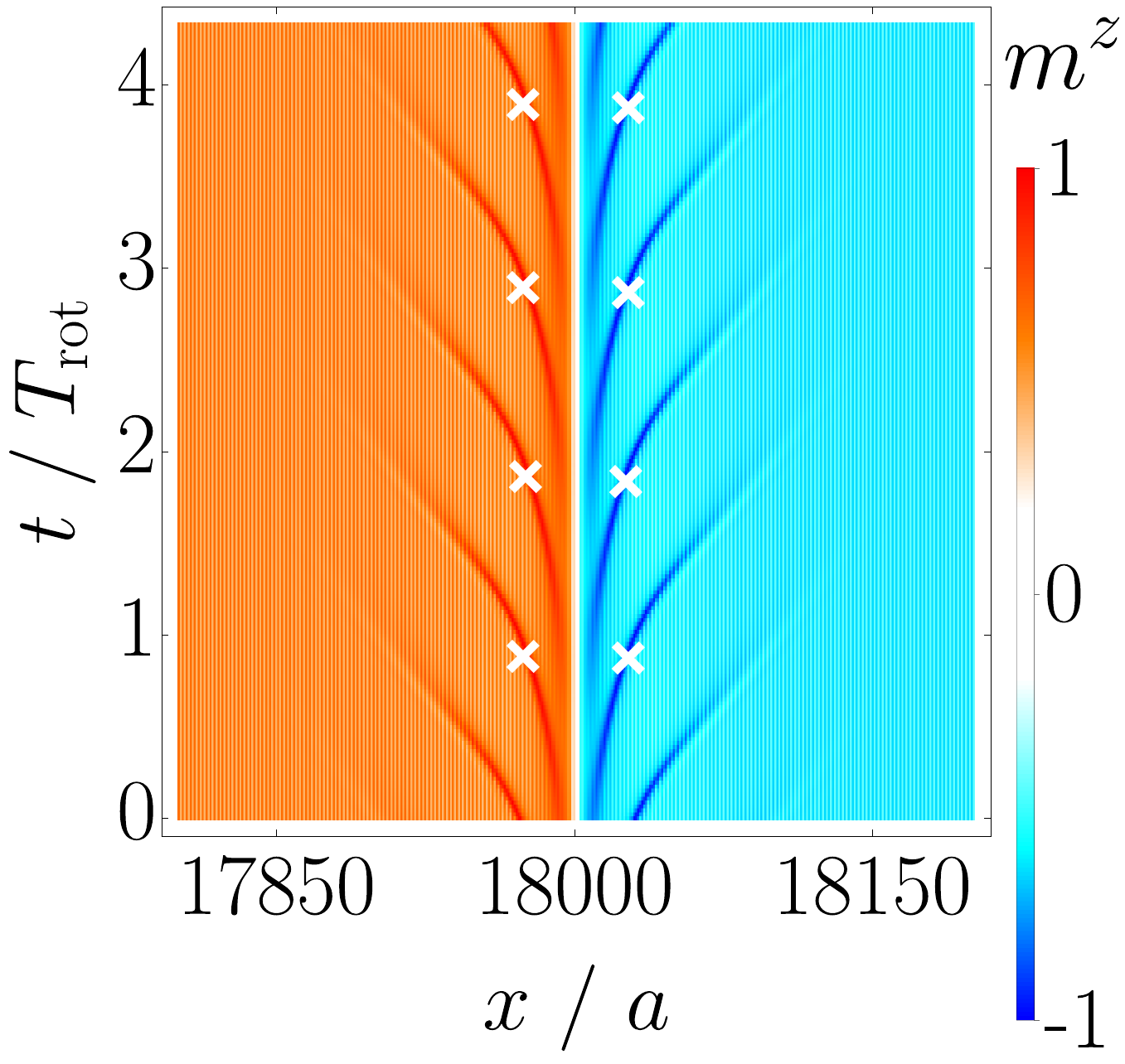}};
    \node[
      above left=-0.5cm and -0.35cm of c] {\large a)};
    \end{tikzpicture} 
    \begin{tikzpicture}
    \draw (0, 0) node[inner sep=0] (d) {\includegraphics[height=0.27\linewidth]{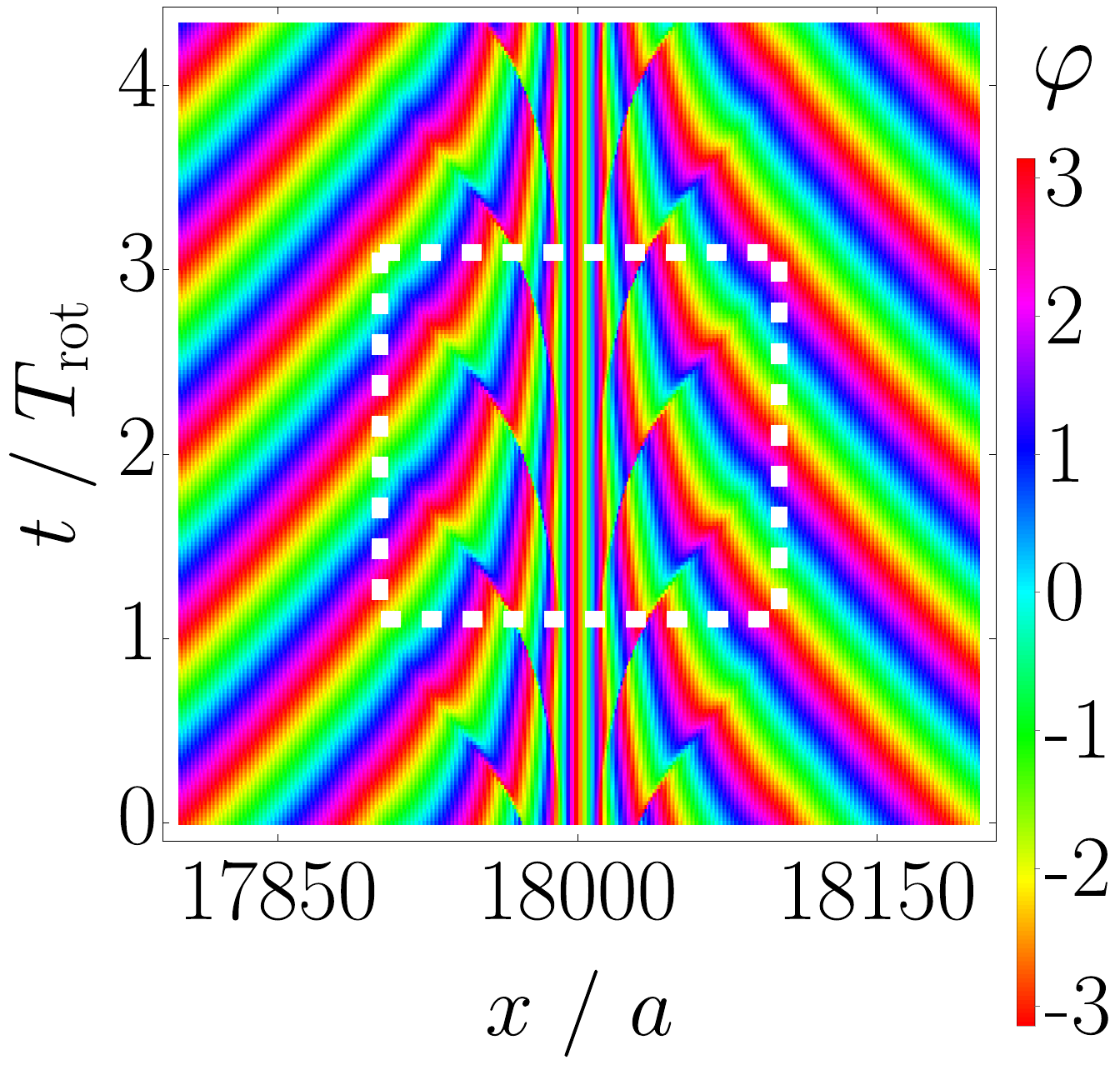}};
    \node[
      above left=-0.5cm and -0.35cm of d] {\large b)};
    \end{tikzpicture}
    \end{minipage}
    \begin{minipage}{0.95\linewidth} 
    \centering 
    \hspace{-7mm}
    \begin{tikzpicture}
    \draw (0, 0) node[inner sep=0] (c) {\includegraphics[height=0.27\linewidth]{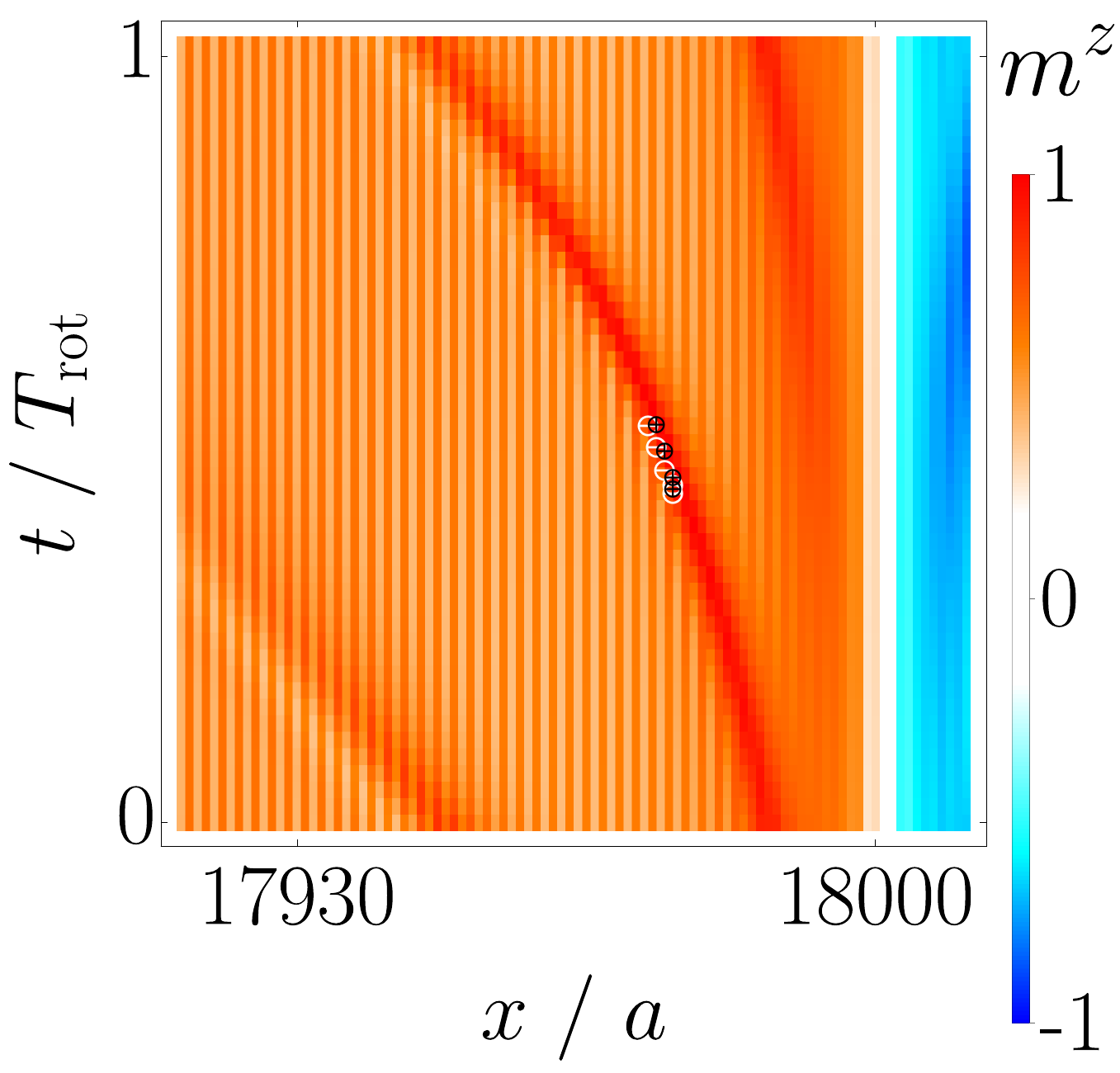}};
    \node[
      above left=-0.5cm and -0.35cm of c] {\large c)};
    \end{tikzpicture} 
   \begin{tikzpicture}
    \draw (0, 0) node[inner sep=0] (d) {\includegraphics[height=0.27\linewidth]
    {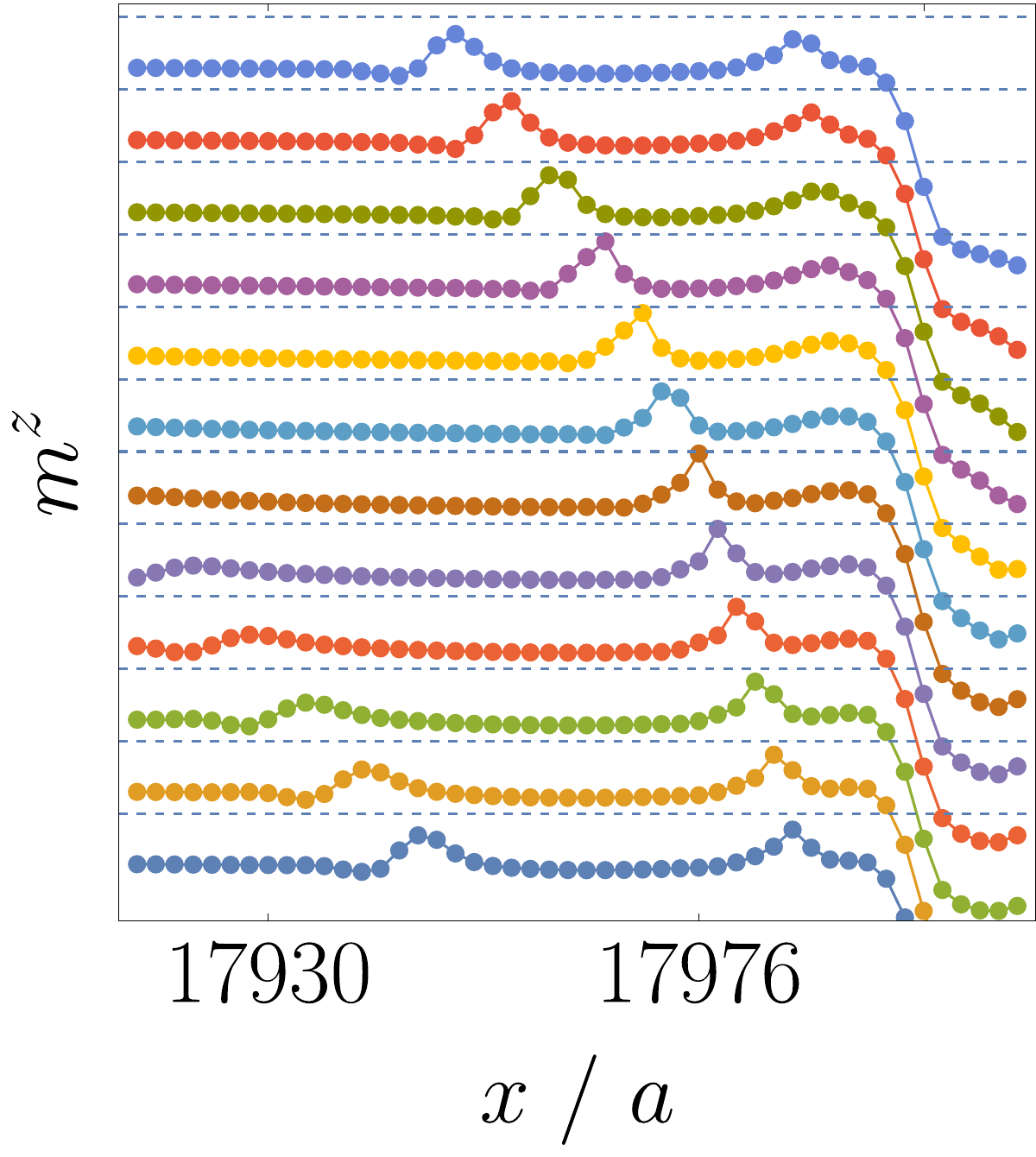}};
    \node[
      above left=-0.5cm and -0.35cm of d] {\large d)};
    \end{tikzpicture} 
    \end{minipage}
%\vspace*{4cm}
    \caption{Non-moving domain wall obtained for a larger oscillating field. Panel a) shows the magnetisation $m^z(x,t)$ and panel b) the phase $\varphi(x,t)$, describing the orientation of the staggered in-plane magnetisation. For a non-moving domain wall, topology enforces the emergence of two space-time vortices (or, equivalently, phase slips) per rotation period (white crosses in panel a)). The magnetisation reaches $\pm 1$ at the (elongated) vortex cores and triggers the emission of sharply defined pulses in both directions. Panel c) shows a zoom to the centre of such an elongated vortex core. Here, white and black dots denote the position of space-time vortices with winding number $\pm 1$. In the shown example, there are 4 white and 5 black dots. Thus, the total winding number is $-1$.  Panel d) directly shows $m^z(x)$ for  $t=n T_\text{rot}/12$, $n=0,\dots,11$.  For better visibility, the curves are shifted relative to each other. The thin dashed lines denote $m^z=1$. At the location of the vortices (brown curve in the middle, at $x=17{,}976 \, a$), the magnetisation reaches $1$.
    Parameters: $J=1, \Delta=0.8,\delta_2=-0.6,\delta_4=1,g_1=1,g_2=0.1,\alpha=0.1,B_0=0.42,\omega=3.6$ with $T_\text{rot}\approx 81$ for a system of 40,000 spins.
   }
    \label{fig:localisedDomain}
\end{figure}

An example of such a localised domain wall is shown in Fig.~\ref{fig:localisedDomain}.
One sees that the domain wall periodically emits very sharply defined instanton-like bursts of magnetisation. This is also a consequence of the dynamical frustration: as the sense of rotation of $\varphi$ is opposite on both sides of the domain, tension necessarily builds and gets suddenly released.
Close to the crosses in Fig.~\ref{fig:localisedDomain}, the magnetisation reaches $\pm 1$ and the phase describing the orientation of spins in the $xy$-plane becomes ill-defined. Below, we will show that topology enforces  singularities in $\varphi$.

To see this, it is useful to compute the winding number of the angle $\varphi(x,t)$ along a contour $C$ in space-time, $W=\frac{1}{2 \pi} \oint_C \boldsymbol\nabla_q \varphi \, d\boldsymbol q$ with $\boldsymbol q=(t,x)$, $\boldsymbol\nabla_q=(\partial_t,\partial_x)$. Choosing for $C$ a rectangular-shaped contour (dashed white line in Fig.~\ref{fig:localisedDomain}b)) extending over $N$ oscillation periods $T_\text{rot}=\frac{2 \pi}{\wrot}$ in the time direction, we find for large $N$ the winding $W=\pm 2 N$, with opposite signs for down-up and up-down domain walls. Here we used that $\partial_t=-|\wrot|$ for an up-domain and $\partial_t=|\wrot|$ for a down domain. Note that the spatial parts of the contour cancel each for a periodic solution with $\vb{S}_i(t)=\vb{S}_i(t+NT_\text{rot})$.
Thus, topology enforces that $\varphi(x,t)$ must have singularities, or space-time vortices \cite{He2017}, with a net winding number of two per oscillation period.
In the context of superconductivity, these space-time vortices are  called phase slips. Alternatively, one can also characterise the resulting magnetisation profile as a fractionally charged defect \cite{delser2023fractional} in space-time.

Numerically, we find that two vortices with identical winding numbers are located a bit to the left and to the right side of the domain wall. The space-time vortices send out sharp pulses of spin excitations, exactly spaced by an oscillation period $T_\text{rot}$. At the vortex core, the spins point in the $z$-direction, $m^z=\pm 1$. Unexpectedly, the vortex cores are not points in space-time, but show an unusual  elongated shape. This is shown in Fig.~\ref{fig:localisedDomain}c), which is a zoom of Fig.~\ref{fig:localisedDomain}a) to the core of a vortex. Using the discretised data underlying Fig.~\ref{fig:localisedDomain}b), we calculate the local winding number from four neighbouring data points of $\varphi$ on the space-time lattice by checking whether the  four phase differences (projected on interval from $-\pi$ to $\pi$) add up to $-2 \pi$, $0$, or $2 \pi$. Black and white dots in Fig.~\ref{fig:localisedDomain}c) denote vortices with winding number $\pm 1$. Several of them align in a line but the net winding is always $-1$, as enforced by topology.

\subsection{Topology of a Moving Domain Wall}
\label{App:movingDWs_topo}
Above, we have shown that a localised domain wall will necessarily produce periodically singular spin configurations characterised by a space-time vortex (phase-slip) in $\varphi$. Thus, any smooth, non-singular $xy$-spin configuration will enforce a motion of the domain wall. 

Here we discuss constraints arising from topology for such a domain wall. We can use that far from the domain wall, the magnetisation is constant in space and also $\partial_x^2 \varphi=0$. Furthermore, the $xy$-magnetisation on the right and left side rotate with opposite speeds, $\varphi_{R/L}=\pm \wrot t$. Thus we can make the ansatz  
\begin{align}
\varphi_{R/L}(x,t)=\varphi_{R/L}^0 + (\partial_x \varphi_{R/L}) x \pm \wrot t,
\end{align}
which is valid sufficiently far away from the domain wall on the left and right sides of it. Next, we calculate the winding number for a long loop oriented parallel to a moving domain wall with velocity $v$. Such a loop in distance $d$ can, e.g., be parameterised by $(t,x)=(\lambda N T_\text{rot},v N T_\text{rot} \lambda+d)$ for $0\le \lambda<1$,  $(t,x)=(N T_\text{rot},v N T_\text{rot}+(3-2\lambda)d$ for $1\le \lambda<2$, $(t,x)=(N T_\text{rot} (3-\lambda),v N T_\text{rot} (3-\lambda) -d
)$ for $2<\lambda<3 $, and $(t,x)=(0,d (2\lambda-7))$ for $3<\lambda<4$, closing the loop with $T_\text{rot}=\frac{2 \pi}{|\wrot|}$, $N \in \mathbb{Z}, N\gg 1$. 
A smooth spin configuration can only be obtained for a vanishing winding number, 
\begin{align}
    W=\frac{N}{2 \pi} \left(2 \wrot T_\text{rot} +v T_\text{rot} (\partial_x \varphi_R-\partial_x \varphi_L) \right)=0 \quad \Rightarrow \quad \partial_x \varphi_R-\partial_x \varphi_L=-\frac{\wrot}{v}.\label{eq:vGrad}
\end{align}
where $\wrot$ is the rotation speed on the right side of the domain wall, see above. 
The wavelength corresponding to the phase changes is given by $\frac{2 \pi}{\lambda_{R/L}}=\partial_x \varphi_{R,L}$. Therefore, equation \eqref{eq:vGrad} can also be written in the more intuitive form
\begin{align}
 |v|=\frac{|\Delta \lambda|}{T_\text{rot}}, \quad \frac{1}{\Delta \lambda}=\frac{1}{\lambda_L}-\frac{1}{\lambda_R}.
\end{align}
Topology enforces that the domain wall has to move by $\Delta \lambda$ within a rotation period.
The difference of gradients to the left and right side of the domain wall has to be inversely proportional to the velocity of the domain wall, as long as $\varphi(x,t)$ configurations are non-singular. This result is based only on topology and on properties far away from the domain wall.  As the difference of gradients has to  be finite, topology enforces a finite domain wall velocity $v$ for vortex-free $\varphi(x,t)$ solutions. To determine the value of $v$, one needs an extra equation, see main text and App.~\ref{app:DomainWallVelocity}.

\begin{figure*}[ht!]
    \begin{tikzpicture}
    \draw (0, 0) node[inner sep=0] (a) {
    \includegraphics[width=0.99\linewidth]{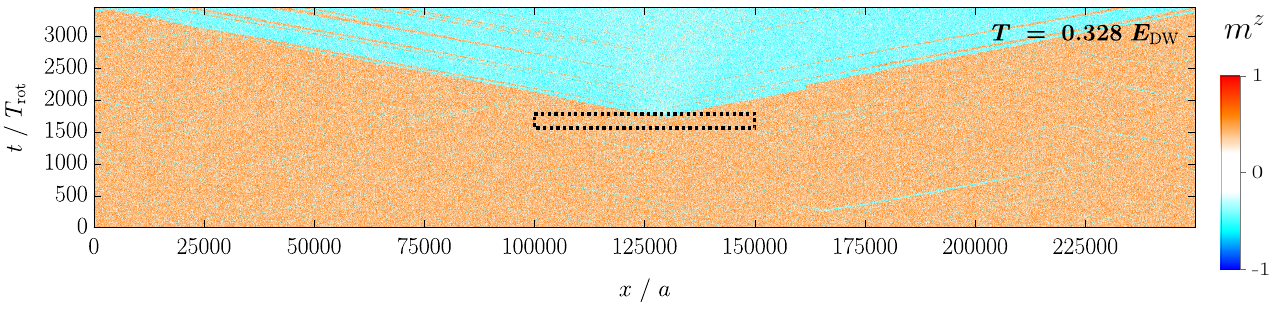}};
    \node[
      above left=-0.5cm and -0.35cm of a] {\large a)};
    \end{tikzpicture}\\
    \begin{tikzpicture}
    \draw (0, 0) node[inner sep=0] (b) {
    \includegraphics[width=0.99\linewidth]{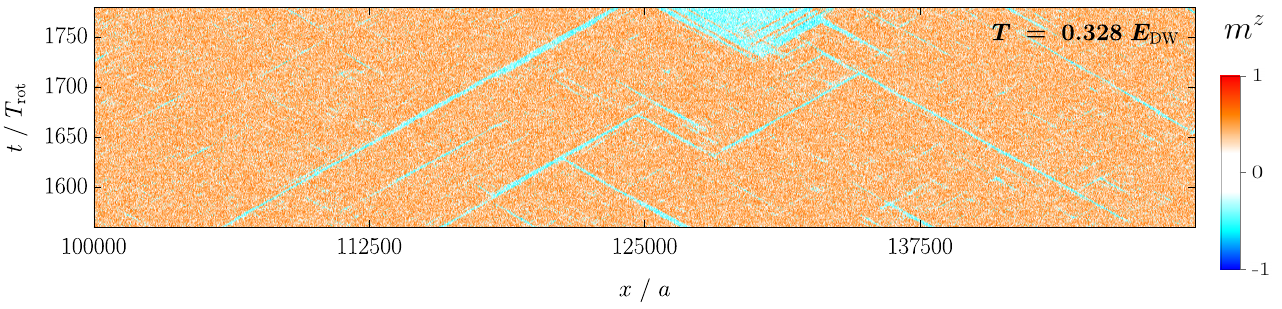}};
    \node[
      above left=-0.5cm and -0.35cm of b] {\large b)};
    \end{tikzpicture}
    
    \caption{Magnetisation as a function of $x$ and $t$ starting from ferrimagnetic initial conditions for the given temperature. Panel b) shows a zoom-in on the region marked by a black dashed line in panel a).
Note that for the same noise level, an equilibrium system has a correlation length of only 20 sites. Two types of defects are visible. The blue lines in panel b) are pairs of parallel moving domains with a  width of up to a few hundred sites. They do not affect the long-ranged order. The simulation also shows a single defect of a different type: a pair of domain walls moving in opposite directions, which is able to revert the magnetisation.   
    Parameters: $J=1, \Delta=0.8, \delta_2=-0.6, \delta_4=1, \alpha=0.1, g_1=1, g_2=0.1, \omega=3.6, B_0=0.15$,  $T_\text{rot}=612$, $T=0.328 \, E_\text{DW}$ for a system of 250,000 spins.}
    \label{fig:magNoise_app}
\end{figure*}

\subsection{Hydrodynamics of Winding Numbers}\label{app:hydro}
The discussion in the preceding two sections shows that the presence or absence of space-time vortices strongly affects the dynamics of the system. As space-time vortices are rare for small driving amplitudes, $\partial_x \varphi$ is an approximately conserved quantity. Therefore, one can introduce the 
 topological charge density
\begin{align}
\rho_\text{top}=\frac{\partial_x \varphi}{2 \pi}.
\end{align}
For smooth $\varphi(x,t)$ and periodic boundary conditions, the topological charge $Q_\text{top}=\int \rho_\text{top}\,dx$ is a conserved winding number which encodes how frequently the AFM magnetisation winds in the $xy$-plane.  The integer $Q_\text{top}$ changes by $\pm 1$ at the location of a space-time vortex but such events are absent for, e.g., the parameters chosen in Fig.~\ref{fig:worldlines} of the main text.

Using this quantity, we can reformulate some of our  findings. Taking the derivative of the first line of Eq.~\eqref{eq:GL} in the methods section, we find that the topological current, defined by $\partial_t \rho_\text{top}+\partial_x j_\text{top}=0$ for non-singular configurations of $\varphi(x,t)$, is given by
\begin{align}\label{currentMagCouling}
j_\text{top}=\frac{\gamma}{2 \pi} m^z-\frac{\rho_s}{ \alpha} \partial_x \rho_\text{top}-\frac{1}{2 \pi \alpha}\left( \dot m^z+\xi_\varphi\right).
\end{align}
Far from a domain wall, where $\partial_t m^z=\partial_x^2\varphi =0$, we therefore obtain a topological current $j_\text{top}\approx \frac{\gamma}{2 \pi} m^z$ of opposite sign to the left and right side of the domain wall. This necessarily leads to an accumulation of topological charge
close to the domain wall. For a moving domain wall, this explains the accumulation of $\partial_x \varphi$ associated with the motion of the domain wall, see Fig.~\ref{fig:movingDW}c) of the main text. For the localised domain wall, the extra topological charge is removed by phase-slips (space-time vortices) in a quantised way.

The accumulation of topological charge and its interplay with the domain wall motion is also the main source of interaction effects. Consider two right- and left-moving domain walls which annihilate. Each of them leaves a trace of positive and negative topological charge behind them, see Fig.~\ref{fig:worldlines} of the main text. After annihilation, this imbalance of $\rho_\text{top}$ can decay by diffusion, but this takes a very long time of order $T_d=\alpha L_d^2/\rho_s$, where $L_d$ is the typical distance of domain walls. For $L_d \gg \rho_s/(\alpha v)$, this time scale is much larger than $L_d/v$, the typical time needed for the next domain walls to arrive at the collision point. While the speed of the domain wall is not affected by a constant $\rho_\text{top}$, interactions with gradients $\partial_x \rho_\text{top}$ are strong. This can be understood qualitatively from the fact that  phase gradients also induce a spin current, Eq.~\eqref{eq:spincurrent} in the main text. Thus, extra magnetisation  $m_z$ accumulates locally whenever $\partial_x \rho_\text{top}$ is finite.
An excess of `up' magnetisation  then suppresses incoming down-domains. 

\NEW{The `active Ising model' introduced by Solon and Tailleur~\cite{activeSpin,activeSpin2} describes actively moving particles with spin which move preferentially either to the left and the right depending on the spin orientation. Therefore, this model also features a current which is proportional to the magnetisation as in Eq.~\eqref{currentMagCouling}. Nevertheless, the dynamics in the active Ising model is characterised by a phase separation of ordered and disordered regions, for which we do not find any evidence in our system. Similarly, the mechanism for a reversal of magnetisation seems to be completely different in the two models.}

\section{Resilience of the Ordered Phase to Noise}\label{app:resilience}
As discussed in the main text, even when the magnet is driven by a very weak oscillating field, its correlation length grows by many orders of magnitude and ordered domains show a remarkable resilience to sizeable noise.
There are three main reasons for this: (i) The formation of a domain wall with the opposite direction of rotation is suppressed by the dynamical frustration of the domain wall. (ii) The system has an efficient self-healing mechanism and is stable with respect to pairs of domain walls moving in a parallel direction. (iii) Pairs of domain  walls moving in antiparallel directions are very rare, as their nucleation is strongly suppressed by an effective attractive interaction of the domain walls.

In Fig.~\ref{fig:magNoise_app}, we show a simulation of 250,000 spins, lasting over $3{,}100 \, T_\text{rot}\approx 1.9\cdot 10^{6} J^{-1}$, or more than $10^6$ oscillation periods of the external magnetic field $B_z(t)$. The simulations are in a regime where the correlation length is larger than the system size. At the same noise level, the equilibrium system has a correlation length of less than 20 sites.

An enlarged view, Fig.~\ref{fig:magNoise_app}b), of the area within the dashed rectangle in Fig.~\ref{fig:magNoise_app}a) shows two types of defects. Type I defects are  pairs of domain walls which move parallel to each other with velocity $\pm v$, where $v$ is the single-domain velocity. They are visible as blue streaks in Fig.~\ref{fig:magNoise_app}b). They occur relatively frequently and can have a width of up to a few hundred sites. Nevertheless, they do not destroy the long-ranged order and are often eliminated when they meet one another.

More important are type II defects, which can be viewed as a pair of domain walls moving in {\em opposite directions}. They occur so rarely that there is only a single such event in our very long simulation.
At this defect, a spin-down domain (light blue) is nucleated within a spin-up (orange) domain, which is rapidly growing to macroscopic scales with velocity $2 v$.
At higher noise level, these events are more frequent, see Fig.~\ref{fig:magNoise}b) of the main text.

While the driven system has a very efficient self-healing mechanism for type I defects, this is not the case for type II defects, which efficiently nucleate a phase with reversed magnetisation.

\begin{figure}
    \begin{minipage}{0.95\linewidth}
    \centering 
    \begin{tikzpicture}
    \draw (0, 0) node[inner sep=0] (c) {\includegraphics[height=0.27\linewidth]
    {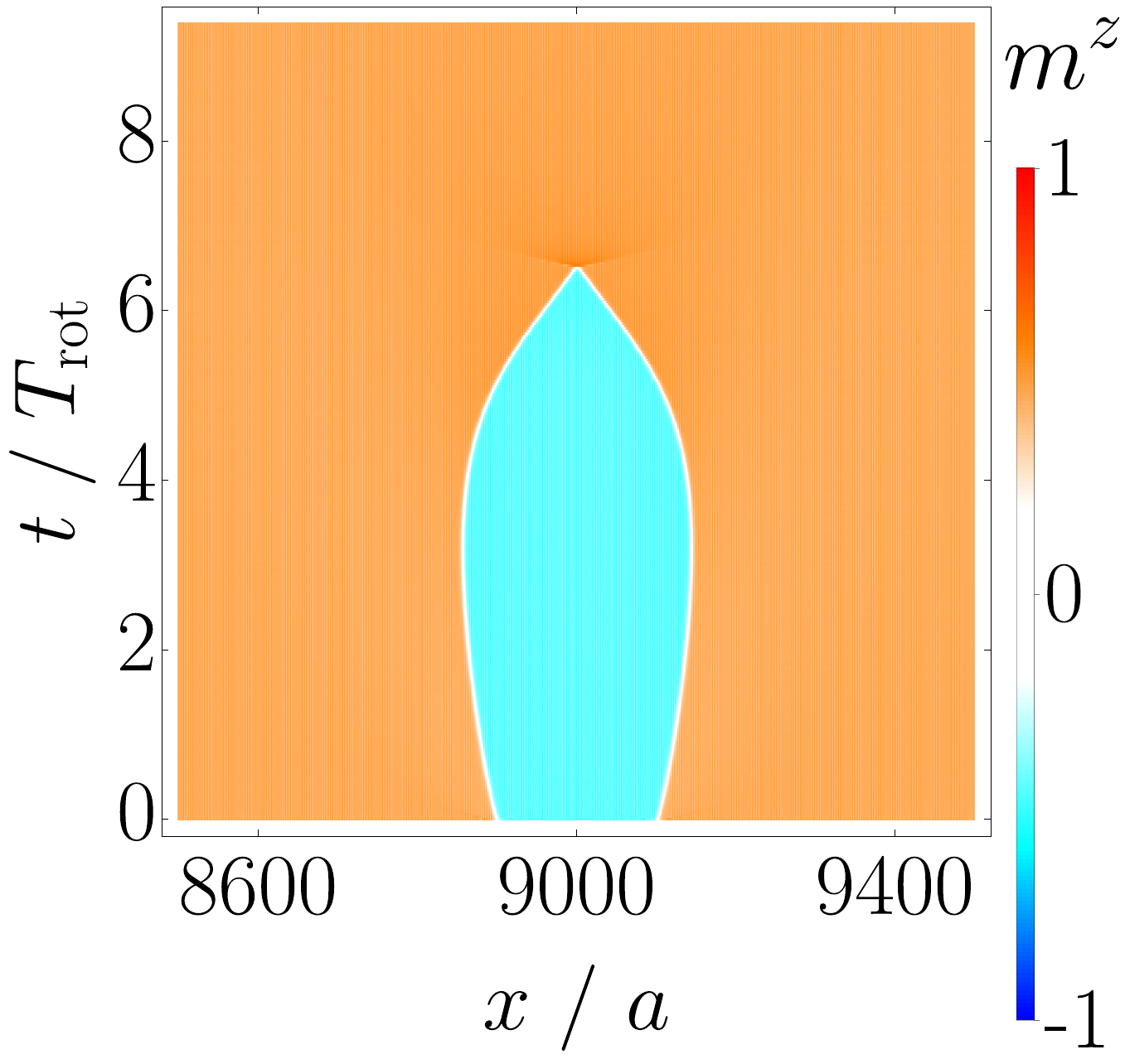}};
    \node[
      above left=-0.5cm and -0.35cm of c] {\large a)};
    \end{tikzpicture}   \hspace{0.4mm}  
    \begin{tikzpicture}
    \draw (0, 0) node[inner sep=0] (d) {\includegraphics[height=0.27\linewidth]{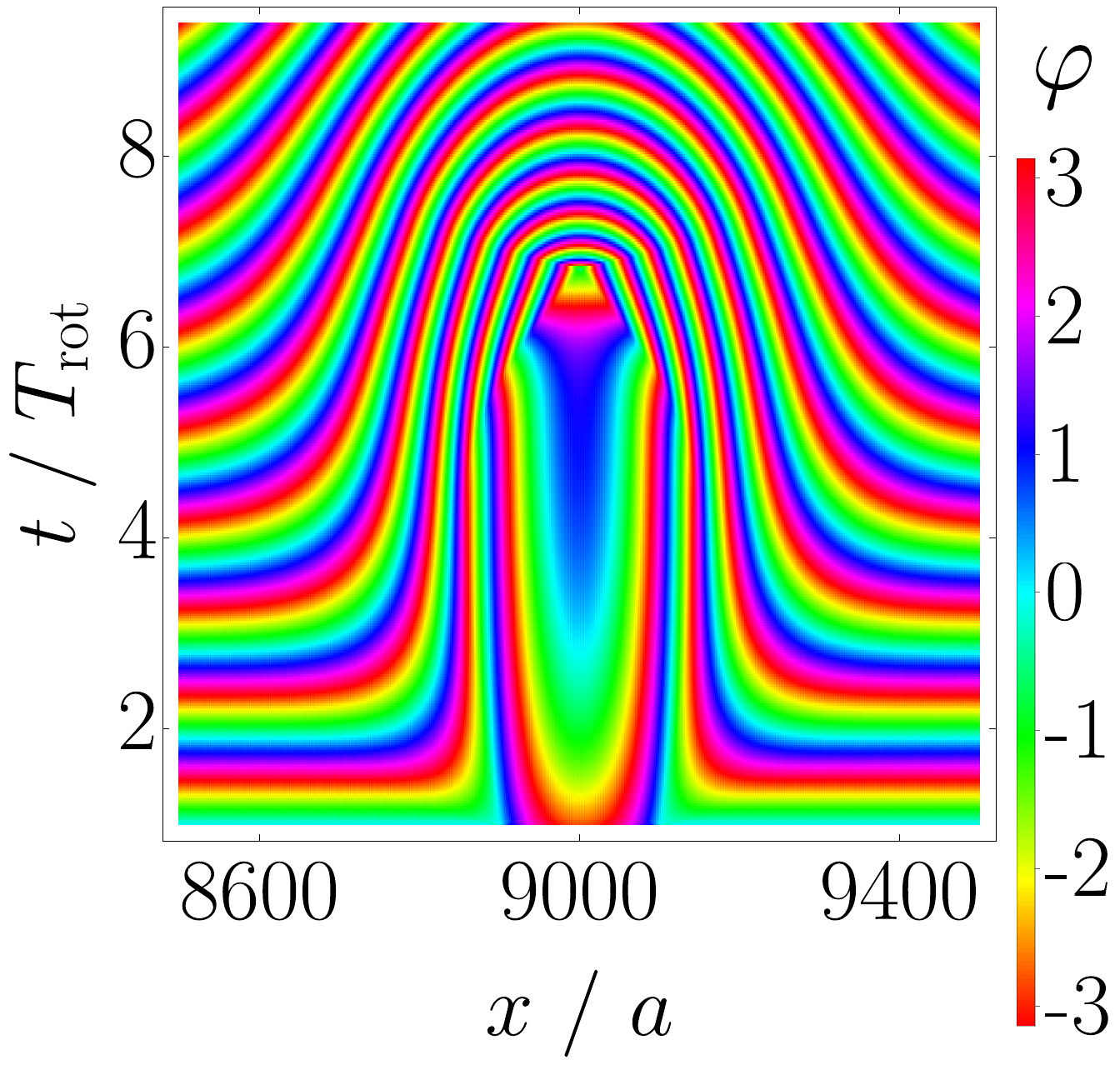}};
    \node[
      above left=-0.5cm and -0.35cm of d] {\large b)};
    \end{tikzpicture}
    \end{minipage} 
    \begin{minipage}{0.95\linewidth}
    \centering   
    \begin{tikzpicture}
    \draw (0, 0) node[inner sep=0] (c) {\includegraphics[height=0.27\linewidth]{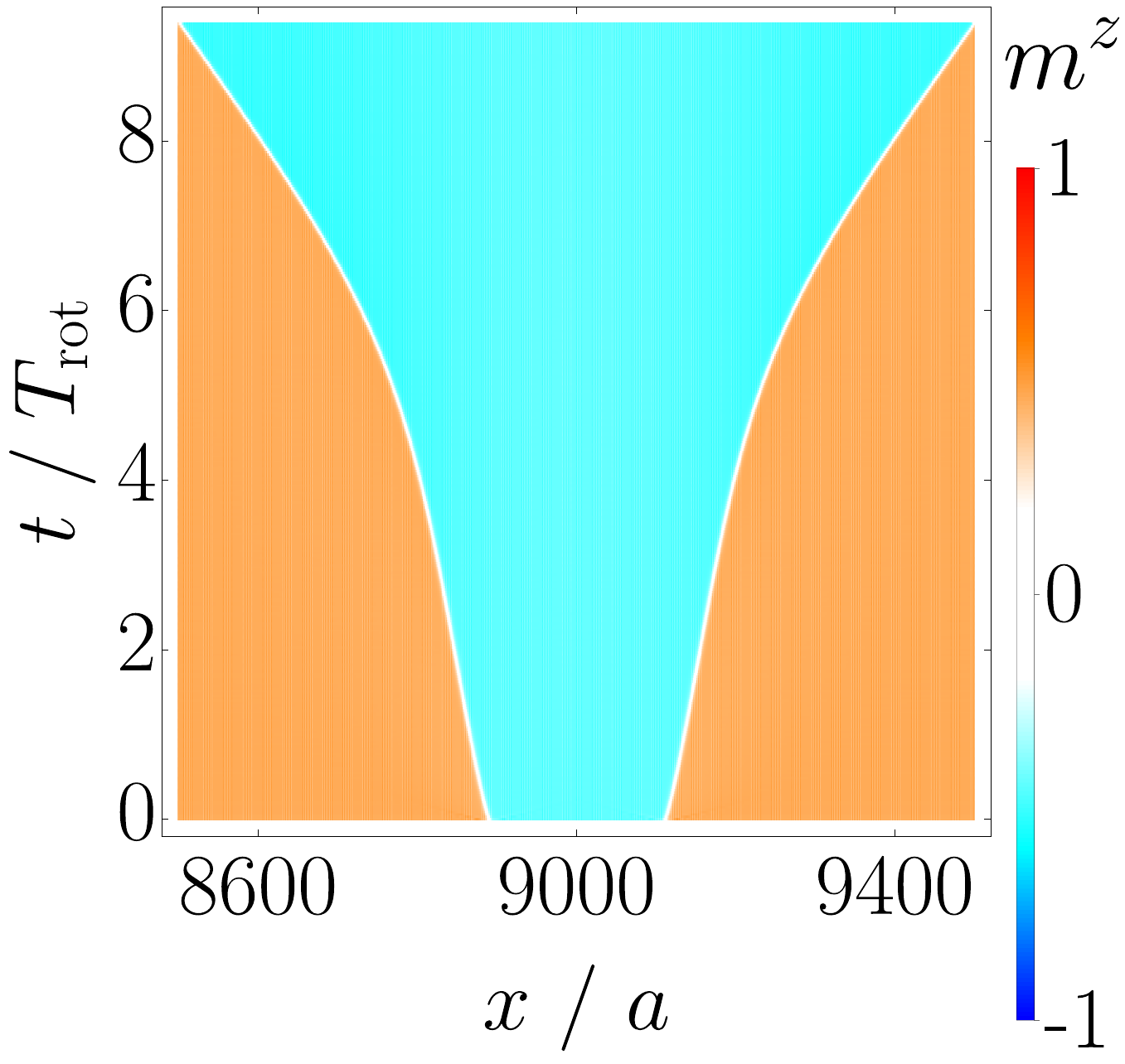}};
    \node[
      above left=-0.5cm and -0.35cm of c] {\large c)};
    \end{tikzpicture}        
    \hspace*{0.1mm}  \begin{tikzpicture}
    \draw (0, 0) node[inner sep=0] (d) {\includegraphics[height=0.27\linewidth]{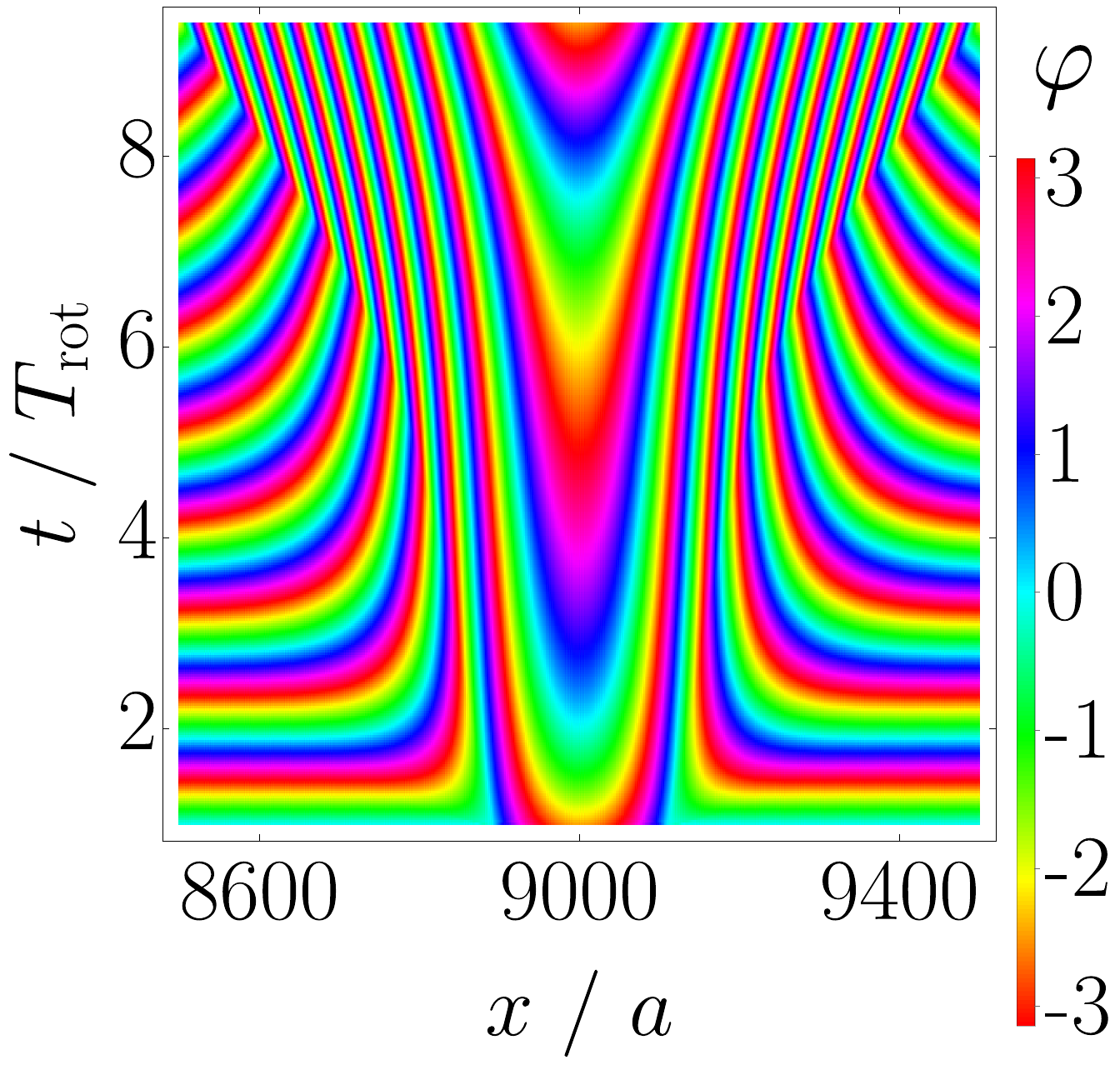}};
    \node[
      above left=-0.5cm and -0.35cm of d] {\large d)};
    \end{tikzpicture}   
    \end{minipage}
    \caption{Simulation of a type II defect in the absence of thermal noise. As an initial state, we consider a spin-down domain of width $L$ embedded in a spin-up domain. The orientation of the $xy$-order is chosen according to Eq.~\eqref{eq:phiprofile}, with $\alpha=0.4$ and $\delta_0=0.17$. The left and right panels show $m^z(x,t)$ and $\varphi(x,t)$, respectively, for $L=200$ (panels a) and b)) and $L=220$ (panels c) and d)). The smaller domain shrinks and is self-annihilating, while the larger domain continues to grow. Parameters: same as in Fig.~\ref{fig:magNoise_app} at $T=0$ for systems of 20,000 spins.}
    \label{fig:healingmechanism}
\end{figure}

As type II defects are so rare, there must be a mechanism which suppresses their creation. We will now argue that this is due to an {\em effective attraction of domain walls moving in opposite directions}, which is able to eliminate type II defects as long as their width is small.

\begin{figure}
    \centering
     \includegraphics[width=0.45 \linewidth]{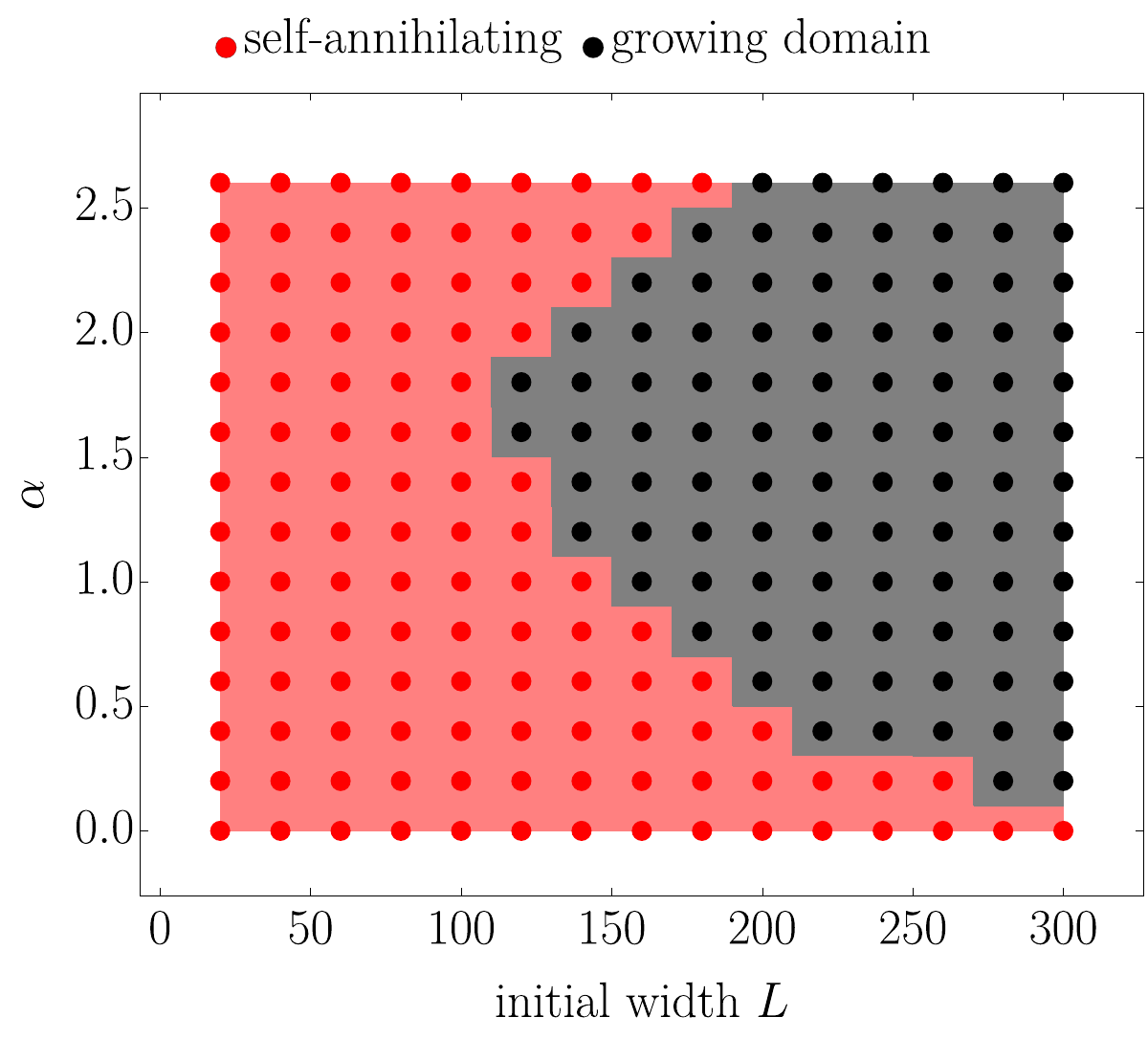}
    \label{fig:enter-label}
    \caption{Phase diagram of type II defects as a function of  $L$ and $\alpha$, which parametrise the initial width of the domain wall and the initial profile of $\varphi(x)$, see Eq.~\eqref{eq:phiprofile}. In the red region, we find solutions where the domain always shrinks and is self-annihilating, see Fig.~\ref{fig:healingmechanism}(a,b). Only the domains in the black region with a large initial width are able to grow, see Fig.~\ref{fig:healingmechanism}(c,d). The fact that only very large domains can grow explains why growing domains are rarely created by thermal noise. Parameters: same as in Fig.~\ref{fig:healingmechanism}.
}\label{fig:phasediagram_shrinkingDW}
\end{figure}

To show that, we have performed a set of simulations of a system {\em without} noise. We have initialised the system with a down-domain of width $L$ embedded in an up-domain. To trigger a motion in opposite direction, we choose as an initial condition for the orientation of the staggered $xy$-order,
\begin{align}\label{eq:phiprofile}
\varphi= \left(\frac{L}{4}-\frac{x^2}{L}\right) \alpha \delta_0 \quad \text{ for} \quad -\frac{L}{2}<x<\frac{L}{2},
\end{align}
 and $\varphi=0$ for $|x|\ge L/2$. Therefore, the gradient $\partial_x \varphi$ jumps at the domain wall boundaries by $\alpha \delta_0$. Here $\delta_0=0.17$ is the jump of the gradient obtained for a single moving domain wall in the long-time limit, see Fig.~\ref{fig:movingDW}c) of the main text.
The amplitude of the jump is directly related to the initial velocity of the domain wall, see Eq.~\eqref{eq:vGrad}.
The profile and signs are chosen such that  for  $\alpha>0$, a motion of the left domain wall to the left and of the right domain wall to the right are expected. 
In Fig.~\ref{fig:healingmechanism}, two such domains are shown. The larger one, Fig.~\ref{fig:healingmechanism}(c,d), with a width of 220 sites, indeed grows in size over time. In contrast, the smaller one, Fig.~\ref{fig:healingmechanism}(a,b), with a width of 200 sites, initially expands, but then starts to shrink and finally vanishes after $6\,T_\text{rot}$. In Fig.~\ref{fig:phasediagram_shrinkingDW},
we denote in red the parameters for which we obtain shrinking domains and in black those for which we find growing ones. The data shows that for small $L$, all domain walls are self-annihilating. Thermal fluctuations have to conspire for a long time in a  spatial interval of $O(100)$ sites to create the initial conditions for a growing type II defect.
It is therefore extremely difficult and rare to create a type II defect which grows linearly in time, as shown in Fig.~\ref{fig:magNoise_app}.  This suggests an exponentially small rate by which such defects are created. 

The origin of the collapse of smaller domains is an effective attractive interaction mediated by the inhomogeneous $\varphi(x,t)$ profile. In the centre of the domain, $\partial_x^2 \varphi<0$. Accordingly, spin currents are flowing towards the centre, $\partial_x j_z\approx \rho_s \partial_x^2\phi <0$, see Eq.~\eqref{eq:spincurrent} of the main text.  Thus, the system would like to increase the magnetisation in the center, $\partial_t m^z \approx -\partial_x j_z>0$ which favours  a {\em shrinking of the spin-down domain} in the center. As  $\partial_x^2 \varphi \propto \frac{1}{L}$ this mechanism is most effective for small domains, which explains why only domains larger than a critical size can expand.

In conclusion, the correlation length of the driven system is so large because frequently occurring type I defects (parallel moving domain walls) are self-healing. At the same time, the much more dangerous type II defects (anti-parallel moving domain walls) are strongly suppressed by an attractive interaction mediated by $\partial_x^2\varphi$. Only domains exceeding $O(100)$ sites are able to grow for the chosen parameters.

\section{Activation of the Goldstone Mode}\label{app:GostoneModeActivation}
In this section we show how driving a ferrimagnetic chain with a weak oscillating magnetic field activates the rotational Goldstone mode in the $xy$-order of the spins. We will calculate analytically the rotation frequency of the AFM $xy$-order, $\wrot$, for small amplitudes of $B_z(t)$. This serves as a particular example of a general phenomenon, expected to occur in any driven magnetic system as long as the drive breaks enough symmetries. Recent studies of other magnetic textures with `activated' translational and rotational Goldstone modes include helimagnets, skyrmions and skyrmion lattices \cite{Nina_screw,shimizu2023,Nina_jellyfish,SkyrmionGoldstoneExp}. A simplified version of the analysis presented here, valid near the critical point $m_0=0$, has been published in \cite{ZelleUniversalPhaseTransition}.

Our starting point is a ferrimagnetic chain ground state. On account of the chain's periodicity, we have $\vb{S}_i=\vb{S}_{i+2n}$ for any $n\in\mathbb{Z}$. We expect this assumption to remain valid once the system is driven, as long as the drive is weak enough that no finite-$k$ instabilities are excited \cite{Nina_screw}. Thus, it is sufficient to define two species of spin, even and odd, with $\vb{S}_e=\vb{S}_{2n}$, $\vb{S}_o=\vb{S}_{2n+1}$, and solve for their coupled dynamics. We parameterise $\vb{S}_{e/o}$ using spherical polar angles $\theta_{e/o}$ and $\phi_{e/o}$,
\begin{equation}\label{eq:spinParametrisation}
\vb{S}_{e}(t)=
\begin{pmatrix}
\sin(\theta_{e})\cos(\phi_{e})\\
\sin(\theta_{e})\sin(\phi_{e})\\
\cos(\theta_{e})
\end{pmatrix}, \quad 
\vb{S}_{o}(t)=
\begin{pmatrix}
\sin(\theta_{o})\cos(\pi+\phi_{o})\\
\sin(\theta_{o})\sin(\pi+\phi_{o})\\
\cos(\theta_{o})
\end{pmatrix}.
\end{equation}
Note that $\varphi_i$ in the main text refers to $\phi_e$ for even $i$ and $\phi_o$ for odd $i$.
In the absence of a drive, $\theta_{e/o}$ and $\phi_{e/o}$ are obtained by solving the Euler-Lagrange equations derived from the static Hamiltonian, \cref{eq:discreteFerrimagnetHamiltonian} in the main text with $B_0=0$. We find $\theta_{e/o}=\theta_0=\arccos(m_0)$, where $m_0=\sqrt{\frac{-\delta_2-2J(1-\Delta)}{\delta_4}}$, and $\phi_e=\phi_0$, $\phi_o=\phi_0+\pi$. Here, the constant $\phi_0$ can take any value, as it corresponds to the rotational Goldstone mode of the system.

Turning on the oscillating $B$-field, $B_0>0$, we now expect $\theta_{e/o}(t),\phi_{e/o}(t)$ to become functions of time. Assuming $B_0/J\ll 1$, we can expand them perturbatively around their equilibrium values,
\begin{equation}\label{eq:ferrimagnetAnglesPerturbativeExpansion}
\begin{aligned}
\theta_{e/o}(t)&=\theta_0+\epsilon\theta^{(1)}_{e/o}(t)+\epsilon^2\theta^{(2)}_{e/o}+\mathcal{O}(\epsilon^2),\\
\phi_{e/o}(t)&=\phi_0+\epsilon\phi^{(1)}_{e/o}(t)+\epsilon^2\omega_{\text{rot},e/o}t+\mathcal{O}(\epsilon^2).\\
\end{aligned}
\end{equation}
Here, $\epsilon$ is a bookkeeping parameter we introduce for doing perturbation theory, which we also place in front of the driving field term, $B_z(t)\to \epsilon B_0\cos(\omega t)$. Generically, we expect the first order response to be purely oscillatory, with $\theta^{(1)}_{e/o}(t)$, $\phi^{(1)}_{e/o}(t)\sim e^{\pm i\omega t}$ oscillating at the same frequency as the driving frequency. By the same logic, at second order, the response should either be oscillatory again, with twice the frequency, $e^{\pm 2 i\omega t}$, or DC, $e^{0 i\omega t}$. More interestingly, the $\mathcal{O}(\epsilon^2)$ response can also grow \emph{linearly} in $t$ if the mode in question is a Goldstone mode of the system. 
Thus, we add the $\omega_{\text{rot},e/o}t$ terms, which physically describe a rotation in time of the $xy$-order around the $S^z_i$-axis, to our ansatz \cref{eq:ferrimagnetAnglesPerturbativeExpansion} at second order in $\epsilon$. The goal of our calculation is to obtain an analytical expression for the rate of rotation $\omega_{\text{rot},o/e}$ and to find out how it depends on the system and drive parameters.

To this end, we first cross the LLG, \cref{eq:LLG} in the methods section, with $\vb{S}_{e/o}$, and then project it onto $\frac{\partial \vb{S}_{e/o}}{\partial \theta_{e/o}}$, $\frac{\partial \vb{S}_{e/o}}{\partial \phi_{e/o}}$, thus obtaining four coupled equations of motion for $\theta_{e/o},\phi_{e/o}$,
\begin{align}
\dot{\phi}_{e/o}\sin(\theta_{e/o})-\alpha\dot{\theta}_{e/o}&=\SignM{-}\frac{\delta H}{\delta \theta_{e/o}}, \label{eq:LLGevenOddSpinsThetaEq}\\
-\sin(\theta_{e/o})\dot{\theta}_{e/o}-\alpha\dot{\phi}_{e/o}\sin^2(\theta_{e/o})&=\SignM{-}\frac{\delta H}{\delta \phi_{e/o}}. \label{eq:LLGevenOddSpinsPhiEq}
\end{align}
Due to the $\vb{S}_{i}=\vb{S}_{i+2n}$ symmetry, we only need to consider a sub-Hamiltonian consisting of any two adjacent sites in the infinite chain \cref{eq:discreteFerrimagnetHamiltonian} in the main text. We denote this two-site Hamiltonian as $\mathcal{H}_{\text{2 sites}}$. Splitting it into its static and driven components, $\mathcal{H}_{\text{2 sites}}=\mathcal{H}_{\text{stat}}+\mathcal{H}_{\text{drive}}$, we have
\begin{equation}\label{eq:freeEnergyDiscrete}
    \begin{aligned}
\mathcal{H}_{\text{stat}}&=J\Bigg[-2\big(\sin(\theta_e)\sin(\theta_o)\cos(\phi_e-\phi_o)+\Delta\cos(\theta_e)\cos(\theta_o)\big)\Bigg]\\
        &+\frac{1}{4}\Bigg[\left(\delta_2+\frac{\delta_4}{2}\right)\left(\cos(2\theta_e)+\cos(2\theta_o)\right)+\frac{\delta_4}{8}\left(\cos(4\theta_e)+\cos(4\theta_o)\right)\Bigg],\\
        \mathcal{H}_{\text{drive}}&=-\epsilon B_1(t)\left(g_e\cos(\theta_e)+g_o\cos(\theta_o)\right).
    \end{aligned}
\end{equation}
Note that for an undriven static system, $\dot{\theta}_{e/o}=\dot{\phi}_{e/o}=0$, so that \cref{eq:LLGevenOddSpinsThetaEq,eq:LLGevenOddSpinsPhiEq} are equivalent to energy minimisation, $\frac{\partial H}{\partial \theta_{e/o}}=\frac{\partial H}{\partial \phi_{e/o}}=0$. We now substitute the spin parametrisation \cref{eq:spinParametrisation}, together with the perturbative expansion \cref{eq:ferrimagnetAnglesPerturbativeExpansion}, into \cref{eq:LLGevenOddSpinsThetaEq,eq:LLGevenOddSpinsPhiEq}, and solve perturbatively in powers of $\epsilon$. At linear order in $\epsilon$, the $\theta^{(1)}_{e/o}(t),\phi^{(1)}_{e/o}(t)$ fields decompose into two $\pm \omega$ Fourier frequency components,
\begin{equation}
\begin{aligned}
\theta^{(1)}_{e/o}(t)&=\tilde{\theta}_{e/o}e^{i\omega t}+\tilde{\theta}^*_{e/o}e^{-i\omega t},\\
\phi^{(1)}_{e/o}(t)&=\tilde{\phi}_{e/o}e^{i\omega t}+\tilde{\phi}^*_{e/o}e^{-i\omega t},
\end{aligned}
\end{equation}
where the coefficients of the $e^{\pm i\omega t}$ are complex conjugates of each other, ensuring that $\theta^{(1)}_{e/o}(t),\phi^{(1)}_{e/o}(t)$ are real. We now solve for the $\tilde{\theta}_{e/o},\tilde{\phi}_{e/o}$ Fourier coefficients by comparing the coefficients of the $e^{i\omega t}$ terms in the four coupled equations in \cref{eq:LLGevenOddSpinsThetaEq,eq:LLGevenOddSpinsPhiEq}, giving
\begin{equation}\label{eq:FirstOrderFieldsAntiferromagneticChainSols}
\scalebox{.89}{
$\begin{aligned}
&\tilde{\phi}^{(1)}_{e/o}=-\frac{B_0}{2}\Bigg[\frac{i (g_e+g_o)}{i \alpha  \delta_4 \cos (4 \theta_0)-i \alpha  \cos (2 \theta_0) (-2\delta_2-\delta_4+4 (\Delta -1) J)\SignM{-}2 \left(\alpha ^2+1\right) \omega}\\
&+\frac{i\omega(g_{e/o}-g_{o/e})}{2 \left(\alpha ^2+1\right) \omega^2-16 (\Delta +1) J^2+(2 \delta_2+\delta_4) \cos (2 \theta_0) (4 J\SignM{-}i \alpha  \omega)+\delta_4 \cos (4 \theta_0) (4 J\SignM{-}i \alpha  \omega)\SignM{+}4 i \alpha  (\Delta +3) J \omega}\Bigg],\\
&\tilde{\theta}^{(1)}_{e/o}=\sin(\theta_0)\Bigg[-\alpha\tilde{\phi}^{(1)}_{e/o}+\\
&\frac{B_0}{2}\frac{4 J(g_{e/o}-g_{o/e})}{2 \left(\alpha ^2+1\right) \omega^2-16 (\Delta +1) J^2+(2 \delta_2+\delta_4) \cos (2 \theta_0) (4 J\SignM{-}i \alpha  \omega)+\delta_4 \cos (4 \theta_0) (4 J\SignM{-}i \alpha  \omega)\SignM{+}4 i \alpha  (\Delta +3) J \omega}\Bigg].
\end{aligned}$
}
\end{equation}
\emph{Check of \cref{eq:FirstOrderFieldsAntiferromagneticChainSols} using the continuity equation ---} As mentioned in the main text, there is a Noether current associated with the Goldstone mode, $J^{\mu}=\frac{\partial \mathcal{L}}{\partial \partial_{\mu}\phi_e}+\frac{\partial \mathcal{L}}{\partial \partial_{\mu}\phi_o}$, which satisfies the continuity equation $\partial_{\mu}J^{\mu}=0$, with $\mu=\{t,x,y,z\}$, when $\alpha=0$. 
We have $J^t=m^z_e+m^z_o$, and $J^{x,y,z}=0$, as we assumed the ferrimagnetic chain to be translationally invariant. Thus, the continuity equation is just $\dot{m}^z_e+\dot{m}^z_o=0$. This should be satisfied at all orders of $\epsilon$. At first order in $\epsilon$, the equation reads $\sin(\theta_0)\left(\dot{\theta}^{(1)}_e+\dot{\theta}^{(1)}_o\right)=0$. Setting $\alpha=0$ in \cref{eq:FirstOrderFieldsAntiferromagneticChainSols}, we see that $\theta^{(1)}_{e/o}(t)\sim (g_{e/o}-g_{o/e})$ implies that this is indeed satisfied.

\begin{figure}
    \centering
     \includegraphics[width=0.45 \linewidth]{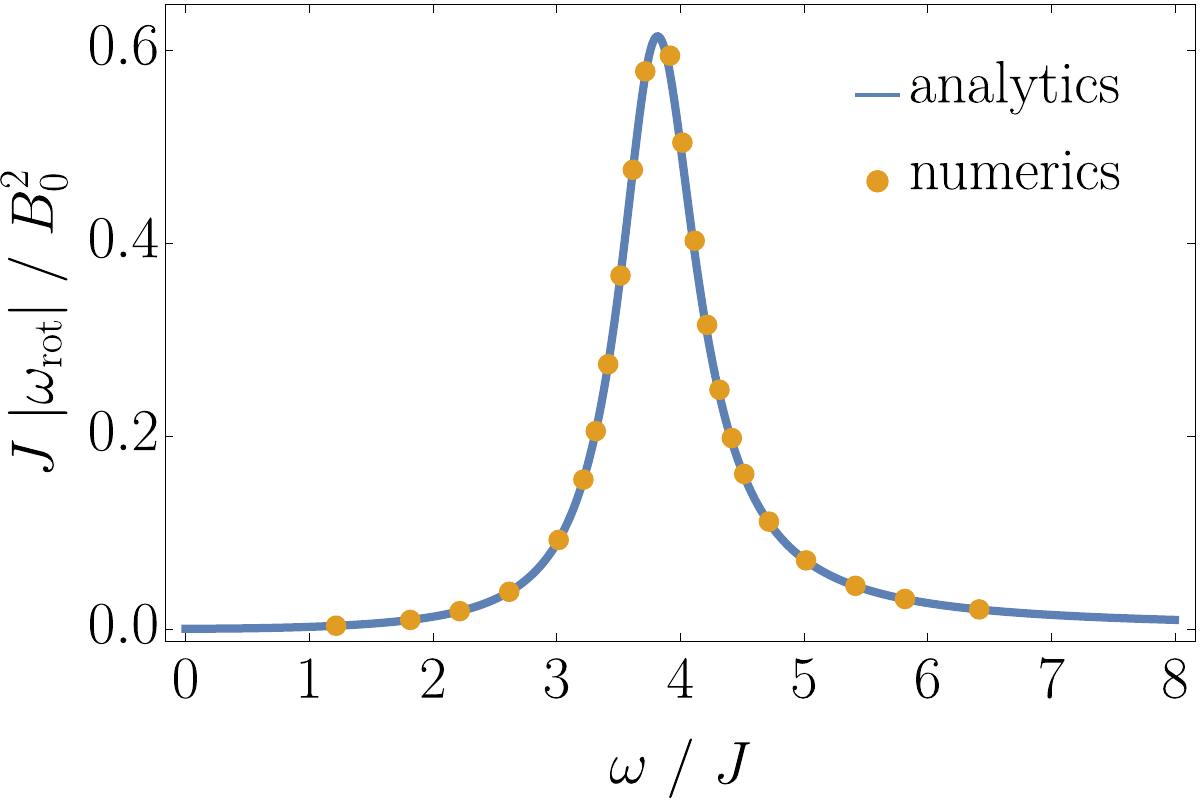}
    \caption{Comparison of numerical and analytical results for the rotation rate of the $xy$-order, $\varphi=\wrot t$, here shown as a function of the frequency $\omega$ of the oscillating magnetic field, $B_z(t)=B_0 \cos(\omega t)$. For small $B_0$, a perfect agreement of analytical and numerical results is obtained. Parameters: $J=1,\Delta=0.8,\delta_2=-0.6,\delta_4=1,g_1=1,g_2=0.1,\alpha=0.1,B_0=0.1$, $T=0$.}\label{fig:omegarotvsomega}
\end{figure}
We now move on to the second order in $\epsilon$ response. Here, we are only interested in obtaining $\omega_{\text{rot},e/o}$. To this end, we average \cref{eq:LLGevenOddSpinsPhiEq} over one period of oscillation $T=2\pi/\omega$ to remove all oscillating terms and consider only the $\mathcal{O}(\epsilon^2)$ terms, yielding
\begin{equation}
\begin{aligned}
\sin^2(\theta_0)\omega_{\text{rot},e/o}=-2\sin(\theta_0)\cos(\theta_0)\left\langle\dot{\phi}^{(1)}_{e/o}\theta^{(1)}_{e/o}+\frac{1}{\alpha}\left(\theta^{(1)}_e+\theta^{(1)}_o\right)\left(\phi^{(1)}_{e/o}-\phi^{(1)}_{o/e}\right)\right\rangle_T.
\end{aligned}
\end{equation}
Substituting in \cref{eq:FirstOrderFieldsAntiferromagneticChainSols}, after some algebra we can see that $\omega_{\text{rot},e}=\omega_{\text{rot},o}=\omega_{\text{rot}}$, which can be written in the more symmetric form
\begin{equation}\label{eq:omegaRotDef}
\omega_{\text{rot}}=-\frac{\cos(\theta_0)}{\sin(\theta_0)}\left\langle\dot{\phi}^{(1)}_e\theta^{(1)}_e+\dot{\phi}^{(1)}_o\theta^{(1)}_o\right\rangle_T.
\end{equation}
In the limit of small $m_0=\cos(\theta_0)$, $\sin(\theta_0)\approx 1$, so $\omega_{\text{rot}}\propto m_0$. By construction, we also have $\wrot\propto B^2_0$. A comparison of this analytical finding to numerical data generated with Mumax3 is shown in \cref{fig:omegarotvsomega}. The agreement is very good.

\section{Flying Domain Wall Velocity in the Presence of Finite Damping}\label{app:DomainWallVelocity}

Having studied the dynamics of the simple ferrimagnetic chain in App.~\ref{app:GostoneModeActivation}, we are now prepared to tackle the challenge of calculating the dynamics of the driven ferrimagnetic domain wall. Far away from the domain wall, the spins rotate in the (anti)clockwise directions at the rate $\pm|\wrot|$, depending on the sign of $m^z$. This creates a dynamical frustration at the centre of the system, which the domain wall resolves by acquiring a horizontal velocity $v$. In the main text, we consider the limit of vanishing damping $\alpha$ and calculate $v$ by integrating the conserved Noether current $\partial_{\mu}J^{\mu}=0$ over all space, and find that $v\sim \sqrt{|\wrot|}$ for $\alpha \to 0$. 

In the presence of finite damping, the Noether current is no longer conserved: instead we have $\partial_{\mu}J^{\mu}=\mathcal{O}(\alpha)$. Nevertheless, the velocity can still be calculated approximately. We present the full calculation in this appendix, and show that a qualitatively new regime emerges where $v\sim|\wrot|/\alpha$, when damping dominates driving, $|\wrot| \lesssim \alpha^2$. 

\subsection{Derivation of the Continuum Description}

Let us set up the calculation. The presence of the domain wall introduces a spatial dependence (in addition to time dependence from the driving) to the fields $\theta_{e/o},\phi_{e/o}$. To simplify the calculation, we make the fields continuous functions of space and define a Hamiltonian density $\mathcal{H}$, such that $H=\int\mathop{dx}\mathcal{H}$. This is a justified approximation as long as the lattice spacing $a$ is much smaller than the domain wall width $\xi_0$. We Taylor expand fields on the sites $x\pm a$ adjacent to site $x$ up to second order in the spatial derivatives,
\begin{equation}
\begin{aligned}
\theta_{e/o}(x\pm a,t)&=\theta_{e/o}(x,t)\pm a\theta'_{e/o}(x,t)+\frac{a^2}{2}\theta''_{e/o}(x,t)+\mathcal{O}(a^3),\\
\phi_{e/o}(x\pm a,t)&=\phi_{e/o}(x,t)\pm a\phi'_{e/o}(x,t)+\frac{a^2}{2}\phi''_{e/o}(x,t)+\mathcal{O}(a^3).
\end{aligned}
\end{equation}
We obtain the Heisenberg contribution to the Hamiltonian density from
\begin{equation}
\begin{aligned}
\mathcal{H}&=\frac{J}{2a}\sum_{\Delta x=\pm a}\left(S^x_e(x)S^x_o(x+\Delta x)+S^y_e(x)S^y_o(x+\Delta x)-\Delta S^z_e(x)S^z_o(x+\Delta x)\right)\\
&\qquad+\Big(S^x_o(x)S^x_e(x+\Delta x)+S^y_o(x)S^y_e(x+\Delta x)-\Delta S^z_o(x)S^z_e(x+\Delta x)\Big).
\end{aligned}
\end{equation}
The continuum version of the Hamiltonian density therefore reads
\begin{equation}\label{eq:continuousFreeEnergy}
\begin{aligned}
\mathcal{H}&=\frac{J}{a}\Bigg[-2\left(\sin(\theta_e)\sin(\theta_o)\cos(\phi_e-\phi_o)+\Delta\cos(\theta_e)\cos(\theta_o)\right)\\
        &\qquad+\frac{1}{2}a^2 \bigg(\sin(\phi_e-\phi_o)\sin(\theta_e)\sin(\theta_o)(\phi_e''-\phi_o'')\\
        &\qquad+\sin(\theta_e)\sin(\theta_o)\cos(\phi_e-\phi_o)\left((\phi_o')^2+(\phi_e')^2\right)\\
        &\qquad+2\sin(\phi_e-\phi_o)\left(\cos(\theta_e)\sin(\theta_o)\theta_e'\phi_e'-\cos(\theta_o)\sin(\theta_e)\theta_o'\phi_o'\right)\\
        &\qquad+\left(\sin(\theta_e)\sin(\theta_o)\cos(\phi_e-\phi_o)+\Delta\cos(\theta_e)\cos(\theta_o)\right)\left((\theta_e')^2+(\theta_o')^2\right)\\
        &\qquad+\left(\Delta\cos(\theta_o)\sin(\theta_e)-\cos(\theta_e)\sin(\theta_o)\cos(\phi_e-\phi_o)\right)\theta_e''\\
        &\qquad+\left(\Delta\cos(\theta_e)\sin(\theta_o)-\cos(\theta_o)\sin(\theta_e)\cos(\phi_e-\phi_o)\right)\theta_o''\bigg)+\mathcal{O}(a^3)\Bigg]\\
        &+\frac{1}{4a}\left(\left(\delta_2+\frac{\delta_4}{2}\right)\left(\cos(2\theta_e)+\cos(2\theta_o)\right)+\frac{\delta_4}{8}\left(\cos(4\theta_e)+\cos(4\theta_o)\right)\right).
\end{aligned}
\end{equation}
Note that the energy terms due to anisotropies and driving $B$-field remain unchanged, as they are local. 

\subsection{Conserved Noether Current}\label{subsubsec:conservedNoetherCurrent}
Before stating and solving the continuous LLG equations resulting from \cref{eq:continuousFreeEnergy}, we derive the Noether current associated with the Goldstone mode of the system. The kinetic energy density of the spins is given by
\begin{equation}
    \mathcal{T}=\frac{1}{2a}\left(\boldsymbol{\omega}^T_e\mathcal{I}_e\boldsymbol{\omega}_e+\boldsymbol{\omega}_o^T\mathcal{I}_o\boldsymbol{\omega}_o\right),
\end{equation}
where $\mathcal{I}$ is the inertia tensor and $\boldsymbol{\omega}_{o/e}=(\omega^x_{o/e},\omega^y_{o/e},\omega^z_{o/e})^T$ is the angular velocity vector. Note that $\dot{\phi}_{o/e}=\omega^z_{o/e}$. The spin angular momentum vector is related to the angular velocity via $\vb{S}=\mathcal{I}\boldsymbol{\omega}$. We also have $\vb{S}=\sign(\gamma)\frac{M_0a^3}{|\gamma|}\vb{m}$, where $\gamma=-|e|g/2m_e$ is the gyromagnetic ratio for an electron with charge $-|e|$, mass $m_e$ and $g$-factor $g$. As we set $\hbar=1$ in our equations, we have $-\vb{m}=\mathcal{I}\boldsymbol{\omega}$. The Lagrangian density is given by $\mathcal{L}=\mathcal{T}-\mathcal{V}$, where $\mathcal{V}$ is the same as \cref{eq:continuousFreeEnergy}. Let us define the symmetric and antisymmetric combinations
\begin{equation} \label{eq:GoldstoneModeVariableDef}
        \phi_{S/A}=\frac{1}{2}\left(\phi_e\pm\phi_o\right), \quad m^z_{S/A}=\frac{1}{2}\left(m^z_e\pm m^z_o\right).
\end{equation}
Note that $\phi_S$ is the Goldstone mode, as $\mathcal{L}$ is independent of it. Next, we calculate the associated Noether current density, $J^{\mu}=\frac{\partial\mathcal{L}}{\partial\partial_{\mu}\phi_S}$. For $J^t$ we have
\begin{equation*}
J^t=\frac{\partial\mathcal{L}}{\partial\dot{\phi}_S}=\frac{\partial\mathcal{T}}{\partial\dot{\phi}_S}=\frac{1}{a}\left(I_e(\dot{\phi}_S+\dot{\phi}_A)+I_o(\dot{\phi}_S-\dot{\phi}_A)\right)=-\frac{2m^z_{S}}{a}.
\end{equation*}
For the spatial components $J^{x,y,z}$, we have $J^{y,z}=0$ and
\begin{equation*}
\begin{aligned}
J^x&=\frac{\partial\mathcal{L}}{\partial\phi'_S}=-\frac{\partial\mathcal{V}}{\partial\phi'_S}\\
&=-\frac{1}{a}Ja^2\left[2\sin(\theta_e)\sin(\theta_o)\cos(2\phi_A)\phi'_S+\sin(2\phi_A)\left(\cos(\theta_e)\sin(\theta_o)\theta_e'+\cos(\theta_o)\sin(\theta_e)\theta_o'\right) \right].
\end{aligned}
\end{equation*}
To leading order in $B_0$, we have $\phi_A=0$, $\sin(\theta_{e/o}(x))=\sin(\theta_0(x))$. Thus, the leading order contribution to $J^x$ reads
\begin{equation}\label{eq:NoetherCurrent}
    J^x\simeq -\frac{1}{a}2Ja^2(1-(m^z)^2)\phi_S',
\end{equation}
where we replaced $\sin^2(\theta_0)=(1-(m^z)^2)$. Substituting $J^{t,x}$ into the continuity equation $\partial_t J^t+\partial_xJ^x=0$, we obtain
\begin{equation}
   \partial_t(m^z_S)+Ja^2\partial_x\left((1-(m^z)^2)\phi_S'\right)=0,
\end{equation}
as stated in the main text. Note that the spin current in \cref{eq:spincurrent} in the main text is defined with a relative minus compared to the Noether current in \cref{eq:NoetherCurrent}, $j^z=-J^x$.

\subsection{LLG Equations for the Continuum Model}
As we mostly use the magnetisation in the main text we want to use the corresponding equations within this section and onwards. This boils down to a sign change on the RHS of \cref{eq:LLGevenOddSpinsThetaEq} and \cref{eq:LLGevenOddSpinsPhiEq} in accordance with the relation shown in subsection~\ref{subsubsec:conservedNoetherCurrent}.
Substituting \cref{eq:continuousFreeEnergy} into the modified \cref{eq:LLGevenOddSpinsThetaEq} and \cref{eq:LLGevenOddSpinsPhiEq}, we obtain the four coupled EoMs for the continuum model,
\begin{align}
\dot{\phi}_{e/o}\sin(\theta_{e/o})&-\alpha\dot{\theta}_{e/o}=-\sin (\theta_{e/o}) \left(\delta_2 \cos (\theta_{e/o})+\delta_4 \cos ^3(\theta_{e/o})\right) \nonumber \\
&+2 J \left(\Delta \sin(\theta_{e/o})\cos(\theta_{o/e})-\cos (\theta_{e/o}) \sin (\theta_{o/e}) \cos (\phi_{e/o}-\phi_{o/e})\right)\nonumber \\
&-a^2 J \Bigg( \nonumber \\
&\theta_{o/e}'' (\Delta  \sin (\theta_{e/o}) \sin (\theta_{o/e})+\cos (\theta_{e/o}) \cos (\theta_{o/e}) \cos (\phi_{e/o}-\phi_{o/e})) \nonumber \\
&+(\theta_{o/e}')^2 (\Delta  \sin (\theta_{e/o}) \cos (\theta_{o/e})-\cos (\theta_{e/o}) \sin (\theta_{o/e}) \cos (\phi_{e/o}-\phi_{o/e})) \nonumber \\
&+2 \theta_{o/e}'\phi_{o/e}' \cos (\theta_{e/o}) \cos (\theta_{o/e}) \sin (\phi_{e/o}-\phi_{o/e}) \nonumber\\
&+\cos (\theta_{e/o}) \sin (\theta_{o/e}) \phi_{o/e}''(x) \sin (\phi_{e/o}-\phi_{o/e}) \nonumber\\
&-\cos (\theta_{e/o}) \sin (\theta_{o/e}) (\phi_{o/e}')^2 \cos (\phi_{e/o}-\phi_{o/e})\Bigg)+\epsilon g_{e/o}B_1(t)\sin(\theta_{e/o}), \label{eq:ThetaEoMModifiedContinuous}\\
-\sin(\theta_{e/o})\dot{\theta}_{e/o}&-\alpha\dot{\phi}_{e/o}\sin^2(\theta_{e/o})=2J\sin(\theta_{e})\sin(\theta_{o})\sin(\phi_{e/o}-\phi_{o/e}) \nonumber \\
&-J a^2\sin (\theta_{e/o}) \Bigg(\sin (\phi_{e/o}-\phi_{o/e}) \left(\sin (\theta_{o/e}) \phi_{o/e}'^2- \theta_{o/e}'' \cos (\theta_{o/e})\right)\nonumber \\
&+2 \theta_{o/e}' \cos (\theta_{o/e}) \phi_{o/e}' \cos (\phi_{e/o}-\phi_{o/e})+ \theta_{o/e}'^2 \sin (\theta_{o/e}) \sin (\phi_{e/o}-\phi_{o/e})\nonumber \\
&+ \sin (\theta_{o/e}) \phi_{o/e}''\cos (\phi_{e/o}-\phi_{o/e})\Bigg). \label{eq:PhiEoMModifiedContinuous}
\end{align}
Despite looking impossible to solve at first glance, $v$ can be extracted from \cref{eq:ThetaEoMModifiedContinuous,eq:PhiEoMModifiedContinuous} using the right ansatzes and approximations. For completeness, we solve them first in the undriven case (static domain wall), before moving on to the driven problem (moving domain wall).

\subsection{Zeroth Order Solution: Static Domain Wall}
In the absence of driving, \cref{eq:PhiEoMModifiedContinuous} is trivially solved by setting $\phi_{e/o}=\text{const.}$ everywhere. \cref{eq:ThetaEoMModifiedContinuous} is symmetric under the swap $\theta_e\leftrightarrow\theta_o$, from which we conclude that $\theta_e=\theta_o=\theta_0$, letting us rewrite it as
\begin{equation}
\begin{aligned}
0&=\sin(\theta_0)\cos(\theta_0)\left(2J\left(\Delta-1\right)-\delta_2-\delta_4\cos^2(\theta_0)\right)\\
&-a^2J\left(\theta_0''\left(\Delta\sin^2(\theta_0)+\cos^2(\theta_0)\right)+(\theta_0')^2\sin(\theta_0)\cos(\theta_0)\left(\Delta-1\right)\right).
\end{aligned}
\end{equation}
By replacing $\cos(\theta_0)=m_0\tilde{m}(\tilde{x})$ and $2J(\Delta-1)-\delta_2=\delta_4 m_0^2$, with the rescaled position coordinate $\tilde{x}=m_0x$, we can write this as
\begin{equation}\label{eq}
\begin{aligned}
0&=\left(\delta_4\tilde{m}(1-\tilde{m}^2)+a^2J\Delta\tilde{m}''\right)m_0^3+\mathcal{O}\left(m_0^5\right).
\end{aligned}
\end{equation}
For small $m_0$ (close to the phase transition \cite{soliton1998}), we can ignore the $m_0^5$ term. In this case, it is easy to check that the ansatz
\begin{equation}
\tilde{m}(\tilde{x})=\tanh\left(\frac{\tilde{x}}{a}\sqrt{\frac{\delta_4}{2J\Delta}}\right)
\end{equation}
solves the equation. Thus, switching back to the original coordinates we have 
\begin{equation}\label{eq:ThetaNearCriticalPoint}
\theta_0(x)=\arccos\left(m_0\tanh\left(x/\xi_0\right)\right)+\mathcal{O}((m_0)^5), \qquad \xi_0=\frac{a}{m_0}\sqrt{\frac{2J\Delta}{\delta_4}}.
\end{equation}
\subsection{Driven Solution: Moving Domain Wall}\label{appD:subsecMovingDomainWallAnsatz}
We now switch on the driving field $B_z(t)$. We modify the ansatz \cref{eq:ferrimagnetAnglesPerturbativeExpansion} to allow for a traveling solution,
\begin{equation}\label{eq:DomainWallAnglesPerturbativeExpansion}
\begin{aligned}
\theta_{e/o}(x,t)&=\theta_0(x-vt)+\Theta_{e/o}(x-vt)+\epsilon\theta^{(1)}_{e/o}(x-vt,t)+\epsilon^2\theta^{(2)}_{e/o}(x-vt)+\mathcal{O}(\epsilon^2),\\
\phi_{e/o}(x,t)&=\Phi_{e/o}(x-vt)+\epsilon\phi^{(1)}_{e/o}(x-vt,t)+\epsilon^2\wtilde t+\mathcal{O}(\epsilon^2).\\
\end{aligned}
\end{equation}
Here, $\theta_0$ is the same function as the one we solved for in the static case, \cref{eq:ThetaNearCriticalPoint}. $\Theta_{e/o}$ and $\Phi_{e/o}$ are new functions, whose exact dependence on $B_0$ we are yet to find out, but which we know vanish when $B_0=0$. The Goldstone mode activation principle still implies that all $\phi_{e/o}$ may acquire a global rotational velocity, described by the $\wtilde t$ term. Note that $\wtilde$ by itself is \emph{not} the same as $\wrot$, as there will also be some rotation coming from the $\Phi(x-vt)$ term. In addition, we still have the $\mathcal{O}(B_0)$ oscillating terms $\theta_{e/o}^{(1)}(x-vt,t),\phi_{e/o}^{(1)}(x-vt,t)$, which tend to the ferrimagnetic chain values \cref{eq:FirstOrderFieldsAntiferromagneticChainSols} far away from the domain wall, $|x|\gtrsim\xi_0$.

\subsubsection{Time-Averaged Equations of Motion}
Next, we solve \cref{eq:ThetaEoMModifiedContinuous,eq:PhiEoMModifiedContinuous} in the presence of driving, $B_0>0$. By time averaging the two equations over $T=2\pi/\omega$, we get rid of any oscillating terms. We transform to a moving frame of reference where $\tilde{x}=x-vt$, and immediately drop the tilde on the $\tilde{x}$ for notational simplicity. Keeping only leading order terms linear in $\Phi_{e/o},\Theta_{e/o}$, as well as the leading order $\mathcal{O}(B_0^2)$ terms resulting from products of the linear response fields $\theta^{(1)}_{e/o},\phi^{(1)}_{e/o}$, we have
\begin{align}
(1-m^2)\bigg(-v\Phi'_{e/o}+\wtilde &+|\wrot|\frac{m}{|m_0|}+f_{e/o}\bigg)+\alpha v\left(-m'+\partial_x\left(\sqrt{1-m^2}\Theta_{e/o}\right)\right)\nonumber\\
&=\sqrt{1-m^2}\Bigg[\nonumber\\
&\delta_4\left(2m^2(m_0^2-2m^2)+3m^2-m_0^2\right)(\Theta_{e/o}+\theta^{(2)}_{e/o})\nonumber\\
&+a^2 J \Bigg(-(\Theta''_{e/o}+\theta^{(2)''}_{e/o})\left(\Delta(1-m^2)+m^2\right)\nonumber\\
&+2(\Delta-1) (\Theta'_{e/o}+\theta^{(2)'}_{e/o})m'm\nonumber\\
&+(\Delta -1)(\Theta_{e/o}+\theta^{(2)}_{e/o})\left(2mm''+\frac{(m')^2}{1-m^2}\right)\Bigg)\Bigg]+g_{e/o},\label{eq:eqThetaLinear}\\
\left(\alpha v+Ja^2\partial_x\right)\left((1-m^2)\Phi'_{e/o}\right) &-v\partial_x\left(m-\left(\sqrt{1-m^2}\Theta_{e/o}\right)\right)\nonumber\\
&= (1-m^2)\left(\alpha\left(\wtilde+|\wrot|\frac{m}{|m_0|}\right)+f_{e/o}\right)\label{eq:eqPhiLinear},
\end{align}
with
% \begin{equation*}
% \begin{aligned}
% g_{e/o}(x)&=\left\langle\sin(\theta_0+\theta^{(1)}_{e/o})\text{RHS}[\text{\cref{eq:ThetaEoMModifiedContinuous}}]\right\rangle_T\\
% (1-m(x)^2)f_{e/o}(x)&=\SignM{\stkout{\ensuremath{\alpha\left(-(1-m(x)^2)\omega_{\text{rot}}\frac{m(x)}{|m_0|}+2m(x)\sqrt{1-m(x)^2}\langle \dot{\phi}^{(1)}_{e}(x)\theta^{(1)}_{e}(x)\rangle\right)}}}\\
% &+\left\langle\text{RHS}[\text{\cref{eq:PhiEoMModifiedContinuous}}]_{\text{products of first order terms}}\right\rangle_T,
% \end{aligned}
% \end{equation*}
\begin{equation*}
\begin{aligned}
g_{e/o}(x)&=\left\langle\sin(\theta_0+\theta^{(1)}_{e/o})\text{RHS}[\text{\cref{eq:ThetaEoMModifiedContinuous}}]\right\rangle_T\\
(1-m(x)^2)f_{e/o}(x)&=\left\langle\text{RHS}[\text{\cref{eq:PhiEoMModifiedContinuous}}]_{\text{products of first order terms}}\right\rangle_T,
\end{aligned}
\end{equation*}
Here, we used $m=\cos(\theta_0(x))$ and $m_0=\cos(\theta_0(x\gg\xi_0))$. $g(x),f(x)$ are both odd functions of $x$. Note also that we multiplied \cref{eq:ThetaEoMModifiedContinuous} with $\sin(\theta_{e/o})$ before time averaging so that the resulting $(1-m^2)$ prefactor of the $\Phi'$ term matches the one in \cref{eq:PhiEoMModifiedContinuous}. We define $\theta^{(2)}_{e/o}(x)$ such that it balances $g_{e/o}(x)$ in \cref{eq:eqThetaLinear}, hence $\theta^{(2)}_{e/o}(x)$ is also an odd function of $x$. Note also that $f_{e/o}(x)$ is bounded to the region of the domain wall, i.e. $f_{e/o}(|x|\gtrsim \xi_0)=0$. This is because in the $|x|\gtrsim\xi_0$ regions, the $\wrot$ term perfectly cancels out all the time-averaged products of first order fields, as explained in App.~\ref{app:GostoneModeActivation}.

\subsubsection{Solving \cref{eq:eqThetaLinear,eq:eqPhiLinear} in the Absence of Damping}\label{subsubsec:DampingAbsent}
For future reference, we present first how to solve \cref{eq:eqThetaLinear,eq:eqPhiLinear} in the absence of damping, $\alpha=0$. The equations read
\begin{align}
(1-m^2)\Bigg(-v(\Phi^{\alpha=0}_{e/o})'+\wtilde &+|\wrot|\frac{m}{|m_0|}+f_{e/o}\Bigg)\nonumber\\
&=\sqrt{1-m^2}\Bigg[\nonumber\\
&\delta_4\left(2m^2(m_0^2-2m^2)+3m^2-m_0^2\right)\Theta^{\alpha=0}_{e/o}\nonumber \\
&+a^2 J \Bigg(-(\Theta^{\alpha=0}_{e/o})''\left(\Delta(1-m^2)+m^2\right)\nonumber\\
&+2(\Delta-1)(\Theta^{\alpha=0}_{e/o})'m'm\nonumber\\
&+(\Delta -1)\Theta^{\alpha=0}_{e/o}\left(2mm''+\frac{(m')^2}{1-m^2}\right)\Bigg)\Bigg],\label{eq:eqThetaLinearNoDamping}\\
Ja^2\partial_x\left((1-m^2)(\Phi^{\alpha=0}_{e/o})'\right) &-v\partial_x\left(m-\left(\sqrt{1-m^2}\Theta^{\alpha=0}_{e/o}\right)\right)=0.\label{eq:eqPhiLinearNoDamping}
\end{align}
\cref{eq:eqPhiLinearNoDamping} can be directly integrated over space to give
\begin{equation}\label{eq:PhiDashNoDamping}
    (1-m^2)(\Phi^{\alpha=0}_{e/o})'=\frac{1}{v}(1-m_0^2)\wtilde+\frac{v}{Ja^2}\left(m-\sqrt{1-m^2}\Theta^{\alpha=0}_{e/o}\right),
\end{equation}
where $\frac{1}{v}(1-m_0^2)\wtilde$ was added as constant of integration. How did we know to add this particular constant? Far enough away from the domain wall, $|x|\gtrsim \xi_0$, we enter back into the ferrimagnetic chain regime, where $\Theta^{\alpha=0}_{e/o}(x)$ and its derivatives vanish. Consequently, the RHS of \cref{eq:eqThetaLinearNoDamping} vanishes, as does $f_{e/o}$, and the equation simplifies to
\begin{equation*}
    (1-m_0^2)\left(-v(\Phi^{\alpha=0}_{e/o})'+\wtilde+|\wrot|\frac{m}{|m_0|}\right)=0.
\end{equation*}
Substituting in \cref{eq:PhiDashNoDamping}, we see that we have two kinds of terms: $m$-dependent and $m$-independent. Balancing the $m$-dependent terms, we have $\frac{v^2m}{Ja^2}=(1-m_0^2)|\wrot|\frac{m}{|m_0|}$, giving us our velocity formula at zero damping, $v^2=Ja^2(1-m_0^2)\frac{|\wrot|}{|m_0|}$. Meanwhile, the remaining $m$-independent terms give us the constant of integration $\frac{1}{v}(1-m_0^2)\wtilde$.

For future reference, we also state the uncoupled EoM for $\Theta^{\alpha=0}_{e/o}$, obtained by substituting \cref{eq:PhiDashNoDamping} into \cref{eq:eqThetaLinearNoDamping},
\begin{align}
(m_0^2-m^2)\Bigg(\wtilde &+|\wrot|\frac{m}{|m_0|}\Bigg)+(1-m^2)f_{e/o}\nonumber\\
&=\sqrt{1-m^2}\Bigg[\Theta^{\alpha=0}_{e/o}\bigg(\delta_4\left(2m^2(m_0^2-2m^2)+3m^2-m_0^2\right)-\frac{v^2}{Ja^2}\bigg)\nonumber\\
&+a^2 J \Bigg(-(\Theta^{\alpha=0}_{e/o})''\left(\Delta(1-m^2)+m^2\right)+2(\Delta-1)(\Theta^{\alpha=0}_{e/o})'m'm\nonumber\\
&+(\Delta -1)\Theta^{\alpha=0}_{e/o}\left(2mm''+\frac{(m')^2}{1-m^2}\right)\Bigg)\Bigg].\label{eq:eqThetaLinearNoDampingDecoupled}
\end{align}
The source term on the LHS of \cref{eq:eqThetaLinearNoDampingDecoupled} vanishes for $|x|\gtrsim\xi_0$, consequently so does $\Theta^{\alpha=0}_{e/o}(|x|\gtrsim \xi_0)$ and its derivatives.

\subsubsection{Solving \cref{eq:eqThetaLinear,eq:eqPhiLinear} in the Presence of Damping: Far-Field Asymptotics}\label{subsubsec:FarFieldBehaviourThetaPhi}

Let us consider \cref{eq:eqThetaLinear,eq:eqPhiLinear} in the far-field limit, $|x|\gg\xi_0$. In this limit, $m\to \pm|m_0|$ and $m',m''\to 0$, and the equations simplify to
\begin{align}
(1-m_0^2)\Bigg(-v\Phi'_{e/o}+\wtilde &+|\wrot|\frac{m}{|m_0|}\Bigg)+\alpha v\sqrt{1-m_0^2}\Theta_{e/o}'\nonumber\\
&=\sqrt{1-m_0^2}\Bigg[2\delta_4m_0^2(1-m_0^2)\Theta_{e/o}-a^2 J\Theta''_{e/o}\left(\Delta(1-m_0^2)+m_0^2\right)\Bigg],\label{eq:eqThetaLinearFarField}\\
\left(\alpha v+Ja^2\partial_x\right) & \left((1-m_0^2)\Phi'_{e/o}\right)+v\sqrt{1-m_0^2}\Theta_{e/o}'\nonumber\\
&=\alpha (1-m_0^2)\left(\wtilde+|\wrot|\frac{m}{|m_0|}\right)\label{eq:eqPhiLinearFarField},
\end{align}
We can solve these by making the following ansatz
\begin{equation}\label{eq:FarFieldPhiTheta}
    (1-m_0^2)\Phi_{e/o}'=\frac{1}{v}(1-m_0^2)\left(\wtilde+|\wrot|\frac{m}{|m_0|}\right)+\Psi_f e^{\kappa x},\qquad \Theta_{e/o}=\Theta_fe^{\kappa x},
\end{equation}
which reduces the coupled equations \cref{eq:eqPhiLinearFarField,eq:eqThetaLinearFarField} to a $2\times 2$ matrix equation,
\begin{equation}
\begin{aligned}\label{eq:matrixEquation}
\begin{pmatrix}
A & B \\
C & D
\end{pmatrix}
\begin{pmatrix}
\Psi_f \\ \Theta_f
\end{pmatrix}&=\begin{pmatrix}
0\\0
\end{pmatrix},
\end{aligned}
\end{equation}
with
\begin{equation*}
\begin{aligned}
A&=-v\sqrt{1-m_0^2},\\
B&=\bigg(\alpha v\kappa-2\delta_4m_0^2(1-m_0^2)+a^2J\kappa^2\left(\Delta(1-m_0^2)+m_0^2\right)\bigg), \\
C&=\left(\alpha v+Ja^2\kappa\right)\sqrt{1-m_0^2},\\
D&=\kappa v.
\end{aligned}
\end{equation*}
Non-trivial solutions of \cref{eq:matrixEquation} occur only when the determinant of the matrix is zero, which gives us a cubic equation in $\kappa$. We are interested in the limit where $v\ll 1$. In this limit, two of the roots tend to a finite value of the same order of magnitude as the inverse length of the domain wall width, $1/\xi_0$, i.e., they decay very quickly away from the centre of the domain wall and are therefore irrelevant in the far field. On the other hand, the third root, given by $\kappa=-\frac{\alpha v}{Ja^2}+\mathcal{O}(\alpha v^3)\ll 1/\xi_0$, is much smaller than the width of the domain wall, and therefore persists a large distance beyond it. This is the root giving the non-trivial far-field behaviour. We now wish to obtain a relation between $\Psi_f$ and $\Theta_f$. Using that $\kappa,\kappa^2$ are much smaller than $J,\delta_2,\delta_4,m_0$, we set them to zero in the first line of \cref{eq:matrixEquation}, giving
\begin{equation}\label{eq:PsiFThetaF}
    v\Psi_f=-2\delta_4m_0^2\sqrt{1-m_0^2}\Theta_f.
\end{equation}
Thus we have $\Theta_f\sim v\Psi_f$. We will use this result in App.~\ref{appB:finiteAlphaSolution}.

\subsubsection{Solving \cref{eq:eqThetaLinear,eq:eqPhiLinear} in the Presence of Damping: Near-Field Approximate Solution}\label{appB:finiteAlphaSolution}
We are finally ready to approximately solve  \cref{eq:eqThetaLinear,eq:eqPhiLinear} in the near field in the presence of damping. Combining the insights gained in App. \ref{subsubsec:DampingAbsent} and \ref{subsubsec:FarFieldBehaviourThetaPhi}, we use the following ansatzes for $\Phi'_{e/o}(x),\Theta_{e/o}(x)$
\begin{equation}\label{eq:ansatzPhiThetaFiniteAlpha}
    \begin{aligned}
        (1-m^2)\Phi'_{e/o}&=\frac{1}{v}(1-m_0^2)\left(\wtilde+|\wrot|\frac{m}{|m_0|}\right)-\frac{v}{Ja^2}\sqrt{1-m^2}\Theta^{\alpha=0}_{e/o}+e^{\kappa x}\psi_{e/o}(x), \\
        \Theta_{e/o}&=\Theta^{\alpha=0}_{e/o}+e^{\kappa x}\vartheta_{e/o}(x).
    \end{aligned}
\end{equation}
where 
\begin{equation}\label{eq:boundaryBehaviourThetaPhiDW}
\psi_{e/o}(|x|\gtrsim \xi_0)=
\begin{cases}
\Psi_f,& \kappa x<0, \\ 
0, & \kappa x>0.
\end{cases}, \quad
\vartheta_{e/o}(|x|\gtrsim \xi_0)=
\begin{cases}
\Theta_f,& \kappa x<0, \\ 
0, & \kappa x>0.
\end{cases}
\end{equation}
Ansatz \cref{eq:ansatzPhiThetaFiniteAlpha} replicates the correct behaviour \cref{eq:FarFieldPhiTheta} in the far field regions. Inside the domain wall, $-\xi_0\lesssim x\lesssim \xi_0$, $\Theta^{\alpha=0}_{e/o}$ balances the source terms which do not explicitly depend on $\alpha$, according to \cref{eq:eqThetaLinearNoDampingDecoupled}. The job of balancing the additional $\alpha$-dependent source term $-\alpha vm'$ on the LHS of \cref{eq:eqThetaLinear} falls to $e^{\kappa x}\psi_{e/o}(x),e^{\kappa x}\vartheta_{e/o}(x)$. Inside the domain wall, $\psi_{e/o}(x)$ and $\vartheta_{e/o}(x)$ are functions which interpolate between $0$ and $\Psi_f$, $\Theta_f$, respectively. As a considerable simplification, we approximate this interpolation to be linear in $x$, 
\begin{equation}\label{eq:PsiVarthetaLinearInterpolation}
\psi_{e/o}(x)=\frac{\Psi_f}{2}\left(1+\sign(v)\frac{x}{\xi_0}\right), \quad \vartheta_{e/o}(x)=\frac{\Theta_f}{2}\left(1+\sign(v)\frac{x}{\xi_0}\right).
\end{equation}
Substituting \cref{eq:PsiVarthetaLinearInterpolation} into \cref{eq:eqThetaLinear} and setting $x=0$, we find
\begin{equation*}
-\frac{v\Psi_f}{2}+\alpha v\left(-\frac{m_0}{\xi_0}+\frac{\Theta_f}{2}\left(\kappa+\frac{\sign(v)}{\xi_0}\right)\right)=\frac{\Theta_f}{2}m_0^2\left(-\delta_4+a^2J(\Delta-1)\frac{1}{\xi_0^2}\right).
\end{equation*}
Using that $\Theta_f\sim v\Psi_f$, we can neglect the term proportional to $\alpha v\Theta_f$ on the LHS, as it is suppressed by a factor $\alpha v\ll 1$ compared to the other terms containing a factor $\Psi_f$. Eliminating $\Theta_f$ using the relation \cref{eq:PsiFThetaF}, we solve for $\Psi_f$,
\begin{equation}
\Psi_f=\frac{2\alpha m_0}{\xi_0}\left[\frac{a^2J(\Delta-1)\frac{1}{\xi_0^2}-\delta_4}{2\delta_4\sqrt{1-m_0^2}}-1\right]^{-1}.
\end{equation} 
Thus, $\Psi_f$ is \emph{independent} of $v$. $\Psi_f$ is however proportional to $\alpha$, which means that the exponential tail $\Psi_fe^{\kappa x}$ vanishes in the limit of zero damping, $\alpha=0$, as expected.
\subsubsection{Evaluation of $v$}
To calculate $v$, we proceed as before by integrating \cref{eq:eqPhiLinear} with the ansatzes \cref{eq:ansatzPhiThetaFiniteAlpha,eq:PsiVarthetaLinearInterpolation} over space. It is sufficient to do this between $-\xi_0\lesssim x\lesssim \xi_0$, as all terms in the equation vanish outside this range. We obtain
\begin{equation*}
\begin{aligned}
    &\alpha v\Psi_f\frac{1}{\kappa}\left(\left(1-\frac{\sign(v)}{\kappa\xi_0}\right)\sinh(\kappa\xi_0)+\sign(v)\cosh(\kappa\xi_0)\right)\\
    &+\left(Ja^2\Psi_f+v\sqrt{1-m_0^2}\Theta_f\right)\left(\sinh(\kappa\xi_0)+\sign(v)\cosh(\kappa\xi_0)\right)=2m_0\left[v-\frac{Ja^2}{v|m_0|}(1-m_0^2)|\wrot|\right].
\end{aligned}
\end{equation*}
Next, we use $\kappa\xi_0\ll 1$ to approximate $\cosh(\kappa \xi_0)\approx 1,\sinh(\kappa\xi_0)\approx \kappa\xi_0$ to leading order, which gives 
\begin{equation*}
    \sign(v)\left[\left(\alpha v\xi_0 +Ja^2\right)\Psi_f+v\sqrt{1-m_0^2}\Theta_f\right]=2m_0\left[v-\frac{Ja^2}{v|m_0|}(1-m_0^2)|\wrot|\right].
\end{equation*}
Once again, we suppress the terms $\alpha v\xi_0 \Psi_f, v\sqrt{1-m_0^2}\Theta_f$, and keep only the dominant $Ja^2\Psi_f$ term. This simplifies the equation to
\begin{equation*}
v^2-\frac{Ja^2}{2m_0}\sign(v)\Psi_fv-\frac{Ja^2}{|m_0|}(1-m_0^2)|\wrot|=0,
\end{equation*}
where we multiplied everything by $v$. As $\Psi_f$ is independent of $v$, this is a quadratic equation in $v$. However, care needs to be exercised in solving it, due the sign change of the linear in $v$ term for negative vs positive $v$. For the parameter range we are interested in, $\frac{Ja^2}{2m_0}\Psi_f$ is always negative. This means that the minimum of the parabola, located at $\sign(v)\frac{Ja^2}{4m_0}\Psi_f$, is never reached. Thus, the regions $v>0$ and $v<0$ only contribute \emph{one} root each. These two roots are given by
\begin{equation}\label{eq:velocityDamped}
v=\pm\left(\sqrt{\left( \frac{\alpha \rho_s }{\xi_0 \eta}\right)^2+\frac{\rho_s |\wrot|}{|m_0|} }-  \frac{\alpha \rho_s}{\xi_0 \eta}\right),
\end{equation} 
with
\begin{equation}
    \eta=2(1-m_0^2)\left[1-\frac{\left(a^2J(\Delta-1)\frac{1}{\xi_0^2}-\delta_4\right)}{2\delta_4\sqrt{1-m_0^2}}\right]=2(1-m_0^2)+\sqrt{1-m_0^2}\left(1+\frac{m_0^2(1-\Delta)}{2 \Delta}\right).
\end{equation}
 $\eta$ interpolates smoothly between $\eta=3$ for a small FM magnetisation, $m_0\to 0$, and $\eta=0$ for $m_0\to 1$, respectively. 
Fig.~\ref{fig:velicityDW} of the main text shows that our approximate formula for $v$, while not exact, describes with a good accuracy the numerically observed velocity with an error of only a few percent for magnetic fields $B_0\lesssim 0.2\,J$. As our analysis is only valid for small  $B_0$, it cannot describe the breakdown of the moving domain wall solution for larger $B_0$.

\section{Further Numerical Results}\label{app:extraResults}
In this section, we display and discuss some extra numerical data to support the conclusions reached in the main text.
\subsection{Short-Ranged Correlations of the $xy$-Order}\label{app:extraResults_shortxyorder}

\begin{figure}
    \centering
    \includegraphics[width=0.64 \linewidth]{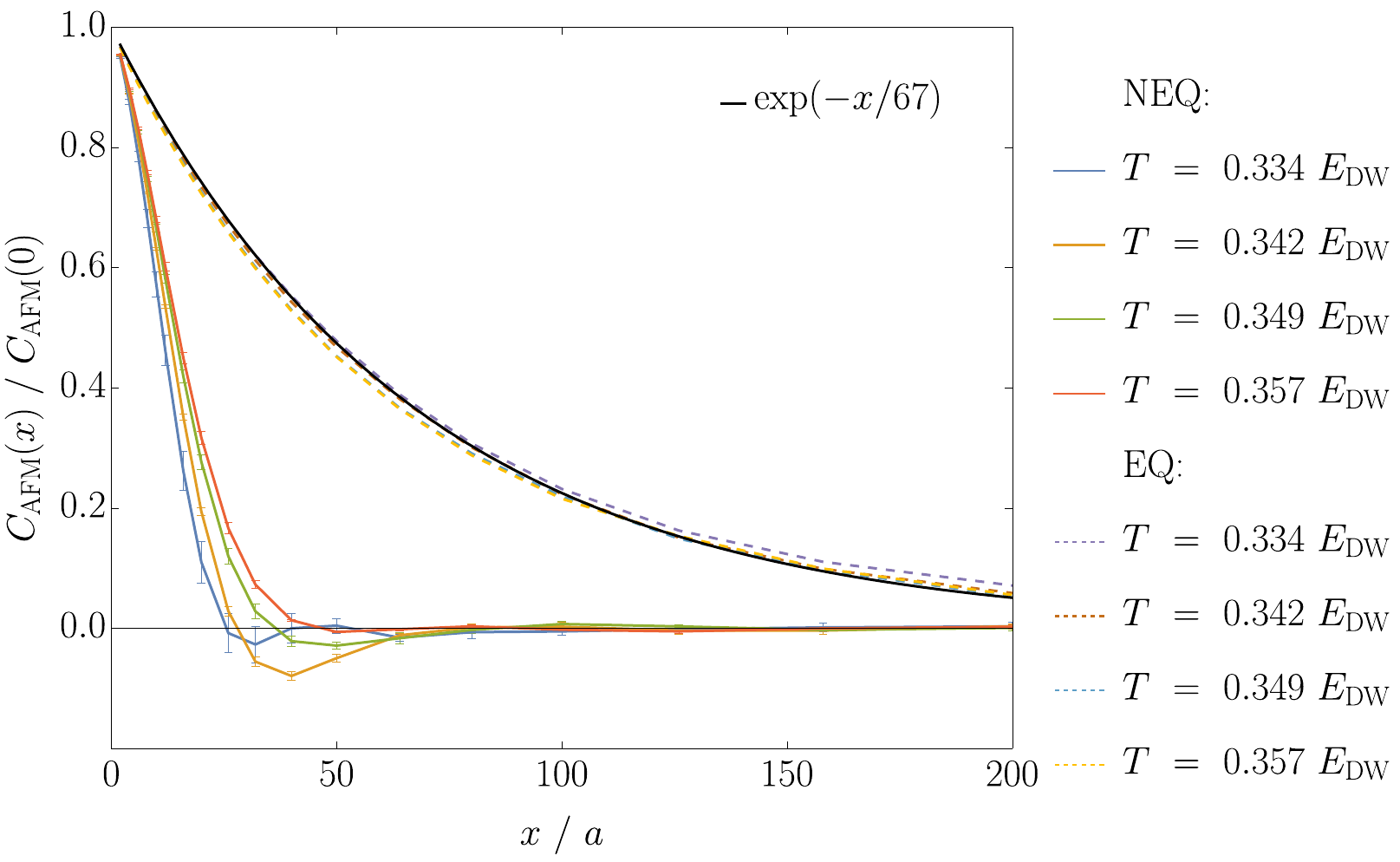}
    \caption{Decay of correlations of the AFM $xy$-order parameter, $C_\text{AFM}(x)=\sum_{j} (-1)^i\langle S^x_j S^x_{j+x/a} + S^y_j S^y_{j+x/a}\rangle$ for integer even $x/a$. Error bars denote the standard deviation of the mean. Parameters as in Fig.~\ref{fig:magNoise_app}. In the equilibrium system, $xy$-correlations decay exponentially on a length scale of about $67$ lattice sites (solid line), in the temperature range investigated here. In the driven system, the correlation length is considerably smaller, only 20-30 sites.}
    \label{fig:finiteTempCorelAFM}
\end{figure}
The discussion in the main text focuses mainly on the giant correlation length of the FM magnetisation, which is enhanced by several orders of magnitude in the driven system. In contrast, the correlation functions associated with the AFM $xy$-order decay faster in the driven system than in the equilibrium case, as shown in  Fig.~\ref{fig:finiteTempCorelAFM}. For the chosen parameters, the $xy$-order decays on a length scale of only 20 or 30 sites, even in a regime where  $m^z$-domains have sizes exceeding $10^5$ sites. The reason is that, embedded in these large $m^z$ domains, there are smaller domains with opposite $m^z$ orientation. Thus, $xy$-spins rotate in opposite directions which reduced $xy$-correlations.

Remarkably, despite this very short correlation length  of the $xy$-order, our analytical formulas for the domain wall velocities, which ignored noise and start from a system with long-ranged $xy$-order, correctly predict the velocity of the domain walls.

\subsection{Dynamics of Ordering}
\label{App:DynamicsofOrdering}

\begin{figure}
    \centering
    \includegraphics[width=0.5\linewidth]{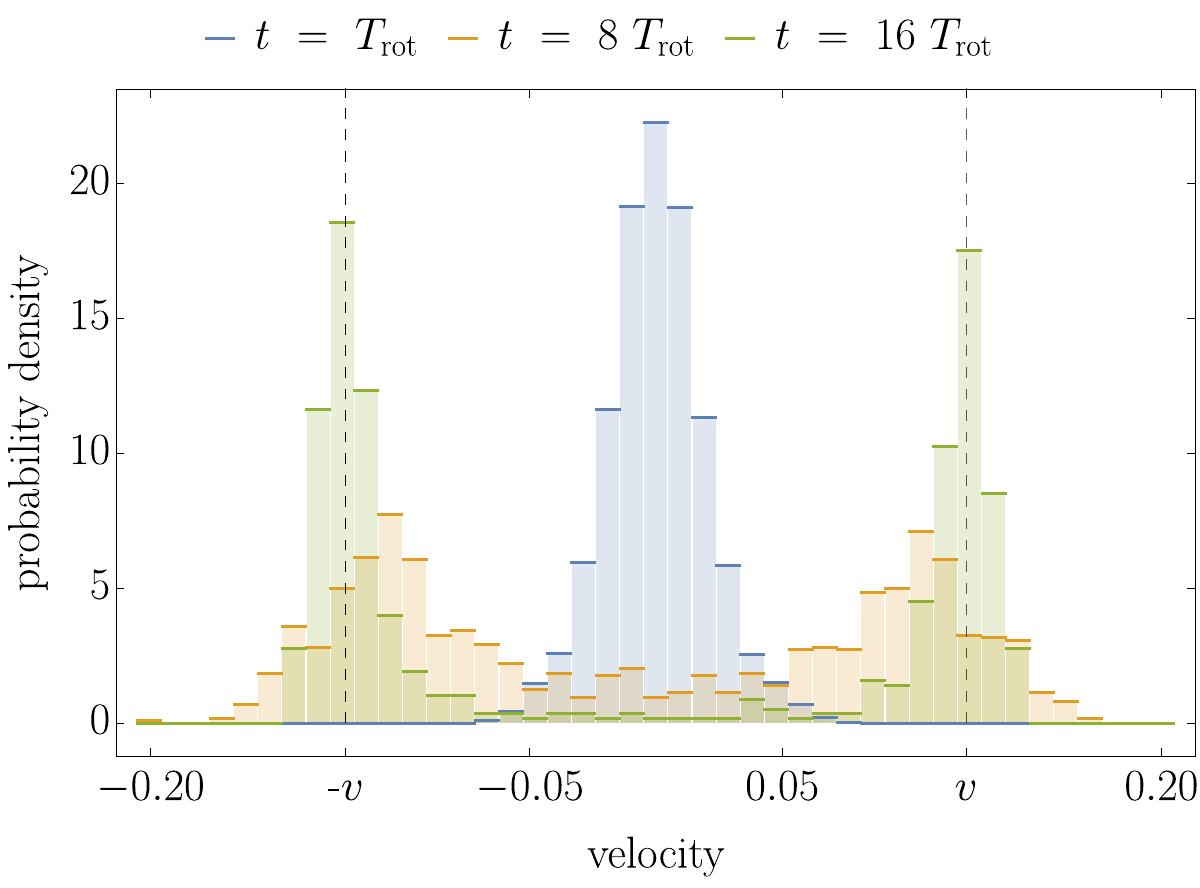}
      \caption{ Distribution of domain wall velocities after a quench from an AFM ordered state into a ferrimagnetic state for three different times after the quench. For long times, the velocity distribution is strongly peaked at the velocity of a single domain wall, $\pm v$. Same parameters as Fig.~\ref{fig:magNoise_app} at $T=0$ for a system of 40,000 spins. }\label{fig:velocityDist}
\end{figure}

Here, we give more information on how long-ranged $m^z$ order develops in the absence of thermal noise after a quench from the AFM phase. This physics is also discussed in Fig.~\ref{fig:velocityDist_and_correlationlength}
of the main text, which shows that at first the correlation length grows very slowly, then speeds up and grows linearly in time for long times. The initial slow growth is explained by the fact that the domain walls need a few rotation periods $T_\text{rot}$ to reach their final velocity. The corresponding velocity distributions of domain walls are are shown in 
Fig.~\ref{fig:velocityDist} for three different times. While for short times, $t \sim T_\text{rot}$, the velocity distribution is centered around zero, the velocity distribution becomes peaked at $\pm v$ in the long-time limit. An analytical theory for $v$ is provided in Sec.~\ref{app:DomainWallVelocity}.

\subsection{Quench from the Ferrimagnetic Phase and Convergence of Results} \label{app:ferriquench}
\begin{figure}
    \centering
    \includegraphics[width=0.64\linewidth]{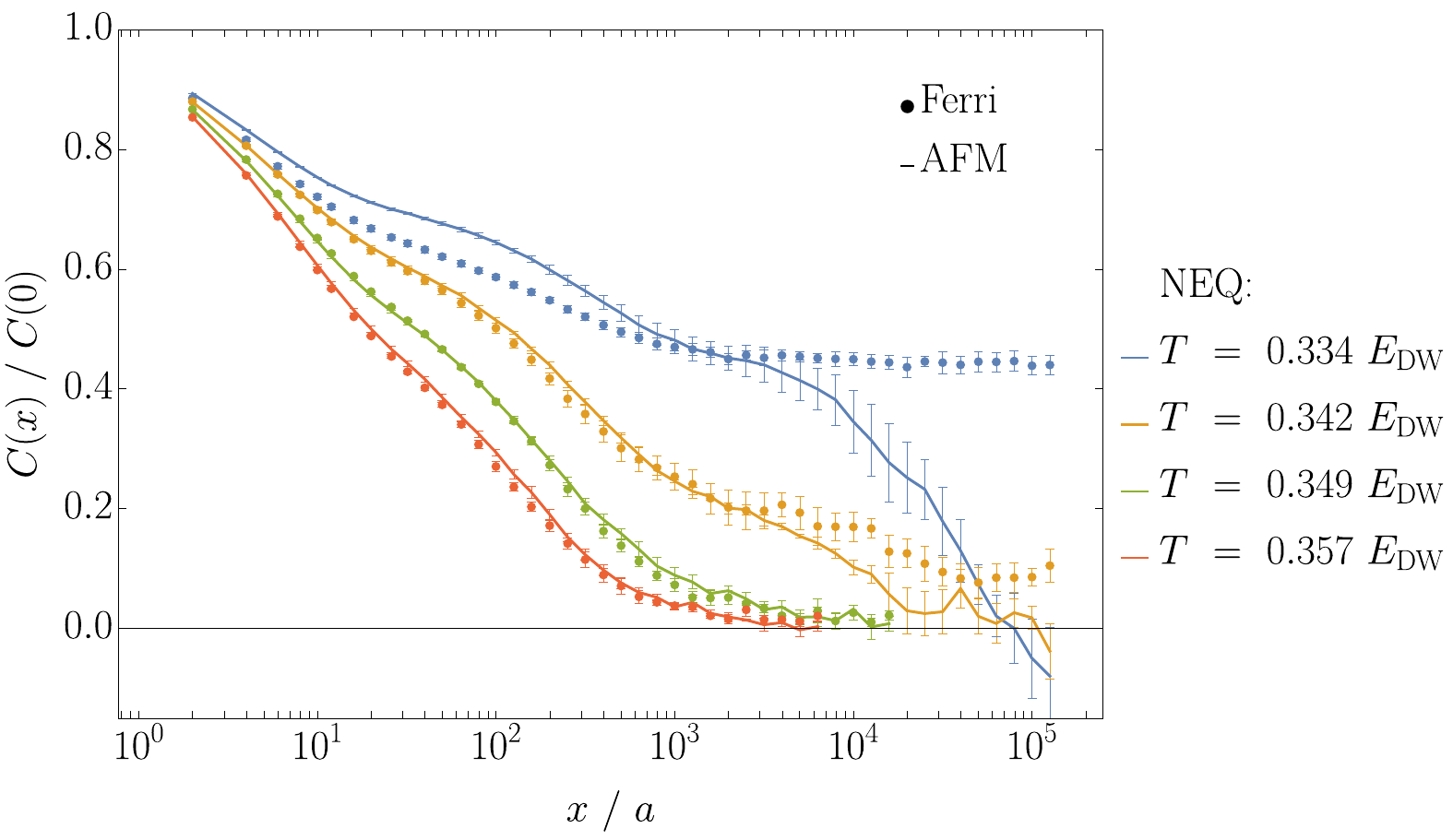}
      \caption{ 
   Comparison of the FM correlation function $C(x)=\langle S^z_i S^z_{i+x/a}\rangle $ measured a time  $t\approx 520{,}000 / J \approx 850 \,T_\text{rot}$ after the quench, for four different temperatures, for a system with 250,000 spins. The solid lines show data from a quench from an initial AFM state (also shown in Fig.~\ref{fig:finiteTempCorel} of the main text). We compare this to data obtained by starting from a perfectly ordered ferrimagnetic state, shown as points. The plot shows that for the three highest temperatures, the correlation function is independent of the initial state within error bars. For the lowest $T$ (blue) this is not the case: the simulated time is shorter than the time required to reach a steady state. Also, the system size is smaller than the correlation length in this case.
   Parameters: As in Fig.~\ref{fig:magNoise_app}. For both initial conditions, averaging was performed over four simulations and error bars show standard deviations of the mean.
   }\label{fig:ferriQuench}
\end{figure}

Our driven system shows order on very long length scales. Therefore the equilibration time needed to reach a steady state also becomes very long.
To investigate whether the finite $T$ correlation functions discussed in the main text reflect steady-state properties, we compare data arising from two initial states in Fig.~\ref{fig:ferriQuench}. The first one, 
a perfectly ordered AFM state with a small randomness in the initial value of $S_i^z=\pm 0.1$, initially produces a very large number of short domains. This is compared to a system which starts from a perfectly ordered ferrimagnetic state, consisting of a single domain which includes all 250,000 spins of our simulation. In the first case, the motion and collisions of the domain walls increases the correlation length until a steady state is reached in the noisy system. In the second case, the noise has to induce more and more defects to reach the same steady state for $t\to \infty$. Fig.~\ref{fig:ferriQuench} shows that for the higher temperatures the system does reach the same steady state within the simulated time. This is, however, not the case for the lowest simulated temperature.

An important characteristic of the correlation functions shown in Fig.~\ref{fig:ferriQuench} is an extremely slow decay of $C(x)$, roughly following \begin{align}\label{eq:logfit}
    C(x)\approx C(0)\left(1-\ln\left[\frac{x}{a}\right]/\ln\left[\frac{\xi_\text{max}}{a}\right]\right).
\end{align}
The slow decay with $x$ arises because there is a very broad distribution of domain sizes. Furthermore, the system is characterised by domains within domains.
A consequence of this fact is that the half-width $\xi_{1/2}$ of $C(x)$, defined by $C(\xi_{1/2})=\frac{1}{2} C(0)$,
{\em can not} be used as a useful measure of domain sizes. Consider, for example, Fig.~\ref{fig:magNoise} of the main text. For the two temperatures shown, $\xi_{1/2}\approx 615$ for $T=0.0334 E_\text{DW}$ and $\xi_{1/2}\approx 35$ for $T=0.349 E_\text{DW}$ as can be read off from Fig.~\ref{fig:ferriQuench}. However, a short inspection of Fig.~\ref{fig:magNoise} of the main text reveals that the true correlations are orders of magnitude larger.
To obtain a more reasonable correlation length and error bars for the inset of Fig.~\ref{fig:finiteTempCorel} of the main text, we proceed in the following way. We first use a fit to Eq.~\eqref{eq:logfit} to determine $\xi_\text{max}$. For the fit, we use values of $x/a$ smaller than $8{,}000$, $20{,}000$, $1{,}000$ and $300$ for $T=0.334, 0.342, 0.349$ and $0.357 \,E_\text{DW}$.
The small cutoff value for the lowest temperature takes into account that the system has not equilibrated at long distances, see Fig~\ref{fig:ferriQuench}. The crudeness of the fit is reflected in generous error bars, which we choose as follows. A lower estimate $\xi^<_\text{max}$ is obtained from $C(\xi^<_\text{max})=0.1\, C(0)$. We estimate $\xi^>_\text{max}$ from the point where $C(x)\approx 0$ within the error bars. As the lowest $T$ (blue curve) is not equilibrated, we can only estimate $\xi^>_\text{max}=\infty$.
With this procedure, we obtain the  estimates shown in Table \ref{tab:corr}.
\begin{table*}
\centering
\begin{tabular}{r|r|r|r}
   $T/E_\text{DW}$  &  $\xi^<_\text{max}/a$ & $\xi_\text{max}/a$ & $\xi^>_\text{max}/a$ \\ \hline
   0.334 & 45,200 & 475,000 & $\infty$  \\
  0.342 & 10,400 & 18,100 & 108,000  \\
  0.349 & 800 & 1,800 & 13,600  \\
  0.357 & 400 & 700 & 5,900  \\
 \end{tabular}\vspace{2mm}
 \caption{Estimates for the correlation length used for the plot in the inset of Fig.~\ref{fig:finiteTempCorel} of the main text. }\label{tab:corr}
\end{table*}

\subsection{Correlations in space-time}\label{app:spacetime}
\begin{figure}
    \centering
    \begin{tikzpicture}
    \draw (0, 0) node[inner sep=0] (a) {
    \includegraphics[height=0.22\linewidth]{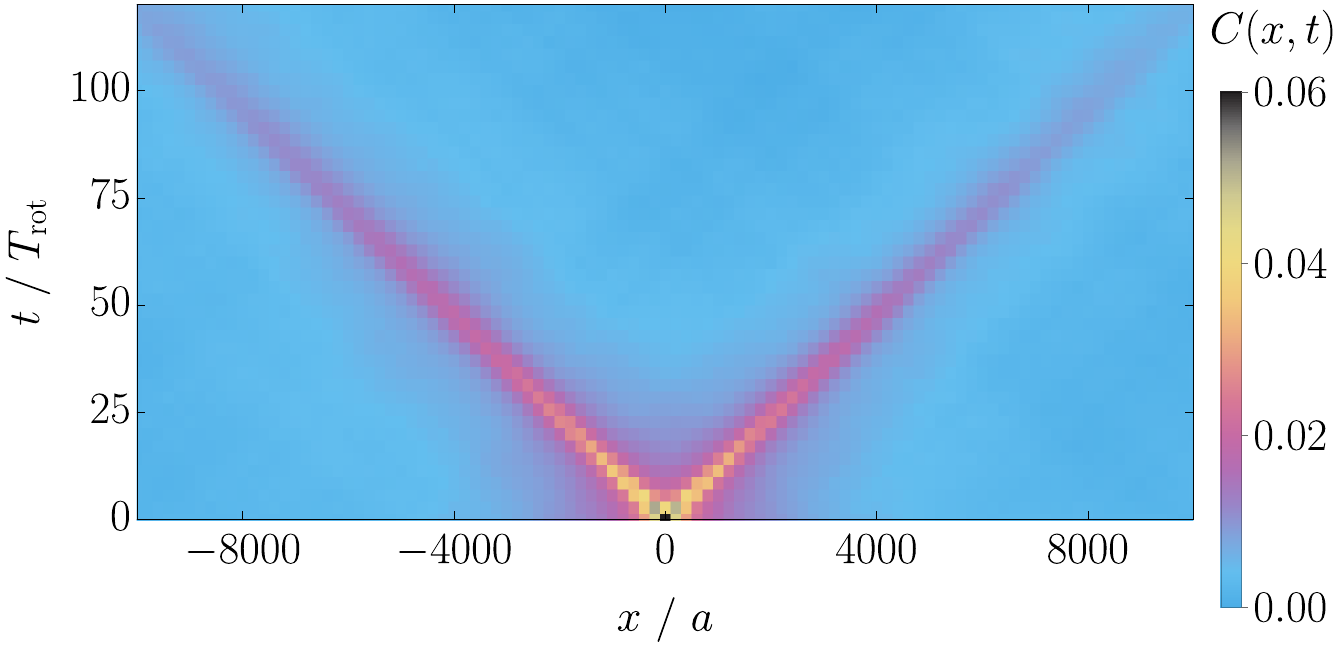}};
    \node[
      above left=-0.5cm and -0.35cm of a] {\large a)};
    \end{tikzpicture}\hfill
    \raisebox{-0.08 cm}{
    \begin{tikzpicture}
    \draw (0, 0) node[inner sep=0] (b) {\includegraphics[height=0.2265\linewidth]{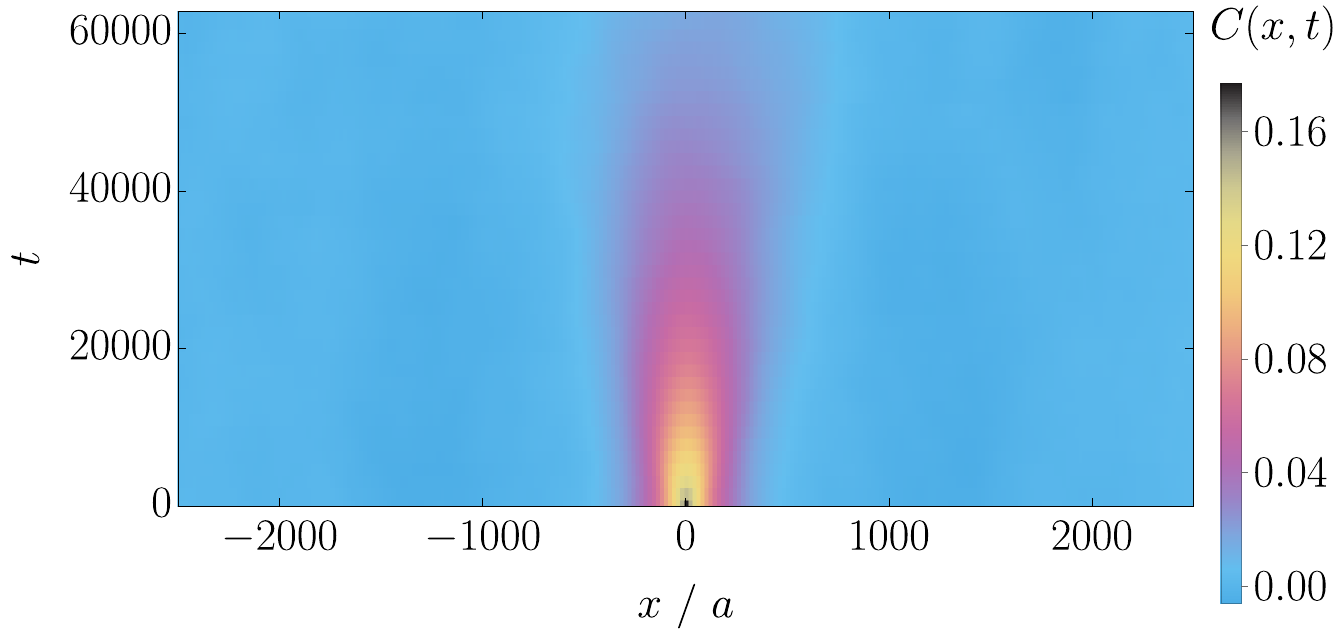}};
    \node[
      above left=-0.5cm and -0.35cm of b] {\large b)};
    \end{tikzpicture}
    }
      \caption{Panel a): Steady-state correlation function $C(x,t)=\left\langle S^z_i(t_0)S^z_{i+x/a}(t_0+t)\right\rangle$ of a driven ferrimagnet as a function of space and time in the presence of thermal noise with  $T=0.349 \, E_\text{DW}$, other parameters as in Fig.~\ref{fig:magNoise_app}.
The data has been averaged over $t_0$ and 5 independent runs. Various cuts through this function are shown in Fig.~\ref{fig:2Dcorrcuts}. \NEWWW{Panel b): Steady-state correlation function for a system in thermal equilibrium. To obtain a similar correlation length, we use a substantially lower temperature, $T=0.160 \, E_\text{DW}$ (system size: $L=40,000$, $B_0=0$ and averaged over 4 runs, other parameters as in panel a)). For large correlation length, the equilibrium system shows diffusive behaviour, see text. Both panel a) and b) show the dynamics on similar time scales ($T_\text{rot}\approx 612$).} }
\label{fig:2Dcorrheatmap}
\end{figure}

Even in the presence of thermal noise, the spatial and temporal correlations of the FM magnetisation
\begin{align}
    C(x,t)=\left\langle S^z_i(t_0)S^z_{i+x/a}(t_0+t)\right\rangle
\end{align}
(averaged over $i$ and a few $t_0$ after the system has equilibrated) are dominated by the fact that domain walls move actively with the velocity $\pm v$. 
This is clearly visible in Fig.~\ref{fig:2Dcorrheatmap}a), which shows the correlation function in the steady state. The correlation function shows pronounced peaks at $x=\pm v t$.

\begin{figure}
    \centering
    \begin{tikzpicture}
    \draw (0, 0) node[inner sep=0] (a) {
    \includegraphics[width=0.46\linewidth]{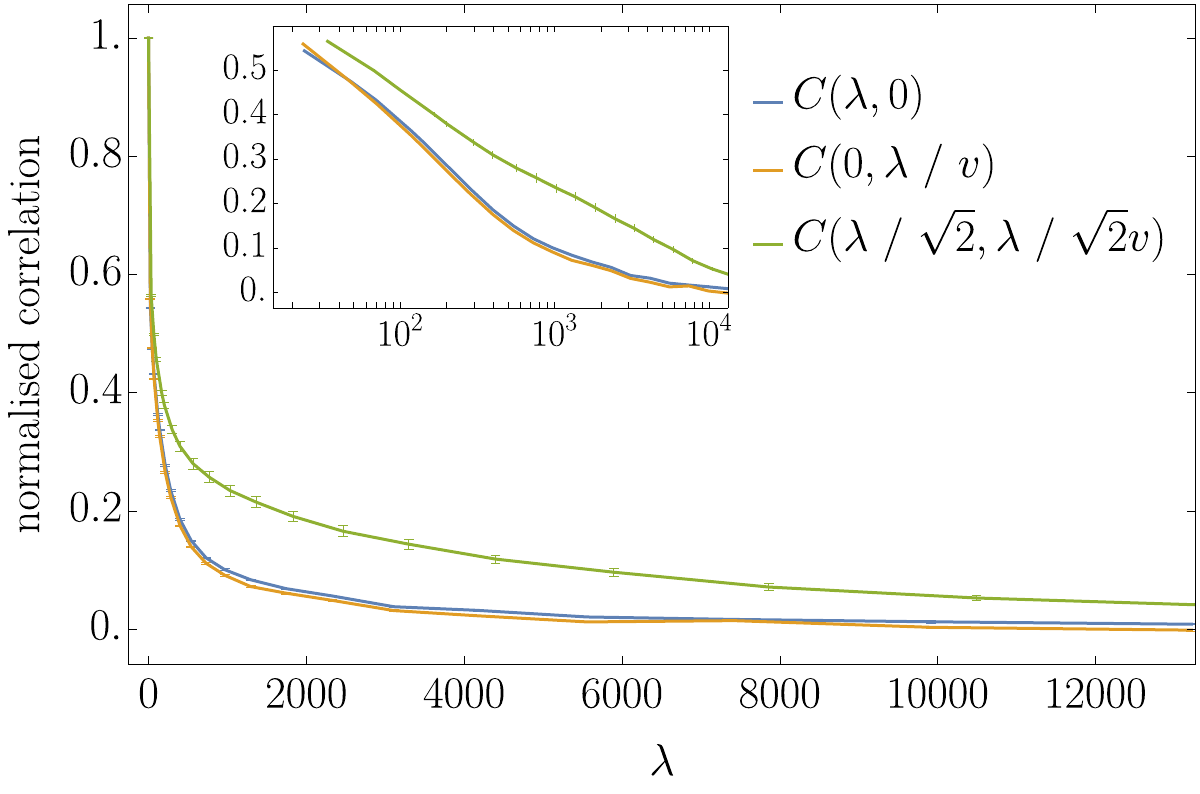}};
    \node[
      above left=-0.5cm and -0.35cm of a] {\large a)};
    \end{tikzpicture}\hfill
    \begin{tikzpicture}
    \draw (0, 0) node[inner sep=0] (b) {\includegraphics[width=0.46\linewidth]{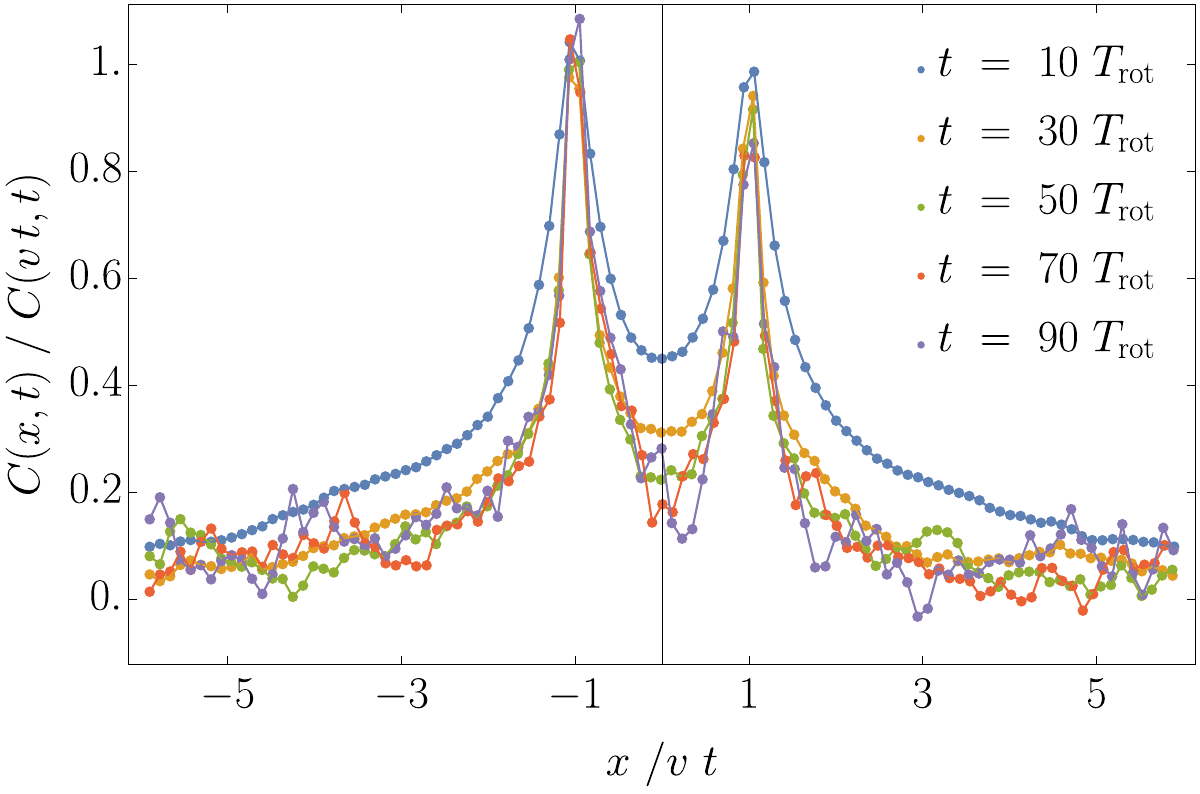}};
    \node[
      above left=-0.5cm and -0.35cm of b] {\large b)};
    \end{tikzpicture}
       
      \caption{Cuts  of the steady-state correlation function $C(x,t)$ shown in Fig.~\ref{fig:2Dcorrheatmap}, in the presence of thermal fluctuations ($T=0.349 \, E_\text{DW}$). Panel a) shows that the equal time correlation function $C(\lambda,t=0)$ is within errors indistinguishable from the local correlation function $C(x=0,t)$ for $t=\lambda/v$. Here, the average velocity of domain walls, $v\approx 0.135 \, Ja$, is about $10\,\%$ faster than the velocity of a domain wall in the noiseless system. A much slower decay of correlations is obtained along the diagonal of  Fig.~\ref{fig:2Dcorrheatmap}, i.e., for $C(\lambda/\sqrt{2},\lambda v/\sqrt{2})$ (green curve, panel a), see also inset which shows the same data with a log scale for $\lambda$), $\lambda=\sqrt{x^2+(v t)^2}$. For all three curves, $\lambda$ measures the distance from the origin, with $\lambda^2=x^2+(v t)^2$. 
      Panel b) shows $C(x,t)/C(v t,t)$ as function of $x/(vt)$. The plot shows that for long times, the width of the peak at $x=v t$ also scales linearly with $t$. }  
\label{fig:2Dcorrcuts}
\end{figure}

Remarkably, the correlation function looks practically identical when the roles of $x$ and $v t$ are exchanged. This is is shown in Fig.~\ref{fig:2Dcorrcuts}a): the $t=0$ and $x=0$ correlation functions are identical within error bars. Furthermore, the width of the peak at $x=\pm vt$ also scales linearly in $t$ for long times, see Fig.~\ref{fig:2Dcorrcuts}b).

The correlation length along the line $x=\pm v t$ is also much longer than in the time or space direction only, see Fig.~\ref{fig:2Dcorrcuts}a). To compare these correlations, it is useful to plot the correlations as function of $\lambda=\sqrt{x^2+(v t)^2}$, see Fig.~\ref{fig:2Dcorrcuts}a). When one moves a distance $\lambda$ along a line parallel to $x=v t$, one crosses no right-moving domains but only left-moving domains (with a factor $\sqrt{2} $ increased density). However, these geometric factors alone cannot explain the increased correlations along $x=\pm v t$, which are roughly a factor five longer for the chosen parameters.

\NEWWW{For comparison, in Fig.~\ref{fig:2Dcorrheatmap}b) we show the correlation function of an equilibrium system. We have reduced the temperature by a factor of 2 to obtain a similar correlation length as the one in panel a). The comparison of panels a) and b) shows the dramatically different dynamics of the two systems. In equilibrium, the dynamics
shows diffusive scaling for systems with a very large correlation length $\xi$. Over time, the spin correlations thus decay on a time scale $\tau \propto \xi^2$. From the half-width of the correlation function, we obtain $\xi\approx 96 $, $\tau \approx 1.2\cdot 10^4$.
For systems with a smaller correlation length (not shown), we also see ballistic features in the correlation function, possibly arising from a coupling to the spin-waves of the $xy$-order.}

\NEW{
\section{Active magnet in two dimensions}
\label{app:2D_results}

In the main text, we provide a comprehensive analysis of the dynamics of a one-dimensional driven ferrimagnet. In this supplement, we show that central results also apply in two dimensions. Domain walls move actively with the same speed as in the one-dimensional model, and their collective dynamics lead to rapid buildup of correlations, characterised by correlation length growing linearly in time.

We generalise the Hamiltonian given in the main text for a one-dimensional spin chain, Eq.~\eqref{eq:discreteFerrimagnetHamiltonian}, to two dimensions by mapping it to a square lattice in the following way,
\begin{equation}        
H=J\sum_{\langle i,j \rangle} \left(S^x_iS^x_j+S^y_iS^y_j-\Delta S^z_i S^z_j  \right)
        +\sum_i \left(\frac{\delta_2}{2} \left. S^z_i \right.^2 +\frac{\delta_4}{4} \left. S^z_i \right.^4 -g_i B_z(t) S^z_i\right),
\end{equation}
where $\langle i,j \rangle$ denotes the sum over nearest neighbouring sites on a square lattice and we use the notation $i=(i_x,i_y)$ to denote positions in 2D. The $g$ factors take the values $g_1$ and $g_2$ on the even and odd sublattices, with  $i_x+i_y$ even or odd, respectively. 

A quantitative difference between the 1D and the 2D model is that each site in 2D has four instead of two nearest neighbours. At the same time, a domain wall in $d$ dimensions is a $d-1$ dimensional object, and thus its codimension is always $1$.
This changes our analytical formulas for the ground-state magnetisation, the rotation frequency of the spins, and the velocity of domain walls. 

For example, the formula for the ground state magnetisation on a $d$-dimensional (hypercubic) lattice reads $m_0=\sqrt{\frac{-\delta_2-2 d J(1-\Delta)}{\delta_4}}$, where $J$ of the 1D formula is replaced by $d J$. At the same time, the formula for the spin stiffness $\rho_s$ is not modified.

\begin{figure}
    \includegraphics[width=\linewidth]{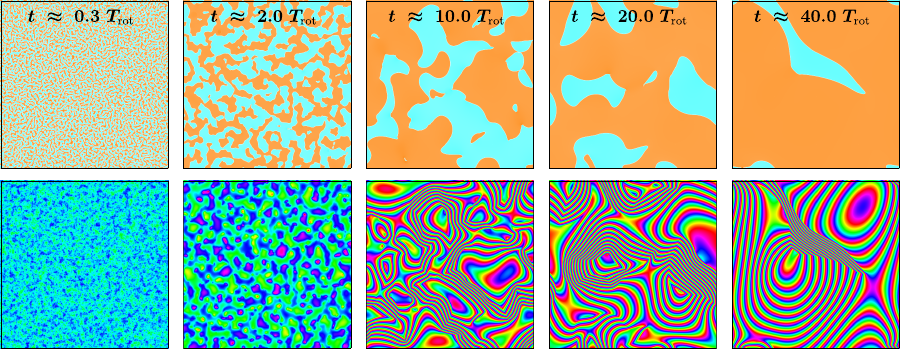}
    \caption{\NEW{Snapshots of a 2D simulation of $2{,}500\cross 2{,}500$ spins after a quench from an AFM-ordered phase to the ferrimagnetic phase at different times, see methods. The top row shows the FM magnetisation and the bottom row displays the phase. We observe the build up of phase gradients behind moving domain walls as in $1D$.
    Parameters: no noise, $J=1, \Delta=0.8,\delta_2=-2\cdot0.6,\delta_4=2\cdot1,g_1=1,g_2=0.1,\alpha=0.1,B_0=\sqrt{2}\cdot 0.15,\omega=2\cdot3.6$ resulting in a rotation period of $T_\text{rot}\approx 612$. } }
    \label{fig:2D_Sim_snapshots}
\end{figure}
To obtain results which are comparable in 1D and 2D, we proceed in the following way: We use the same values for $J$, $\Delta$ and the $g_i$, but double the numerical values for the driving frequency $\omega$ and the anisotropies $\delta_2$ and $\delta_4$, using $\omega^{2D} = 2 \cdot \omega^{1D}$, $\delta_2^{2D}=2\cdot\delta_2^{1D}$, $\delta_4^{2D}=2\cdot\delta_4^{1D}$. The amplitude of the oscillating field, in contrast, is changed only by a factor $\sqrt{2}$, $B_0^{2D} = \sqrt{2}\cdot B_0^{1D}$.

With these changes, the ground state magnetisation, the spin-stiffness and the rotation frequency $\wrot$
are unaffected, $m_0^{2D}=m_0^{1D}$, $\rho_s^{2 D} = \rho_s^{1D}$, $\wrot^{2D}=\wrot^{1D}$. Thus, according to Eq.~\eqref{eq:vFast} in the main text, the velocity of an idealised straight domain wall is also approximately the same, $v^{2D}\approx v^{1D}$, but the formula Eq.~\eqref{eq:velocityFull} of the main text changes, as the domain wall width is reduced by a factor $\sqrt{2}$ for the scaling described above.

With this at hand, we can perform quench simulations with the same initial conditions as stated in the methods section. Snapshots of such simulations are shown in Fig.~\ref{fig:2D_Sim_snapshots}, for a system of $2500 \times 2500 \approx 6\cdot 10^6$ spins. A movie of the dynamics can be found in the supplementary information.

For a more quantitative analysis of the scaling behaviour in the long-time limit, this system size is, however, insufficient, as already after about 20 rotation periods $T_\text{rot}$, magnetic domains become as large as the system size, see Fig.~\ref{fig:2D_Sim_snapshots}. We therefore performed simulations for a larger system of $10{,}000 \times 10{,}000 = 10^8$ spins. Furthermore, we increased the strength of the oscillating field to $B_0=\sqrt{2} \cdot 0.2$ to reduce the distance $v T_\text{rot}\sim 1/B_0$, which allows to simulate more rotation periods before finite-size effects set in.

In Fig.~\ref{fig:2D_corr_scaled_driven}, we show the resulting  equal-time correlation function calculated from $C(x,t)=\Bigl< S^z_j \left( S^z_{j+(x/a,1)}+S^z_{j+(1,x/a)}\right)/2\Bigr>$, where the average runs over the sites $j$ at a fixed time $t$ after the quench. 
We plot the correlation as function of $x/(v t)$, where $v$ is the velocity of domain wall motion obtained from our $1D$ simulations, see Fig.~\ref{fig:velicityDW} in the main text. 

The maximal spread of information occurs with a velocity always smaller than $2 v$, arising from two counter-propagating domain walls. After about 20 rotations, we obtain a universal curve 
\begin{align}
    C(x,t)\approx \tilde C\!\left(\frac{x}{vt}\right)\quad \text{for } t\gtrsim 20 \,T_\text{rot},
\end{align}
characterised by two length scales and an intermediate plateau.
Most importantly, our numerical result shows that as in $1D$, the length scale grows linearly in time,
\begin{align}
    \xi(t) \sim t,
\end{align}
as expected from the analytical argument given in the main text.

We have also checked that in the absence of external driving by an oscillating field, the correlations grow slower. The scaling plot of \ref{fig:2D_corr_scaled_undriven} shows that correlations grow approximately with $t^{0.75}$ in this case. This is much faster than reported previously for Ising models \cite{Domain_Growth_Ising_2D}, where $\xi \propto t^{1/2}$ has been observed, following the Allen-Cahn law \cite{AllenCahn} for domain wall coarsening. While this requires further study, we speculate that the faster growth of correlation arises from long-ranged interaction mediated by the Goldstone mode of the in-plane magnetic order.

\begin{figure}
    \centering
    \includegraphics[width=0.64\linewidth]{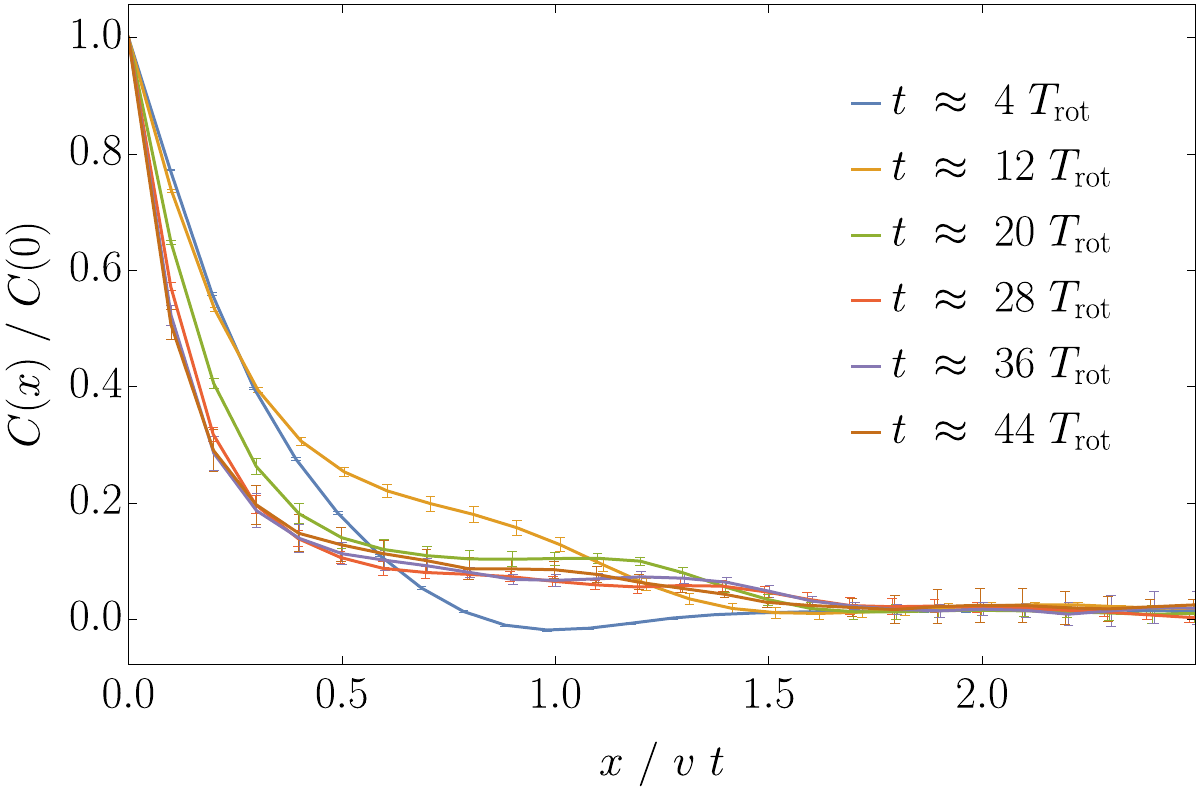}
      \caption{\NEW{Scaling plot of equal-time correlation functions after a quench, see Fig.~\ref{fig:2D_Sim_snapshots}. We show $C(x)=\Bigl< S^z_j \left( S^z_{j+(x/a,1)}+S^z_{j+(1,x/a)}\right)/2\Bigr>$  (averaged over $j=(j_x,j_y)$) for even $x/a$ as a function of $x/(vt)$ with $v\approx 0.163$ being the velocity of a straight domain wall. Parameters as in Fig.~\ref{fig:2D_Sim_snapshots} but with a higher field amplitude $B_0=\sqrt{2}\cdot 0.2$, i.e, $T_\text{rot}\approx 344.7$, averaged over 5 initial states with $10{,}000\cross10{,}000$ spins each. The error bars are the standard deviation of the mean.}}\label{fig:2D_corr_scaled_driven}
\end{figure}

\begin{figure}
    \centering
    \includegraphics[width=0.64\linewidth]{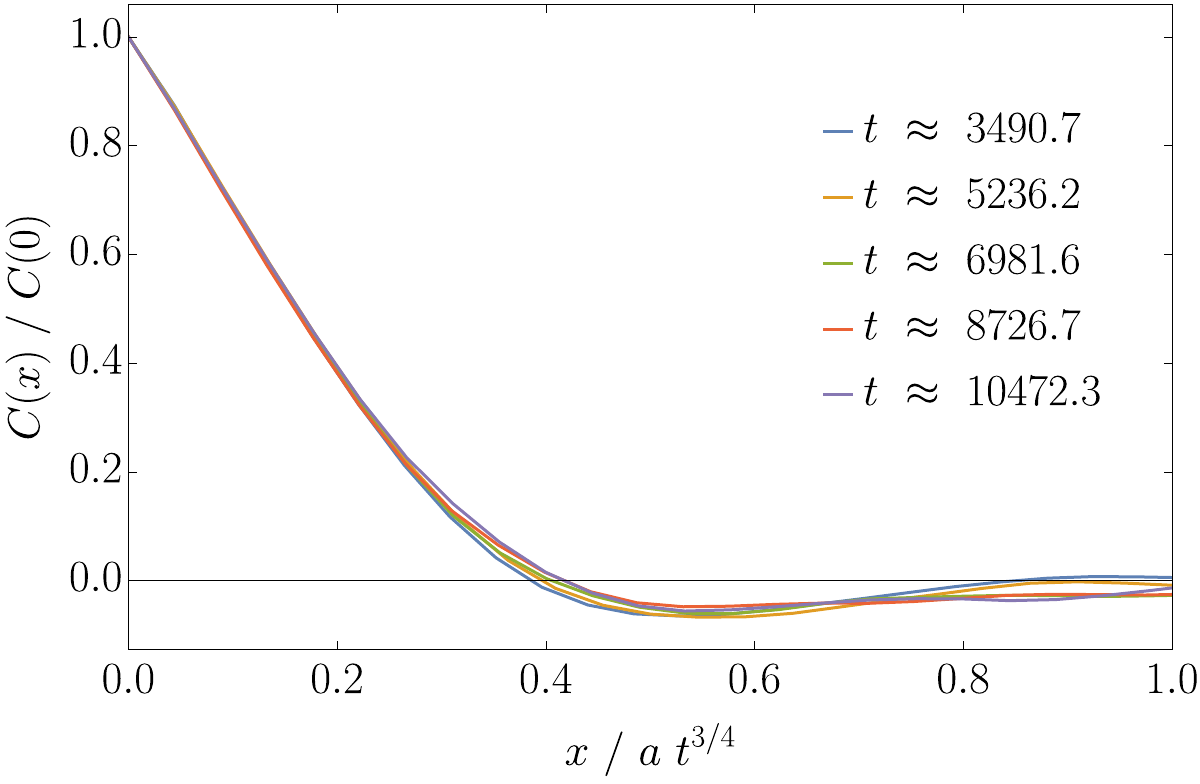}
      \caption{\NEW{Growth of correlations after a quench for an {\em non-driven} system: Scaling plot of the equal-time correlation function $C(x)=\Bigl< S^z_j \left( S^z_{j+(x/a,1)}+S^z_{j+(1,x/a)}\right)/2\Bigr>$  (averaged over $j=(j_x,j_y)$) shown for even $x/a$ as a function of $x/(at^{3/2})$. As before shown for a quench from an AFM-ordered phase into the ferrimagnetic phase in the absence of noise, $T=0$. Parameters and system size as in Fig.~\ref{fig:2D_Sim_snapshots} but without any oscillating external fields: $B_0=0$.}}\label{fig:2D_corr_scaled_undriven}
\end{figure}

}

\twocolumn



\end{document}